\documentclass[preprintnumbers,prl,twocolumn,floatfix,preprintnumbers,amsmath,amssymb,nofootinbib,superscriptaddress]{revtex4}
\usepackage{graphicx}
\usepackage{enumitem}
\usepackage{epsfig}
\usepackage{bm}
\usepackage{amsfonts}
\usepackage{color}
\usepackage{subfigure}
\usepackage{amsmath}
\usepackage{amssymb}
\usepackage{graphicx, comment}
\usepackage{makecell}
\usepackage{float}
\usepackage[normalem]{ulem}
\usepackage{xcolor}

\usepackage[utf8]{inputenc}
\usepackage[english]{babel}

\usepackage[
      colorlinks=true,
      linkcolor=blue,
      urlcolor=magenta,
      filecolor=green,
      citecolor=red,
      pdfstartview=FitV,
      pdftitle={},
        pdfauthor={Ruchika},
        pdfsubject={},
        pdfkeywords={},
        pdfpagemode=None,
        bookmarksopen=true
      ]{hyperref}
\usepackage{color}

\newcommand{\bcm}{}

\usepackage{ulem,xpatch}

\usepackage{amsmath,amssymb}

\newcommand{\Hunit}{km s$^{-1}$ Mpc$^{-1}$}

\newcommand{\Rtwo}{SH0ES'22 }

\date{Accepted XXX. Received YYY; in original form ZZZ}

\begin{document}

\title{2D BAO vs 3D BAO: Hints for new physics?}

%

\author{Ruchika}
\email{ruchika.science@usal.es}
\affiliation{Physics Department and INFN, Università di Roma “La Sapienza”, P.le Aldo Moro 2, 00185 Rome, Italy}
\affiliation{Departamento de Física Fundamental and IUFFyM, Universidad de Salamanca, E-37008 Salamanca, Spain}

\date{\today}

\begin{abstract}
As next-generation telescopes and observational surveys continue to expand the boundaries of our understanding, tensions and discrepancies between observational datasets are becoming increasingly prominent. In this work, we focus on one such discrepancy: the differences between 2D and 3D Baryon Acoustic Oscillation (BAO) measurements. Without extending beyond the standard $\Lambda$CDM framework, we systematically study and highlight this discrepancy in different parameter spaces. This work examines the constraints on fundamental cosmological parameters ($H_0$, $r_d$, $\Omega_m$) derived from Baryon Acoustic Oscillation (BAO) and Type Ia Supernovae (SNIa) data. By analyzing BAO observational datasets from two distinct methodologies (2D and 3D) alongside the Pantheon Plus SNIa sample, we identify a significant systematic difference: 2D BAO measurements consistently yield higher values of $hr_d$ compared to both 3D BAO and DESI analyses. While 2D BAO measurements appear to bridge the Hubble tension by simultaneously accommodating both a higher $H_0$ value (aligning with SH0ES) and a larger sound horizon $r_d$ (matching Planck), this apparent reconciliation comes at the cost of introducing tension with the well-constrained Planck measurement of $\Omega_{m0}h^2$. This behavior arises because of systematically higher values of the product $H_0r_d$ observed in 2D BAO analysis compared to 3D analyses. Therefore, given these systematic differences, we advocate for careful consideration when using 2D BAO measurements to address the Hubble tension, suggesting that understanding the origin of this 2D-3D discrepancy should be a priority for future investigations.
\end{abstract}



\maketitle
\section{Introduction}\label{sec1:intro}
Since the pivotal supernova (SN) discovery of 1997, indicating the onset of the accelerated phase of the universe, cosmologists have diligently pursued measurements of cosmic expansion utilizing various observational probes. Notably, the cosmic microwave background (CMB) \citep{ ade1,ade2,aghanim}, Type Ia supernovae (SN Ia)  \citep{ sn1, sn2, sn3}, and Baryon Acoustic Oscillations (BAO)  \citep{  bao1, bao2, bao3, bao4, bao5} have emerged as key instruments for gauging the Hubble expansion or Hubble Constant. Precision cosmology hinges upon the accurate calibration and interpretation of observational data obtained from diverse probes, including the CMB and the standard distance ladder. Anomalies within these datasets have garnered significant attention, as they may herald breakthroughs in our understanding of fundamental cosmological processes.

Despite general agreement among most probes within a two to three-sigma range, recent analyses, particularly the Planck 2018 \cite{aghanim} and SH0ES 2022 datasets \citep{R22}, have revealed tensions exceeding five sigmas. Several works have been dedicated to careful examination of CMB sky and standard distance ladder. 

Hubble Hunter's Guide \citep{Hhg} lists out many departures from $\Lambda$CDM to solve the cosmological tensions and singles out a solution which increases Hubble expansion rate before recombination by modifying sound horizon at drag epoch $r_d$.
Other efforts have been made to identify anomalies in CMB polarization data  \citep{Matteo} with the help of JWST datasets. Studies like G-Transition hypothesis \citep{Ruchika:2023ugh}, Planck Mass transition  \citep{Giampaolo} assess the requisite signatures within the standard distance ladder. Recent study \citep{Vagnozzi:2023nrq} reports that early time solutions alone are not sufficient to solve Hubble Tension. By dissecting the underlying assumptions and methodologies, these studies \footnote{These studies point to introduce
unknown physical processes, such as modifications
to the expansion history of the Universe, possible interactions between dark energy and matter, or early/late-
time new physics, for discussions in these directions see, \textit{e.g.}, Refs.~\cite{Anchordoqui:2015lqa,Karwal:2016vyq,Benetti:2017juy,Mortsell:2018mfj,Kumar:2018yhh,Guo:2018ans,Poulin:2018cxd,Graef:2018fzu,Agrawal:2019lmo,Escudero:2019gvw,Niedermann:2019olb,Sakstein:2019fmf,Knox:2019rjx,Hart:2019dxi,Ballesteros:2020sik,Jedamzik:2020krr,Ballardini:2020iws,DiValentino:2020evt,Niedermann:2020dwg,Gonzalez:2020fdy,Braglia:2020auw,RoyChoudhury:2020dmd,Brinckmann:2020bcn,Karwal:2021vpk,Herold:2022iib,Gomez-Valent:2021cbe,Cyr-Racine:2021oal,Niedermann:2021ijp,Saridakis:2021xqy,Herold:2021ksg,Odintsov:2022eqm,Aboubrahim:2022gjb,Ren:2022aeo,Adhikari:2022moo,Nojiri:2022ski,Schoneberg:2022grr,Joseph:2022jsf,Gomez-Valent:2022bku,Odintsov:2022umu,Ge:2022qws,Schiavone:2022wvq,Brinckmann:2022ajr,Khodadi:2023ezj,Kumar:2023bqj,Ben-Dayan:2023rgt,Ruchika:2023ugh,Yadav:2023yyb,Sharma:2023kzr,Ramadan:2023ivw,Fu:2023tfo,Efstathiou:2023fbn,Montani:2023ywn,Stahl:2024stz,Vagnozzi:2023nrq,Zhai:2023yny,Garny:2024ums,Co:2024oek,Toda:2024ncp,Giare:2024ytc,Percival:2007yw,Giare:2024akf,Akarsu:2024eoo,Giare:2024smz,Pogosian:2020ded, Staicova:2023jic,Specogna:2025guo,Menci:2024rbq,Adil:2023exv} or Refs.~\cite{Abdalla:2022yfr,DiValentino:2025sru} for recent reviews.} seek to contextualize these anomalies within the broader framework of precision cosmology. 

We recall that Baryon acoustic oscillations are considered to be one of the most powerful probes for measuring the undergoing phase of accelerated expansion of the universe. When combined with the Planck satellite results, one finds that the Universe is spatially flat \cite{aghanim}. BAO is used to infer cosmological distance ratios to understand the fundamentals of our universe, allowing one to perform multiple consistency checks of data and theory. 

The standard BAO analyses from BOSS-eBOSS-DESI \cite{alam,Ata:2017dya, desicollab} measure distance ratios ($D_M/r_d$ and $D_H/r_d$) using correlation-function templates that incorporate theoretical assumptions based on $\Lambda$CDM cosmology. While template dependence is not currently considered a dominant source of systematic uncertainty, alternative approaches have been developed. For instance, the Purely-Geometric-BAO method \citep{Anselmi2, Anselmi1, Anselmi3} aims to minimize the cosmological assumptions in the correlation function analysis. Such approaches could prove valuable for investigating potential systematic effects in future, particularly in the context of current cosmological tensions where subtle modelling dependencies might become increasingly relevant.

In this work, we will analyze and compare the constraints on cosmological parameters using Baryon Acoustic Oscillations from two different teams : the first team uses the 2D BAO methodology \citep{carvalho,Carvalho2017,Carvalho2018}, while the second team employs the 3D BAO methodology \citep{alam,Beutler:2011hx,Ross:2014qpa,Ata:2017dya}. We argue that using a different methodology to estimate cosmological information from the standard ruler should not bring significant change in the final inference of cosmological parameters. Keeping the model fixed to the simplest $\Lambda$CDM model, we focused on the inherent tensions between BAO measurements. This paper scrutinizes the tension between two different approaches to BAO analysis (2D and 3D) applied to the same BOSS dataset, and explores the implications for our understanding of cosmic expansion. We also incorporated the SNe Ia dataset in our study. Taking along with low redshift observational probes like SNe Ia, we question here if 2D BAO can be used equally to study the evolution of the universe and if it can be at all used to check different exotic models that are being proposed to solve cosmological tensions.
We also incorporated the analysis with the DESI dataset to conclude the final remarks.

\section{Baryon Acoustic Oscillation as a standard ruler}\label{sec2:baointo}

Baryon Acoustic Oscillations (BAO) serve as a crucial cosmological standard ruler in the field of cosmology. They provide valuable information for understanding the large-scale structure of the universe and inferring cosmological parameters from observational data. The comoving position of the acoustic peak, which is a characteristic feature of the BAO, is particularly targeted by the cosmological community. This peak arises as a result of acoustic waves travelling through the early universe, leaving a distinct imprint on the distribution of matter. By measuring the characteristic scale of these oscillations in the large-scale structure of the universe, researchers can use BAO as a standard ruler to infer cosmological parameters such as the expansion rate of the universe and the amount of dark energy. In essence, BAO offers a powerful tool for cosmologists to probe the underlying cosmology of the universe by studying the clustering of galaxies and other cosmic structures. This allows them to better understand the nature of dark energy, dark matter, and the overall geometry and evolution of the universe.

In the standard cosmological description, in the early universe before the recombination epoch, baryons and photons were tightly coupled to each other and there was a formation of acoustic waves within the primordial photon-baryon plasma. As the universe expanded and cooled, reaching the epoch of decoupling, known as the drag epoch, the propagation of these acoustic waves ceased. At this pivotal moment, the baryon distribution retained imprints of the acoustic oscillations, manifesting as overdensities separated by a distinct length scale ($r_d \sim 150$ Mpc) known as the sound-horizon comoving length at the drag epoch. As this signature imprinted on matter distribution is governed by early universe physics before and around recombination, it is treated as a standard ruler. It is also well calibrated by CMB observations to very high accuracy \cite{aghanim}. Similarly, \cite{Eisenstein1998} predicted a bao peak in large-scale correlation function around the same comoving galaxy separation (100 $h^{-1}$) Mpc which was later confirmed by SDSS observations \cite{SDSS:2005xqv}.
That is why BAO is used to constrain dark energy behaviour and helps in breaking the residual degeneracies with CMB observations.

The evolution of matter on large scales, including dark matter and baryons, is primarily influenced by gravity. This gravitational interaction leaves a distinctive feature in the 2-point correlation functions (CF) of matter and its observed tracers, such as galaxies. This characteristic scale, a consequence of the physics governing the early universe, serves as a fundamental cosmological standard ruler.

Since the so-called acoustic peak was very closely identified with the correlation function (CF) feature in  BAO, it sparked the initial notion of measuring a comparable length scale across both the early and late universe. The intention behind this approach was to leverage the consistency of this scale throughout cosmic history, thereby harnessing its cosmological implications. However, in the present era of precision cosmology, this excellent intuition is encountering several challenges.

To fit the observational anisotropic power spectrum data from CMB sky, we mostly assume a standard cosmological model (e.g. flat $\Lambda$CDM), and we can calculate the value of $r_d$ as the derived parameter. Using late-time probes such as the galaxy correlation function, $r_d$ needs to be derived again from cosmology-dependent multi-parameter fit.

\subsection{Parameteriszing the Comoving Distance Scale}
Using late-time probes, there are two established methods in the literature for determining the BAO feature through the two-point correlation function. The first method employs the three-dimensional two-point correlation function (3D BAO dataset) \citep{alam,SDSS:2005xqv,Percival:2006gt,BOSS:2012vpn,DESI:2023tpn,desicollab}, which can be analyzed in both configuration space $\xi(s)$ and Fourier space $P(k)$, where $s$ is the comoving radial separation and $k$ is the wavenumber. The second method utilizes the two-dimensional angular correlation function (2D BAO dataset) \citep{carvalho,Cole:2005sx} $\omega(\theta)$, where $\theta$ represents the angular separation between galaxy pairs on the sky. While 3D BAO analysis contains more information and provides the tightest constraints when precise spectroscopic redshifts are available, the 2D angular approach may remain valuable for photometric surveys with larger redshift uncertainties. The angular clustering is less sensitive to radial smearing from photo-z errors. Additionally, angular correlation functions offer computational advantages when analyzing the very large datasets typical of wide-area imaging surveys. Below we describe both methodologies in detail.

\subsubsection{ Two Point Correlation Function}
Among several 3D estimators in literature, the most commonly used is the Two Point Correlation Function (2PCF) \citep{landy}

\begin{equation}
\xi(s) = \frac{DD(s) - 2DR(s) + RR(s)}{RR(s)},
\end{equation}
where DD(s) and RR(s) represent the number of galaxy pairs in real-real and random-random catalogues respectively. The parameter $s$ is chosen assuming a fiducial cosmology and refers to the comoving separation scale at which the two-point correlation function is evaluated to identify the BAO feature. It provides critical insights into the geometry and expansion history of the universe. In flat universe, the expression for $s$ between two galaxies at redshift $z_1$ and $z_2$ is given by
 \begin{equation}
s = \sqrt{ r^{2}(z_{1}) + r^{2}(z_{2}) - 2 r(z_{1}) r(z_{2}) \cos \theta_{12}},
\end{equation}
where $\theta_{12}$ is the angular distance between pair of galaxies at redshift $z_1$ and $z_2$. Choosing the expression for $r$ is what makes it cosmological model dependent. For flat $\Lambda CDM$, $H_0$ and $\Omega_m$ being the Hubble parameter at present and matter density parameter, $r$ is expressed as:
\begin{equation}
r(z_i) =\frac{c}{H_{0}}  \int_{0}^{z_i} \frac{dz}{\sqrt{\Omega_{m}(1 + z)^{3} + (1 - \Omega_{m})}}. 
\end{equation}

The full 3D correlation function can be analyzed directly in terms of both radial and transverse separations, or decomposed into Legendre multipoles or clustering wedges statistics \cite{Kazin_2011,Kazin:2013rxa,alam}. For computational efficiency and ease of interpretation, many studies extract the BAO information through the first few Legendre multipoles or clustering wedges. The Legendre multipoles in configuration space and the power spectrum multipoles are defined respectively as
\begin{flalign}
\xi_\ell(s) \equiv \frac{2\ell + 1}{2} \int_{-1}^{1} L_\ell(\mu_1)\xi(\mu_1, s) \, d\mu, \\
P_\ell(k) \equiv \frac{2\ell + 1}{2} \int_{-1}^{1} L_\ell(\mu_2)P(\mu_2, k) \, d\mu,    
\end{flalign}
where $\xi(\mu_1,s)$ and $P(\mu_2, k)$ are two dimensional correlation function and power spectrum respectively. Here variables $\mu_1$ and $\mu_2$ are the cosine of the angle between the line of sight direction and separation vector $s$ in real and $k$ space respectively.

The relationship between two-dimensional Legendre multipoles in redshift space can be expressed through the mapping between configuration space and Fourier space. For redshift-space distortions (RSD), the multipole moments of the correlation function $\xi_\ell(s)$ and power spectrum $P_\ell(k)$ are related by:
\begin{equation}
\xi_\ell(s) = \frac{i^\ell}{2\pi^2} \int_{0}^{\infty} P_\ell(k) j_\ell(ks) k^2 \, dk,
\end{equation}
where $j_\ell$ is the $\ell^{th}$ order spherical Bessel function. In standard analyses such as \citep{alam}, the monopole ($\ell = 0$), quadrupole ($\ell = 2$), and hexadecapole ($\ell = 4$) moments are typically used, as they provide a nearly complete description of the redshift-space clustering $\xi(\mu, s)$ in the distant observer approximation within the linear regime.

Using the power spectrum and two-point correlation function described above, the BAO scale is measured in redshift space. The observable is the shift in the BAO peak position with respect to fiducial cosmology.  The parallel and perpendicular shifts to the line of sight give bounds on Hubble expansion rate $H(z)$, comoving and angular diameter distance relative to the sound horizon at drag epoch $r_d$ parameter. 

\begin{flalign}\label{eq:alpha}
\alpha_{\perp} =   \frac{D_M(z) r_{d,fid}}{D_M^{fid}(z) r_d} ;   \quad
\alpha_{\parallel} =   \frac{H^{fid}(z) r_{d,fid}}{H(z) r_d}.
\end{flalign}

To illustrate how comoving distance \( D_M(z) \) and the Hubble parameter \( H(z) \) are measured, consider the preferred angular separation of galaxies, \( \Delta \theta \), within an ensemble of galaxy pairs oriented perpendicular to the line of sight. The comoving distance at this redshift is then determined by $D_M(z) = r_d / \Delta \theta.$ When examining the separation vector parallel to the line of sight, if a preferred redshift separation \( \Delta z \) is observed, the corresponding equivalent distance is $D_H = c/H(z) = r_d / \Delta z,$ thereby inferring the Hubble expansion parameter at that redshift. For certain redshift bins with low signal-to-noise ratios, and when both the transverse and line-of-sight components are present, isotropic BAO (Baryon Acoustic Oscillation) measurements are obtained using \( D_V(z) \), where
\begin{flalign}
D_V(z) = \left[ (1+z)^2 D_A^2(z) z D_H \right]^{1/3}.
\end{flalign}
Here, $ D_A(z) = D_M(z)/(1+z)$ is the angular diameter distance, and $D_V(z) $ represents the average of the distances measured perpendicular and parallel to the line of sight of the observer \citep{SDSS:2005xqv}.

The BOSS Survey (Baryon Oscillation Spectroscopic Survey) \citep{alam} assumes a correlation function template using the fiducial cosmology as a flat $\Lambda$CDM model with the following parameters: dimensionless Hubble constant $h = 0.676$, fluctuation amplitude $\sigma_8 = 0.8$, baryon density $\Omega_bh^2 = 0.022$, optical depth $\tau = 0.078$ and spectral tilt $n_s = 0.97$. For this model, the sound horizon at the drag epoch parameter is $r_d = 147.78$ Mpc. These parameter values are within $ 1 \sigma$ of Planck 2015 values from CMB. Crucially, the template also includes two dilation parameters, $\alpha_\perp$ and $\alpha_\parallel$, that allow the BAO peak position to shift in the transverse and radial directions respectively, relative to the fiducial cosmology. These scaling parameters are precisely defined as
\begin{equation}
\alpha_\perp = \frac{D_M(z) \, r_{d,\mathrm{fid}}}{D_{M,\mathrm{fid}}(z) \, r_d}, \quad
\alpha_\parallel = \frac{H_{\mathrm{fid}}(z) \, r_{d,\mathrm{fid}}}{H(z) \, r_d}.
\end{equation}

Even when the study \citep{alam}, was extended beyond $\Lambda$CDM models such as $ow_0w_a$CDM model or $0w$CDM model with extra relativistic species and by incorporating SN Type I-a dataset, the obtained Hubble Constant value was $H_0 = 67.3 \pm 1.0$ \Hunit and $H_0 = 67.8 \pm 1.2 $ \Hunit respectively not shifting from the mean value of standard $\Lambda$CDM model ($H_0 = 67.6 \pm 0.5 $ \Hunit (Planck alone)). And, hence keeping the tension with the standard distance ladder \Rtwo alive.
    
\subsubsection{ Two Point Angular Correlation Function}

The two-point angular correlation function (2PACF) is defined as the excess probability of finding two-point sources in two solid angles $d \Omega_1$ and $d \Omega_2$ with angular separation $\theta$ as compared to a homogeneous Poisson distribution  \citep{carvalho}. To avoid the contribution of radial signal, only narrow redshift shells of very small width $\delta z$ are considered. 
\begin{equation}
\omega(\theta) = \frac{DD(\theta) - 2DR(\theta) + RR(\theta)}{RR(\theta)},
\end{equation}
where $\theta$  is the measured angular separation between the pairs. The acoustic scale position is characterized by $\theta_{\rm FIT}$. When $\delta z = 0$, this fitted scale $\theta_{\rm FIT}$ becomes equivalent to the true BAO scale $\theta$. The 2PACF measurements exhibit multiple peaks, where the genuine BAO feature exists alongside systematic effects. These unwanted systematic signals can be addressed by calculating the expected 2PCF, which in turn allows determination of the expected 2PACF. Fig. 1 of Carvalho et al. \cite{carvalho} demonstrates these multiple features in the correlation curves, where the primary challenge lies in separating the cosmological BAO signal from systematic contributions that arise from redshift-selected galaxy samples. The identification method relies on a key characteristic: a true BAO signature presents as a persistent peak at the specific angular separation $\theta_{BAO}$, while systematic signals, particularly those from galaxy clusters and groups, produce fluctuations across multiple angular scales. These systematic features demonstrate notable instability when subjected to small positional variations in galaxy coordinates. This fundamental difference enables the distinction between the robust BAO peak and transient systematic effects in the final 2PACF measurement.

\begin{flalign}
& \omega_E(\theta, \bar{z}) = \int_{0}^{\infty} dz_{1} \, \phi(z_{1}) \int_{0}^{\infty} dz_{2} \, \phi(z_{2}) \, \xi_{E}(s, \bar{z}), & \\
& \xi_{E}(s, z) = \int_{0}^{\infty} \frac{dk}{2\pi^2} k^2~j_{0}(ks)~b^{2} P_{m}(k, z), &
\end{flalign}
where $\bar z$ is the average redshift of $z_1$ and $z_2$, $j_0$ is the zeroth order Bessel Function and $P_m$ is the matter power spectrum calculated using the fiducial cosmological model $\Lambda$CDM with parameters set to $w_bh^2 = 0.0226$, $w_ch^2 = 0.112$, $100 \Theta = 1.04$, $\tau = 0.09$, $A_s e^9 = 2.2$, and $n_s = 0.96$ for SDSS DR10 galaxies (\citep{carvalho}).\\
Once $\Theta_{FIT}$ is estimated, it can be used directly to put bounds on cosmological parameters and cosmological evolution. The relation describing the measured angle $\theta_{BAO}$ and angular diameter distance is given by
\begin{equation}\label{eq:theta}
    \theta_{BAO} = \frac{r_d}{(1+z)D_A(z)},
\end{equation}
where $D_A(z)$ is the angular diameter distance and $r_d$ is the sound horizon at the drag epoch.

\section{Data and Methodology}\label{sec3:data} 
\indent
In the subsequent sections, we describe the observational datasets utilized in this study.\\
We begin by detailing the BAO data from two distinct teams. Within the framework of flat $\Lambda$CDM, BAO measurements constrain $\Omega_{m0}$ through the redshift evolution of $H(z)r_d$ and $D_M(z)/r_d$, while also constraining the product $H_0r_d$ through these quantities' values at $z=0$.
We then present the Supernova Type Ia (SN Ia) data. These measurements complement the BAO analysis by providing independent constraints on the same cosmological parameters, but through different combinations: SN Ia constrain $\Omega_{m0}$ through the redshift dependence of their apparent magnitudes, while also constraining $H_0$ directly through their absolute magnitude calibration. Given the ongoing tension in the Hubble constant ($H_0$) measurements, which directly affects the absolute magnitude ($M_B$) used for calibrating SN Ia, we adopt a comprehensive approach. Rather than relying solely on the $M_B$ derived from local distance ladder measurements, we compile a diverse array of possible $M_B$ values from various probes to facilitate a complete analysis.
For the sake of brevity, we designate the theta measurements of BAO, as tabulated in Table \ref{tabledata:PACF} simply as "BAO Data - $2D$" and data given in Table \ref{tabledata:PCF} as "BAO Data - $3D$". Furthermore, upon the inclusion of the BAO Dark Energy Spectroscopic Instrument (DESI) dataset, we explicitly refer to it as "BAO Data: $DESI$", despite its categorization within the BAO Data-3D framework.\\~\\

\textbf{BAO Data - 2D}:
 We employed a dataset comprising 12 Baryon Acoustic Oscillation (BAO) measurements, denoted as $\theta_{\textrm{BAO}}(z)$. The determination of the BAO feature involves measuring angles between galaxy pairs on the sky, which are direct observables independent of cosmological assumptions. The cosmological model dependence enters only through the theoretical templates derived from $\Lambda$CDM used to identify and characterize the BAO signal in the angular correlation function \citep{carvalho,sanchez}, rather than through any particular choice of cosmological parameters. Once identified, these angular measurements constrain the absolute scale of BAO, denoted as $r_d$, when combined with the angular diameter distance ($D_A$) to the respective redshift. The relationship between $\theta_{\textrm{BAO}}$, $D_A$, and $r_d$ is given by Equation (\ref{eq:theta}). While we treat these measurements as independent in our analysis, it is important to note that some correlation between measurements at different redshifts may exist due to the common fitting procedures used to extract the BAO signal. A complete covariance analysis of these correlations, which could potentially affect the precision of our constraints, will be addressed in future work. Table \ref{tabledata:PACF} presents the compiled BAO dataset, encapsulating these measurements for further analysis.\\

\begin{table*}
\centering
\begin{tabular}{| c | c | c | c || c | c | c | c |}
\hline
$\Bar{z}$ & $\theta_{\textrm{BAO}}(z)[^{\circ}]$ & $D_A(z)/r_d$ & Reference & $\Bar{z}$ & $\theta_{\textrm{BAO}}(z)[^{\circ}]$ & $D_A(z)/r_d$ & Reference \\ \hline
0.45 & $4.77 \pm 0.17$ & $8.28 \pm 0.30$ & SDSS DR10 \citep{carvalho} & 
0.57 & $4.59 \pm 0.36$ & $7.95 \pm 0.62$ & SDSS DR11 \citep{Carvalho2017} \\ 
0.47 & $5.02 \pm 0.25$ & $7.76 \pm 0.39$ & SDSS DR10 \citep{carvalho} & 
0.59 & $4.39 \pm 0.33$ & $8.21 \pm 0.62$ & SDSS DR11 \citep{Carvalho2017} \\ 
0.49 & $4.99 \pm 0.21$ & $7.71 \pm 0.32$ & SDSS DR10 \citep{carvalho} & 
0.61 & $3.85 \pm 0.31$ & $9.24 \pm 0.74$ & SDSS DR11 \citep{Carvalho2017} \\ 
0.51 & $4.81 \pm 0.17$ & $7.89 \pm 0.28$ & SDSS DR10 \citep{carvalho} & 
0.63 & $3.90 \pm 0.43$ & $9.01 \pm 0.99$ & SDSS DR11 \citep{Carvalho2017} \\ 
0.53 & $4.29 \pm 0.30$ & $8.73 \pm 0.61$ & SDSS DR10 \citep{carvalho} & 
0.65 & $3.55 \pm 0.16$ & $9.78 \pm 0.44$ & SDSS DR11 \citep{Carvalho2017} \\ 
0.55 & $4.25 \pm 0.25$ & $8.70 \pm 0.51$ & SDSS DR10 \citep{carvalho} & 
2.225 & $1.77 \pm 0.31$ & $10.04 \pm 1.76$ & SDSS QS \citep{Carvalho2018} \\ 
\hline
\end{tabular}
\caption{BAO measurements from angular separation of pairs of galaxies (denoted as BAO dataset: 2D throughout the analysis). The column \( D_A(z)/r_d \) is calculated using Eq.~\eqref{eq:theta}.}
\label{tabledata:PACF}
\end{table*}

\begin{table*}
\centering
\begin{tabular}{| c | c | c || c | c | c |}
\hline
${z}$ & Anisotropic Constraint & Reference & ${z}$ & Isotropic Constraint  & Reference\\
\hline

0.38& $D_A$/$ r_d$ = 7.42, $D_H$/$r_d$ = 24.97  &  BOSS DR 12 [\citep{alam}] & 0.106 & $D_V$/$ r_d$ = 2.98 $\pm$ 0.13 & 6dF [\citep{Beutler:2011hx}] \\
0.51 & $D_A$/$ r_d$ = 8.85, $D_H$/$r_d$ = 22.31 &  BOSS DR 12 [\citep{alam}] & 0.15 & $D_V$/$ r_d$ = 4.47 $\pm$ 0.17 & MGS [\citep{Ross:2014qpa}] \\
0.61 & $D_A$/$ r_d$ = 9.69, $D_H$/$r_d$ = 20.49 & BOSS DR 12 [\citep{alam}] & 1.52 &  $D_V$/$ r_d$ = 26.1 $\pm$ 1.10 & eBOSS quasars [\citep{Ata:2017dya}] \\
2.40 & $D_A$/$ r_d$ = 10.76, $D_H$/$r_d$ = 8.94 & BOSS DR 12 [\citep{alam}] &  & &  \\
\hline
\end{tabular}
\caption{BAO measurements from volumetric measurements (denoted as BAO dataset: 3D throughout the analysis). The correlation matrix corresponding to anisotropic constraints is presented in the Appendix \eqref{eq:covmatrix}.}
\label{tabledata:PCF}
\end{table*}

\begin{figure*}
\vspace{0.1cm}
\centering
\resizebox{500pt}{120pt}{\includegraphics{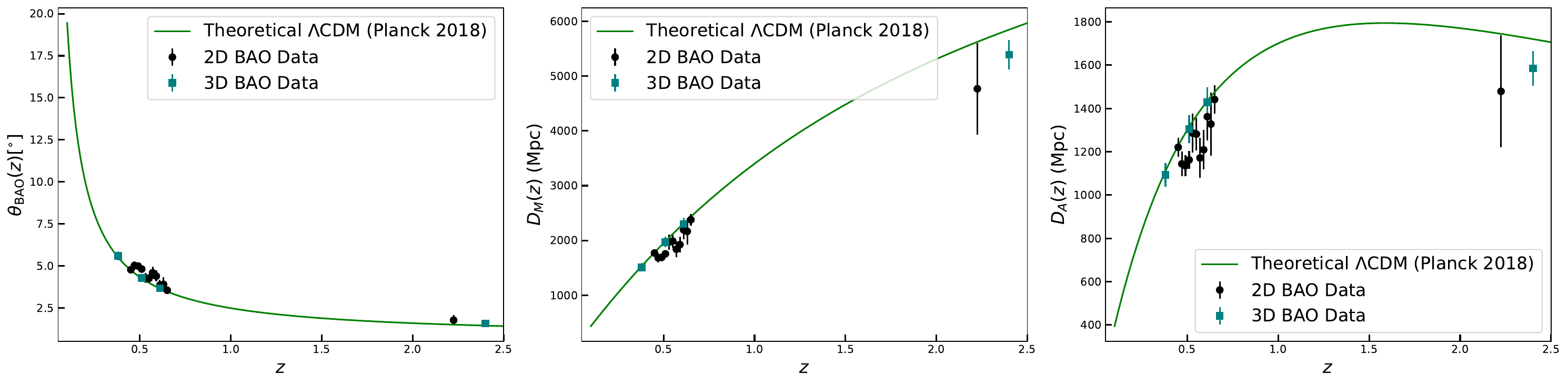}}
\caption {\label{fig:datapoints} This figure compares 2D and 3D BAO observational data with the theoretical predictions of the $\Lambda$CDM model. The left panel shows the evolution of the angular scale, $\theta(z)$, as a function of redshift, the middle panel shows the evolution of the comoving distance, $D_M(z)$, with redshift, and the right panel shows the evolution of the angular diameter distance, $D_A(z)$, as a function of redshift,  all computed using the Planck (2018) derived sound horizon. The 2D BAO data points are shown in black, while the 3D BAO data points are shown in teal colour. The theoretical predictions from the $\Lambda$CDM model, calculated using the Planck 2018 cosmological parameters, are shown in green for all three panels.}
\end{figure*}

\textbf{BAO Data - 3D}:
For the three-dimensional (3D) Baryon Acoustic Oscillations (BAO) data, we amalgamate isotropic BAO measurements obtained from various surveys. These include the 6dF galaxy survey at redshift $z = 0.106$ \citep{Beutler:2011hx}, the SDSS DR7-MGS survey at an effective redshift $z = 0.15$ \citep{Ross:2014qpa}, and measurements from the SDSS DR14-eBOSS quasar samples at redshift $z = 1.52$ \citep{Ata:2017dya}. Additionally, BAO measurements using Lyman-alpha samples in conjunction with quasar samples at redshift $2.4$ from the SDSS DR12 are incorporated \citep{Bourboux:2017cbm}. Furthermore, anisotropic BAO measurements from the BOSS DR12 galaxy sample at redshifts $0.38$, $0.51$, and $0.61$ along with the covariance matrix \citep{evslin} are utilised. The compiled BAO dataset is presented in Table \ref{tabledata:PCF}. Henceforth, we collectively refer to these datasets as "$3D$ BAO" data. 

We intentionally begin with the anisotropic BAO measurements from the BOSS DR12 to establish an important baseline: the 3D BAO methodology and its results have remained essentially unchanged over time. By starting with legacy BAO data that has been extensively analyzed in the literature, we demonstrate that the tensions we identify are not artifacts of any particular dataset but between 2D and 3D methodologies. As shown in Figure \ref{fig:comparefourparams}, when we subsequently apply the same analysis to DESI DR1, the results remain consistent, confirming that these discrepancies persist.  

The primary objective of this work is to investigate potential discrepancies between 2D and 3D BAO data. To achieve this, we begin by identifying the redshift ranges where both 2D and 3D BAO measurements are available. Subsequently, we compare various observables derived from these datasets against the predictions of the standard $\Lambda$CDM model predicted using Planck 2018 cosmological parameters. This analysis is summarized in Figure \ref{fig:datapoints}.\\
Figure \ref{fig:datapoints} provides a comparative visualization of 2D and 3D BAO observational data alongside the theoretical predictions of the $\Lambda$CDM model.\\
1) The left panel depicts the evolution of the angular scale, $\theta(z)$, as a function of redshift.\\
2) The middle panel illustrates the redshift evolution of the comoving distance, $D_M(z)$.\\
3) The right panel shows the redshift dependence of the angular diameter distance, $D_A(z)$.\\
These distances in the middle and right panels are derived using the Planck 2018 sound horizon, which serves as a reliable standard ruler since it depends only on early universe physics, not on late-time cosmological evolution.\\
The 2D BAO data points are represented by black markers, while the 3D BAO data points are shown in teal in this figure. The theoretical predictions of the $\Lambda$CDM model, computed using the Planck 2018 cosmological parameters, are displayed as green curves across all panels. This comparison highlights the consistency—or lack thereof—between the observational data and the model predictions, thereby shedding light on any underlying discrepancies.\\

\textbf{SN Type-I:} In addition to the Baryon Acoustic Oscillations (BAO) data, we used the expansive "Pantheon Plus Sample" of Type-I Supernovae (SN-Ia), comprising a total of 1701 supernovae spanning the redshift range from 0.01 to 2.26 \citep{Scolnic}. This comprehensive dataset incorporates the SH0ES distance anchors, utilizing the host cepheid galaxies for calibration \citep{sntable}.

In our analysis, we adopted a range of Hubble Constant values obtained from different techniques. One such instance involves the adoption of a fixed value for the absolute magnitude, $M_{B}$ = -19.214 $\pm$ 0.037 magnitudes. This determination arises from a synthesis of geometric distance estimates derived from Detached Eclipsing Binaries in the Large Magellanic Cloud (LMC) \citep{piet}, the MASER NGC4258 \citep{reid}, and recent parallax measurements of 75 Milky Way Cepheids with Hubble Space Telescope (HST) photometry \citep{riess21} GAIA Early Data Release 3 (EDR3) \citep{lind20a,lind20b}. Notably, this represents the most precise and contemporary model-independent assessment of the Absolute Magnitude to date. Other adopted values of $M_{B}$ used in our analysis are given in Table \ref{Tab:prior_H0}.

Utilizing the Absolute Magnitude of standard candles, such as Type-Ia Supernovae (SNe-Ia), one can readily deduce their distance given their observed apparent magnitude or flux. The relationship between the apparent magnitude of SNe-Ia and their relative distance modulus is expressed by the equation:
\begin{equation}\label{dl}
{D}_{L}=10^{(\mu - 25)/5}  \text{ Mpc}.
\end{equation}
Here, $\mu = m_b - M_B$ represents the distance modulus, where $m_b$ denotes the apparent magnitude of SNe-Ia and $M_B$ signifies the absolute magnitude of SNe-Ia.

\subsection{Model and Methodology}\label{sec:model}
We aim to compare the constraints on cosmological parameters derived from 2D and 3D Baryon Acoustic Oscillations datasets. To maintain clarity and focus specifically on comparing these BAO datasets, we deliberately restrict our analysis to the $\Lambda$CDM framework. While the standard $\Lambda$CDM model typically employs six parameters ($\Omega_{b0} h^2$, $\Omega_{c0} h^2$, $\theta_*$, $\tau$, $A_s$, and $n_s$), we adopt a modified four-parameter approach better suited to our BAO-focused analysis.

Instead of using the full six-parameter set, which primarily describes both the early and late universe, we concentrate on the late-universe parameters that BAO and SN Ia directly constrain: the matter density parameter ($\Omega_{m0}$), the Hubble constant ($H_0$), and the sound horizon at drag epoch ($r_d$). Additionally, we include the absolute magnitude of Type Ia supernovae ($M_B$) as our fourth parameter, necessary for the SN Ia calibration. This reduction from six to four parameters is possible because our analysis focuses on low-redshift measurements that are insensitive to the early-universe parameters ($\tau$, $A_s$, and $n_s$), while the remaining parameters can be reparameterized in terms of our chosen set.

Notably, we treat $r_d$ as a free parameter rather than deriving it from early-universe physics (which would involve $\Omega_{b0}h^2$ and $\Omega_{c0}h^2$)\footnote{
Typically, $\Omega_{b0}h^2 = 0.0224 \pm 0.0001$ from Planck measurements, with small variations depending on the specific analysis \cite{aghanim}.\\
As noted in \citet{Eisenstein1998}, $r_d$ depends on both $\Omega_{m0}h^2$ and $\Omega_{b0}h^2$. The approximate formula is:
\begin{equation}\label{eq:rd}
r_d \approx 55.154 \times (\Omega_{m0}h^2)^{-0.25} \times (\Omega_{b0}h^2)^{-0.125} \text{ Mpc}.
\end{equation}
}. This approach allows us to directly compare BAO datasets without making assumptions about early-universe physics, and to explore potential tensions between early and late-universe measurements. Using the combined dataset (BAO + Pantheon Plus) described in Section \ref{sec3:data}, we fit these four parameters: $\Omega_{m0}$, $H_0$, $r_d$, and $M_B$.

In the $\Lambda$CDM cosmology framework, within a spatially flat FLRW universe, the present scale factor $a_0$ is normalized to 1.0. The Hubble parameter at redshift $z$ is given by:
\begin{equation}
\frac{H^{2}(z)}{H_{0}^{2}} = \Omega_{m0}(1+z)^{3} + (1 - \Omega_{m0}),
\end{equation}
where the subscript $0$ denotes present-day values and $\Omega_{m0}$ represents the present-day matter density. Other relevant quantities are defined as follows:
\begin{flalign}\label{eq:defdis}
& D_M(z) = \int_0^z \frac{c~d\bar{z}}{H(z)}, &\\
& D_A(z) = (1+z)^{-1}~D_M(z), & \\
& D_L(z) = (1+z)~D_M(z), & \\
& D_V(z) = ((1+z)~D_A(z))^{2/3} \left( \frac{c~z}{H(z)} \right ), ^{1/3}&
\end{flalign}
where $D_M$, $D_A$ and $D_L$ are comoving distance, angular diameter distance and luminosity distance respectively.\\
We applied a uniform prior distribution on $\Omega_{m0}$ and $r_d$ as [0.1, 0.9] and [130, 170] respectively as outlined in Table \ref{Tab:prior_H0}. Rather than relying solely on a single calibration for the absolute magnitude $M_B$ of the SN sample, we opted to consider multiple values of $M_B$ (or equivalently, $H_0$), a practice also observed in various studies such as \cite{lemos}. For conversion, we employed the relation:

\begin{equation}
M_B= 5(\textrm{log}~ H_0 - \alpha_{B}- 5),
\end{equation}
where $\alpha_{B}$ is the observed intercept of B band apparent magnitude-redshift relation for SNe in the Hubble flow \cite{R22}. Presently, we assumed its value to be $0.71273 \pm 0.00176$ obtained independently of CMB and BAO. We employed Gaussian priors on $M_B$ derived from various measurements of the Hubble Constant, including those from ACT+WMAP CMB \cite{aiola}, Planck CMB + Lensing \cite{aghanim}, SH0ES collaborations \cite{Reiss2019,Reiss2021}, values derived from Masers \cite{Masers}, the Tully Fisher \cite{TFR} Relation, and BOSS DR12+BBN \cite{BOSS}, as detailed in Table \ref{Tab:prior_H0}. We also use the $H_0$ value from a recent SN study \cite{Efstathiou} (denoted as SH0ES 2021$_{a}$), slightly higher than SH0ES 2021 values due to different period ranges and photometric samples. For generating Markov Chain Monte Carlo (MCMC) chains based on the aforementioned dataset, we utilized the publicly available code EMCEE \citep{ForemanMackey:2012ig}. The obtained results are presented in Table \ref{Tab:res}.

\begin{table}
\centering
\begin{tabular}{|c|c|c|}
\hline
Measurement of $M_B$ & Prior (Gaussian$(\mu, \sigma^2)$) & $H_0 \pm \sigma_{H_0}$ \\
\hline
Planck CMB + Lensing & $(-19.422, 0.019^2)$ & $67.3 \pm 0.5$ \\
ACT + WMAP CMB & $(-19.414, 0.036^2)$ & $67.6 \pm 1.1$ \\
BOSS DR12 + BBN & $(-19.385, 0.070^2)$ & $68.5 \pm 2.2$ \\
SH0ES 2021$_a$ & $(-19.214, 0.039^2)$ & $73.2 \pm 1.3$ \\
SH0ES 2021 & $(-19.241, 0.040^2)$ & $74.1 \pm 1.3$ \\
Masers & $(-19.220, 0.089^2)$ & $73.9 \pm 3.0$ \\
SH0ES 2019 & $(-19.217, 0.042^2)$ & $74.0 \pm 1.4$ \\
Tully Fisher & $(-19.159, 0.075^2)$ & $76.0 \pm 2.6$ \\
\hline
\end{tabular}
\caption{Priors on $M_B$ and corresponding $H_0$ (in units of  Km/s/Mpc)  measurements. For other cosmological parameters such as $\Omega_{m0}$ and $r_d$, we use a flat prior as [0.1,0.9] and [130, 170] respectively.}
\label{Tab:prior_H0}
\end{table}
\subsection{Cosmological Chisq Analysis}

To infer information about the cosmological parameters within the framework of the flat $\Lambda$CDM model, we conducted a Chisq analysis utilizing the observables from the BAO 2D and BAO 3D datasets.

For the BAO 2D data, the observable is $\theta_{\textrm{BAO}}(z)$ as defined in Equation (\ref{eq:theta}). The corresponding chi-squared function is given by:

\begin{equation}\label{eq:chiPACF2}
\chi^2_{\textrm{2D BAO}} = \sum_{i} \left[ \frac{{\theta_{\textrm{BAO}}(z_i)}^{\textrm{obs}} - \theta_{\textrm{BAO}}(z_i)^{\textrm{model}}}{\sigma_{(\theta_{\textrm{BAO}}(z_i))}} \right]^2,
\end{equation}
where the index $i$ labels the observations. The superscript “obs” corresponds to the observed value of $\theta_{\textrm{BAO}}$ at redshift $z_i$
and the superscript “model” corresponds to the value of $\theta_{\textrm{BAO}}$
calculated from the theoretical model at the same redshift (here model parameters are $H_0$, $\Omega_{m0}$, $r_d$)\footnote{Note that Equation (\ref{eq:chiPACF2}) assumes uncorrelated measurements between different redshift bins.}.
For the BAO 3D data, we have both anisotropic and isotropic observables. Equation (\ref{eq:chiPCF-aniso}) describes the anisotropic component, where $D_i$ represents the best fit of observables as $D_A(z_i)/r_d$ and $D_M(z_i)/r_d$ respectively. The covariance matrix for this dataset can be found in Table 1 of \citep{evslin}.

\begin{multline}\label{eq:chiPCF-aniso}
    \chi^2_{\textrm{3D~BAO}} = \\
    \sum_{i,j} [ D(z_i)^{\rm{obs}} - D(z_i)^{\rm{model}} ] \text Cov^{-1}_{i,j} [ D(z_j)^{\rm{obs}} - D(z_j)^{\rm{model}} ] .
\end{multline}

Further, for the isotropic component of the BAO 3D data, the observable is $D_V(z_i)/r_d$, where $D_V(z_i)$ is the volume-averaged distance. The Chi-squared definition for this component is provided below in Equation (\ref{eq:chiPCF-iso}):
 
\begin{multline}\label{eq:chiPCF-iso}
    \chi^2_{\textrm{3D~BAO}} = \\
    \sum_{i} \left[ \frac{{D_V(z_i)/r_d}^{\rm{obs}} - {D_V(z_i)/r_d}^{\rm{model}}}{\sigma_{({D_V(z_i)/r_d})}} \right]^2.
\end{multline} 

For minimising the supernova cosmology parameters, the $\chi^2$ function is defined as in Equation (\ref{eq:chiPACF}), where $\mu_i$ represents each supernova distance, with $i$ ranging from 1 to 1701. $\mu_{b,i}^{\textrm{model}}$ denotes the predicted distance modulus, estimated using cosmological parameters describing the expansion history (here, $\Lambda$CDM).

\begin{multline}\label{eq:chiPACF}
    \chi^2_{\textrm{SNe-Ia}}(H_0, \Omega_{m0}, M_B) = \\
    \sum_{i,j} [ \mu_{b,i}^{\rm{obs}} - \mu_{b,i}^{\rm{model}} ] \text Cov^{-1}_{i,j} [ \mu_{b,j}^{\rm{obs}} - \mu_{b,j}^{\rm{model}} ].
\end{multline}
And, $\text{Cov}^{-1}_{i,j}$ describes the covariance matrix for supernova data. The total chi-square is then defined as the sum of the chi-square contributions from the Pantheon Plus sample and the BAO sample, where "BAO" encompasses both BAO datasets used for comparison.
\begin{multline}
\chi^2_{\textrm{Total}}(H_0, \Omega_{m0}, M_B,r_d) = \\
 \chi^2_{\textrm{SNe-Ia}}(H_0, \Omega_{m0}, M_B) + 
    \chi^2_{\textrm{BAO}}(H_0, \Omega_{m0}, r_d).  
\end{multline}\label{eq:chitotal}

\section{Results and Discussion}\label{sec4:result2d3d}

\subsection{Tension between BAO 3D Data and BAO 2D Data?}\label{sec4}

\subsubsection{Cosmological constraints on the product $r_d * h$ }

In BAO analyses, the quantities constrained differ between 2D and 3D methodologies. The 2D BAO methodology constrains the ratio $D_M(z)/r_d$, where $D_M(z)$ is the comoving distance and $r_d$ is the sound horizon at the drag epoch. At low redshifts (z → 0), this ratio approaches $cz/(H_0 r_d)$, effectively constraining the combination $H_0 r_d$. In contrast, 3D BAO analyses directly constrain both $D_A(z)/r_d$ and $H(z)r_d$ through measurements of transverse and radial clustering, respectively. When combined with supernova (SNe) data, which constrain the value of the Hubble constant, the degeneracy between \( r_d \) and \( H_0 \) is broken, allowing for independent estimates of \( r_d \).
In our analysis, we found that the mean value of the product \( r_d \cdot H_0 \) is higher when using 2D BAO data compared to the 3D BAO methodology.\footnote{This is a direct consequence of the difference in $\theta_{\textrm{BAO}}(z)[^{\circ}]$ or $D_A(z)/r_d$ around z $\approx$ 0.55 (refer to Figure \ref{fig:datapoints}).}

\begin{flalign}\label{eq:rdh3D}
& \Omega_{m0} = 0.314 \pm 0.0132 & \quad \text{BAO: 3D + PP} \\ 
& r_d h = 100.17 \pm 2.69 \, \textrm{Mpc} & \quad \text{(Cal: SH0ES 21)}
\end{flalign}

\begin{flalign}\label{eq:rdh2D}
&  \Omega_{m0} = 0.331 \pm 0.018 & \quad \text{BAO: 2D + PP}\\
&  r_dh = 106.02 \pm 3.25 \textrm{ Mpc} & \quad \text{(Cal: SH0ES 21)}
\end{flalign}

\begin{flalign}\label{eq}
& \Omega_{m0} = 0.315 \pm 0.0134 & \quad \text{BAO: 3D + PP}\\
& r_dh = 100.14 \pm 1.57  \textrm{ Mpc} & \quad \text{(Cal: CMB)} 
\end{flalign}

\begin{flalign}\label{eq}
& \Omega_{m0} = 0.330 \pm 0.0178 & \quad \text{BAO: 2D + PP}\\
& r_dh = 105.95 \pm 2.21 \textrm{ Mpc} & \quad \text{(Cal: CMB )} 
\end{flalign}
 
The calibrations mentioned in the above equations refer to the calibration of Type Ia supernovae (SNe Ia) based on the measurements of $M_B$ (absolute magnitude of SNe Ia) derived from SH0ES 2021 and Planck CMB + Lensing data. These are indicated as ``Cal: SH0ES 21'' and ``Cal: CMB'' in the equations., respectively, as detailed in Table \ref{Tab:prior_H0}.\\
Regardless of the calibration used, the mean value of the product \( r_d \cdot H_0 \) derived from 2D BAO data remains higher than that from 3D BAO data. This trend holds across all calibrations considered in this paper, as shown in Figure \ref{fig:resrdhcompareandom0countour}. We also plot green and orange bands representing constraints from the CMB \cite{aghanim} (temperature, polarization, and lensing: $r_d \cdot h = 98.82 \pm 0.82$ Mpc; $\Omega_{m0}= 0.3153 \pm 0.0073$)  \textrm{and} DESI ($r_d \cdot h = 101.8 \pm 1.3$ Mpc; $\Omega_{m0}= 0.295 \pm 0.015$) \citep{desicollab}. Despite the differences between the 2D and 3D methodologies, the results from both BAO analyses are generally consistent within a $1.5\sigma$ interval across all SNe calibrations, except for the case where the SNe calibration is derived from Planck CMB + Lensing, which shows a tension of $2.3\sigma$ \footnote{This is primarily due to the tension between the SNe data and the Planck data. Since our analysis assumes the $\Lambda$CDM model, which rules out the possibility of new physics in the local universe, calibrating SNe with the Planck-derived $M_B$ may not be the most appropriate approach.}.

\subsubsection{Effect on the sound horizon at the drag epoch: The Standard Ruler}

In this section, we present the results of our analysis and discuss the tension observed between the two BAO datasets: 2D and 3D. Specifically, we focus on the cosmological parameter $r_d$, which represents the sound horizon at the drag epoch. Previous studies have confirmed a high correlation between $H_0$ and $r_d$. For instance, the Planck $\Lambda$CDM observations yield a derived value of $H_0 = 67.27 \pm 0.60$ \Hunit, corresponding to a sound horizon of $r_d = 147.05 \pm 0.30$ Mpc. However, along with many other studies \citep{Bernal:2016gxb,CosmoVerse:2025txj,Knox:2019rjx}, we in one of our studies \citep{evslin} have found that when $H_0$ is constrained from low-redshift studies, its value tends to be around $ \sim 73$ \Hunit, requiring $r_d$ to be $\sim 137$ Mpc, leading to a tension of more than $2.5$ sigmas with Planck, regardless of the behaviour of dark energy.

It is crucial to note that this discrepancy arises when using the 3D BAO data. We got similar results as one can see in Table \ref{Tab:res}. While using 3D BAO data and Pantheon Plus with SH0ES $2021_a$ calibration, we got constraints on $r_d$ as $134.30 \pm 2.41$ Mpc corresponding to $h$ constraints $0.75 \pm 0.012$. Conversely, when utilizing the 2D BAO data, we obtained $h$ as $0.745 \pm 0.0129$, corresponding to $r_d$ of $142.22 \pm 3.37$ Mpc, which is compatible with the Planck value of $r_d$ within one sigma (see figure \ref{fig:comaparerdlowandhighz}).

We have extensively examined this dataset to further understand the cosmological tensions and the impact of the BAO dataset. To facilitate a comprehensive discussion, we have divided our analysis into two subsections:\\

\begin{table*}[htb]
\centering
\footnotesize
\setlength{\tabcolsep}{2pt}
\begin{tabular}{|c|c|c|c|c|c|c|}
\hline
& \multicolumn{3}{c|}{Panth Plus + BAO Data : 3D} & \multicolumn{3}{c|}{Panth Plus + BAO Data : 2D} \\
\hline \hline
Measurements & $r_d$ (Mpc) & $H_0$ & $\Omega_{m0}$ & $r_d$ (Mpc) & $H_0$ & $\Omega_{m0}$ \\
\hline
SH0ES 2021$_a$ & $134.30 \pm 2.41$ & $74.63 \pm 1.18$ & $0.314 \pm 0.0132$ & $142.22 \pm 3.37$ & $74.56 \pm 1.29$ & $0.331 \pm 0.0179$ \\ \hline
SH0ES 2021 & $135.78 \pm 2.73$ & $73.78 \pm 1.32$ & $0.314 \pm 0.0132$ & $143.93 \pm 3.56$ & $73.64 \pm 1.35$ & $0.331 \pm 0.0181$ \\ \hline
Masers & $136.53 \pm 4.46$ & $73.47 \pm 2.37$ & $0.314 \pm 0.0133$ & $143.03 \pm 6.2$ & $74.20 \pm 2.96$ & $0.3306 \pm 0.0178$ \\ \hline
SH0ES 2019 & $134.58 \pm 2.67$ & $74.48 \pm 1.32$ & $0.314 \pm 0.0133$ & $142.36 \pm 3.62$ & $74.49 \pm 1.46$ & $0.3308 \pm 0.018$ \\ \hline
Tully Fisher & $134.05 \pm 3.04$ & $74.89 \pm 1.78$ & $0.313 \pm 0.0132$ & $139.05 \pm 4.91$ & $76.28 \pm 2.44$ & $0.331 \pm 0.0179$ \\
\hline \hline
Planck CMB + Lensing & $147.44 \pm 1.89$ & $67.92 \pm 0.62$ & $0.315 \pm 0.0134$ & $156.36 \pm 2.90$ & $67.77 \pm .65$ & $0.330 \pm 0.0178$ \\ \hline
BOSS DR12 + BBN & $145.02 \pm 4.82$ & $69.14 \pm 2.21$ & $0.314 \pm 0.0135$ & $153.92 \pm 5.55$ & $68.89 \pm 2.22$ & $0.331 \pm 0.0182$ \\ \hline
ACT + WMAP CMB & $146.89 \pm 2.84$ & $68.17\pm 1.16$ & $0.315 \pm 0.0133$ & $155.86 \pm 3.64$ & $67.99 \pm 1.15$ & $0.3312 \pm 0.0179$ \\
\hline
\end{tabular}
\caption{Constraints on parameters for $\Lambda$CDM using both BAO Data Sets and various Pantheon Plus calibrations $M_B$. The error bars represent the $1\sigma$ confidence interval. The parameter $\Omega_m$ is constrained to $\approx 0.314 \pm 0.013$ using Panth Plus + BAO Data (3D), while using Panth Plus + BAO Data (2D), the constraint is $\approx 0.331 \pm 0.018$. These results remain consistent regardless of changes in the Supernova Absolute Magnitude calibration.}
\label{Tab:res}
\end{table*}

\begin{figure*}
\vspace{0.1cm}
\resizebox{240pt}{180pt}{\includegraphics{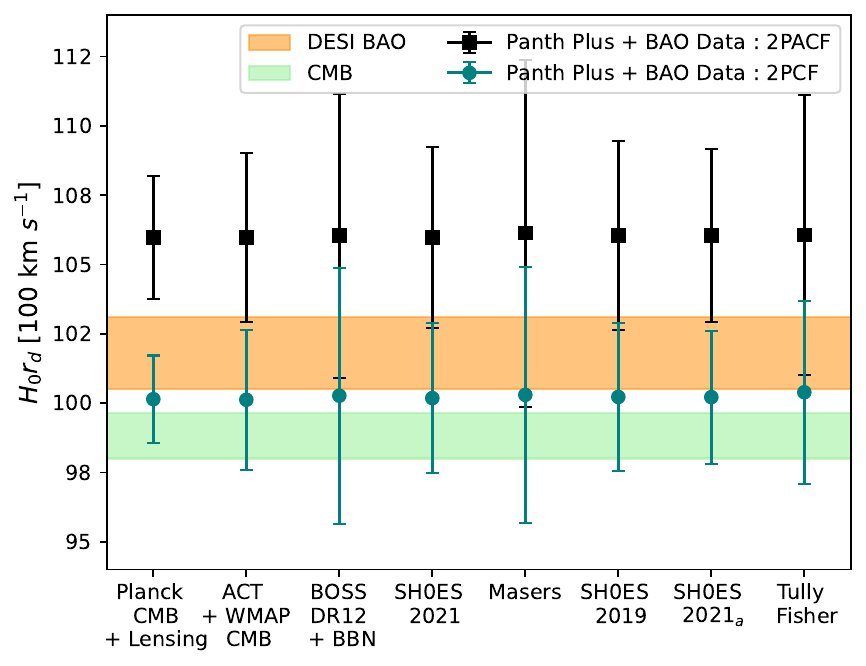}}
\vspace{0.1cm}
\resizebox{240pt}{180pt}{\includegraphics{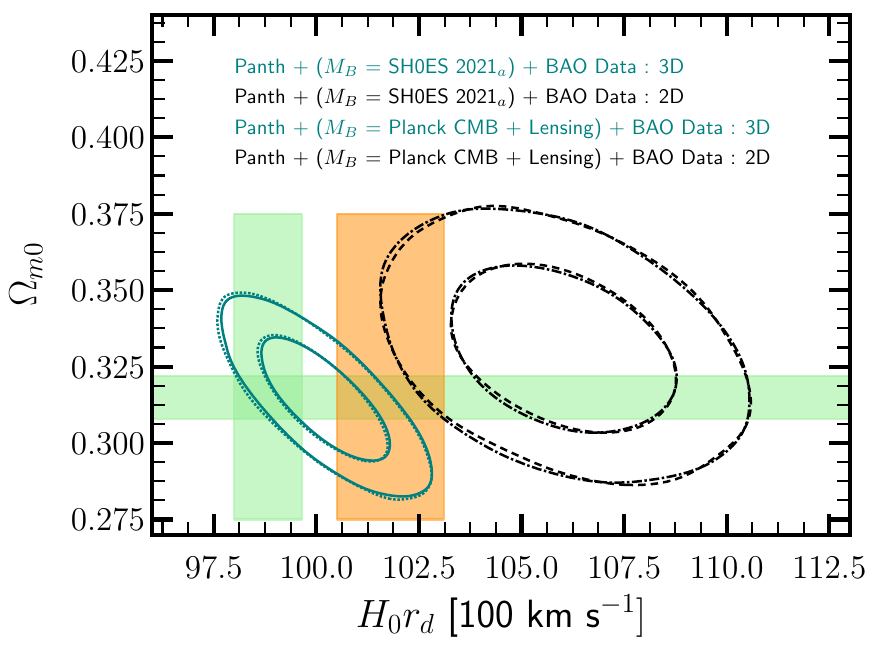}}
\caption {\label{fig:resrdhcompareandom0countour} Left: This plot shows the values of the product of the Hubble parameter at present and sound horizon at the drag epoch $H_0r_d$ obtained while utilising various calibrations for the Pantheon Plus sample and two different BAO datasets. Right: This plot shows the contours of $H_0r_d$ and $\Omega_{m0}$ matter density at present obtained for various calibrations for the Pantheon Plus sample when combined with two different BAO datasets. The orange and light green bands used for comparison are the results from DESI ($r_dh = 101.8 \pm 1.3$ Mpc) and CMB temperature, polarization and lensing ($r_dh = 98.82 \pm 0.82$ Mpc and $\Omega_{m0} = 0.315 \pm 0.007$) respectively. We used the notation $r_dh$ $\equiv$ $H_0r_d$/(100 \Hunit) in both plots.}

\end{figure*}

\textbf{a) Pantheon Plus Calibration with $M_B$ Compatible with Low-Redshift Experiments }\\~\\
Our analysis involved utilizing five different measurements of $H_0$ from various low-redshift studies. We estimated the absolute magnitude of Type-Ia Supernovae ($M_B$) from these measurements to calibrate our Pantheon Plus sample. As illustrated in Table \ref{Tab:res}, the upper-left portion, utilizing the Pantheon Plus and BAO Data: 3D, yielded an estimated $r_d \sim  135$ Mpc, exhibiting more than three sigma tension with Planck's estimation of $r_d$. Conversely, when employing the 2D BAO Data alongside the same Pantheon Plus sample and the same calibration of $M_B$, we obtained $r_d$ elevated to around $ 142$ Mpc, aligning with Planck's $r_d$ within one sigma (see Figure \ref{fig:resrdhcompareandom0countour})\footnote{In the left panel of Figure \ref{fig:resrdhcompareandom0countour}, the values of $H_0$ and $r_d$ are calculated separately, and the product $H_0 r_d$ is manually computed using the entries from Table IV. In this process, the uncertainties are propagated assuming no correlation between $H_0$ and $r_d$, which typically results in larger and less precise error bars. For instance, for the 3D BAO dataset plus SH0ES 2021$_a$ calibration, the derived $r_dh$ is $100.21 \pm 1.03$ Mpc when directly inferred from the MCMC posterior, but the manually computed value is $100.21 \pm 2.39$ Mpc, clearly showing the impact of neglecting correlations. Similarly, for the 3D BAO dataset plus CMB lensing calibration, the derived $r_dh$ is $100.14 \pm 1.04$ Mpc, compared to $100.14 \pm 1.58$ Mpc from manual multiplication.\\
In contrast, the right panel of Figure \ref{fig:resrdhcompareandom0countour} shows results obtained directly from the posterior chains, where the full covariance between $H_0$ and $r_d$ is taken into account. This method accurately reflects the joint probability distribution, leading to more robust and stable error bars. Notably, the contours for $H_0 r_d$ and $\Omega_{m0}$ remain consistent across different SNe calibrations, as they properly include all correlations.\\
Therefore, the apparent discrepancy in error bars between the two panels—particularly the smaller error bars for Planck in the left panel compared to SH0ES—arises solely from differences in how uncertainties are computed. The right panel presents the statistically rigorous result, while the left serves as an illustrative comparison using uncorrelated propagation.}

From this analysis, we infer that while the tension in $H_0$ between Planck and SH0ES is confirmed, there is no tension in $r_d$ when using the 2D data. With $H_0$ around $73$ \Hunit, we achieve an $r_d$ of roughly $142$ Mpc, compatible with Planck \footnote{We designate as our \textbf{\textit{baseline analysis}} this scenario where $H_0$ aligns with SH0ES measurements while $r_d$ remains compatible with Planck predictions. This concordance emerges specifically when employing 2D BAO data in conjunction with SNe calibrations derived from low-redshift experiments.}. To ensure the robustness of our results independent of the calibration of the Pantheon Plus sample, we consider calibrations of $M_B$ from various experiments. Notably, our findings from all low-redshift experiments converge, indicating mutual consistency. \\~\\

\textbf{b) Pantheon Plus Calibration with $M_B$ Compatible with High-Redshift Experiments}

We calibrated the Pantheon Plus sample using $M_B$ calibration from three $H_0$ measurements for high redshifts, namely Planck CMB + Lensing, BOSS DR12 + BBN, and ACT+WMAP CMB. We found that with the Pantheon Plus calibrated to high-redshift experiments, we obtained $h$ value around $0.67$ and $r_d$ close to $147$ Mpc while using the BAO 3D data. However, when utilizing the BAO 2D data with the same calibration, we obtained the same $H_0$ value but $r_d \sim 156$ Mpc, which exhibits tension with Planck's value of $r_d$. 
Since 2D BAO measures a higher product of $hr_d$ than 3D BAO (Equation \ref{eq:rdh2D}, \ref{eq:rdh3D}, and Figure \ref{fig:resrdhcompareandom0countour}), a higher $r_d$ corresponding to higher $h$ can be allowed (also see Figure \ref{fig:comaparerdlowandhighz} of Appendix).

\section{Comments on Cosmological Model Used}\label{sec5:ede}
While keeping in mind that BAO constrains the product $H(z)r_d$ and not $H_0$ and $r_d$ individually, the presented results (Figure \ref{fig:resrdhcompareandom0countour}-left plot) reveals less than 1.5$\sigma$ tension within two independent BAO datasets: 2D BAO and 3D BAO datasets. But if we look at contour $H_0r_d$-$\Omega_{m0}$ (Figure \ref{fig:resrdhcompareandom0countour}-right plot), the tension in $H_0r_d-\Omega_m0$ contours is more than two sigma. We can clearly see that there is no tension between $\Omega_{m0}$ of 2D BAO, 3D BAO and CMB estimated $\Omega_{m0}$. The discrepancy appeared in 2D and 3D BAO data when exploring $\Omega_{m0}$-$H_0r_d$ contour plane.

\begin{table}
\centering
\resizebox{0.5\textwidth}{!}{%
\begin{tabular}{|c|c|c|}
\hline
\hline
& \multicolumn{1}{c|}{Panth Plus + BAO : 3D} & \multicolumn{1}{c|}{Panth Plus + BAO : 2D} \\
\hline \hline
Measurements & $\Omega_{m0} h^2$ & $\Omega_{m0} h^2$ \\
\hline
SH0ES 2021$_a$ & $0.175 \pm 0.008$ & $0.184 \pm 0.011$ \\ \hline
SH0ES 2021 & $0.171 \pm 0.009$ & $0.180 \pm 0.011$ \\ \hline
Masers & $0.170 \pm 0.012$ & $0.182 \pm 0.017$ \\ \hline
SH0ES 2019 & $0.174 \pm 0.009$ & $0.184 \pm 0.012$ \\ \hline
Tully Fisher & $0.176 \pm 0.010$ & $0.193 \pm 0.015$ \\ \hline \hline
Planck CMB + Lensing & $0.145 \pm 0.006$ & $0.152 \pm 0.008$ \\ \hline
BOSS DR12 + BBN & $0.150 \pm 0.011$ & $0.157 \pm 0.013$ \\ \hline
ACT + WMAP CMB & $0.146 \pm 0.008$ & $0.153 \pm 0.009$ \\ \hline
\end{tabular}
}
\caption{Constraints on derived parameter $\Omega_{m0} h^2$ for $\Lambda$CDM model for both BAO Data Sets and various Pantheon Plus calibrations $M_B$. The error bars quoted are at $1\sigma$ confidence interval.}
\label{Tab:resomh2}
\end{table}

To improve visualization and highlight the underlying trends in the right plot, we used the same colour for the 2D and 3D BAO datasets across  SNe calibrations, while varying the line style. This approach makes it clear that the error bars on $H_0 r_d$ and $\Omega_{m0}$ remain nearly identical, or exactly the same when changing the SNe calibration. The main effect on the results is entirely driven by the choice of BAO dataset (2D or 3D), rather than the SNe calibration.

\begin{figure}
\vspace{0.1cm}
\centering
\resizebox{240pt}{180pt}{\includegraphics{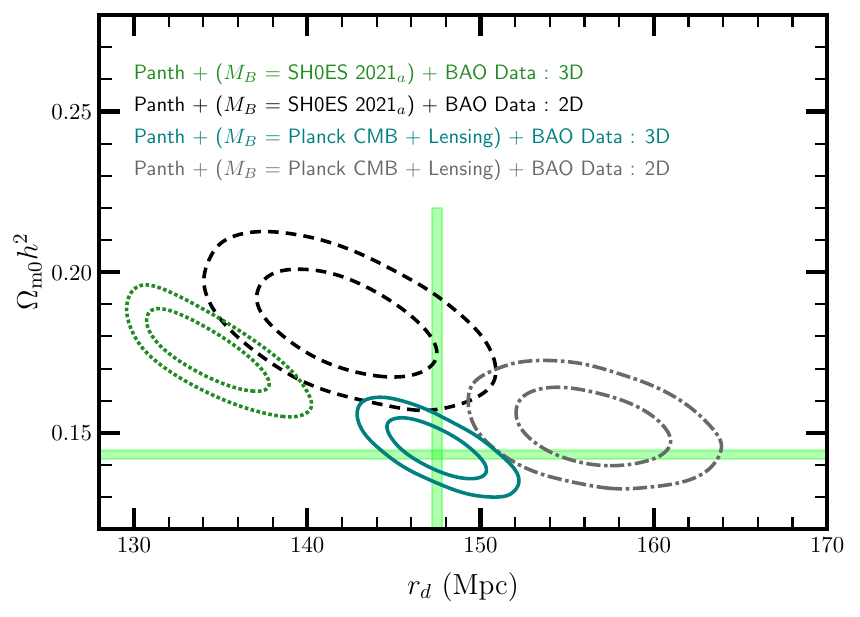}}
\caption {\label{fig:rdom0h2countour} This plot shows all four combinations as Fig. \ref{fig:resrdhcompareandom0countour} but in countour space of $(r_d, \Omega_{m0}h^2)$. Green bands shows constraints from Planck 2018 results ($\Omega_{m0}h^2= 0.1432 \pm 0.0013$; $r_d = 147.05 \pm 0.30$ Mpc). }
\end{figure}

The observed tension in $H_0 r_d$ between the 2D and 3D BAO datasets in the $\Omega_{m0} - H_0 r_d$ plane \footnote{The derived $H_0 r_d$ from posterior chains has smaller error bars compared to manually propagating uncertainties, as it accounts for correlations between $H_0$ and $r_d$. Furthermore, the SNe calibration does not significantly affect the results, as the error bars on $H_0 r_d$ and $\Omega_{m0}$ remain consistent even when the SNe calibration changes.} suggests that either systematic uncertainties in the BAO measurements are playing a significant role or that the assumptions of the $\Lambda$CDM model may not fully capture the underlying physics.

We find that the cosmological parameter $\Omega_{m0}$ remains consistent with the Planck 2018 results, regardless of the combination of data—Pantheon Plus (all $M_B$ calibrations) + (BAO 2D or BAO 3D)—used (see Table \ref{Tab:res}). While the tension in the $\Omega_{m0} - H_0 r_d$ plane is visible (Figure \ref{fig:resrdhcompareandom0countour}-right plot), it is not immediately clear what compensates for the higher values of $H_0 r_d$ when $\Omega_{m0}$ remains constant and compatible with Planck. To better understand this behavior and enable a direct comparison with Planck constraints, we now focus on the parameter $\Omega_{m0}h^2$, which is directly constrained by Planck. We present constraints on $\Omega_{m0}h^2$ for both 2D and 3D BAO datasets, combined with all the SNe calibrations used in this analysis. Furthermore, we examine the correlation between $\Omega_{m0}h^2$ and $r_d$, as shown in Figure \ref{fig:rdom0h2countour} and presented in Table \ref{Tab:resomh2}, to gain deeper insights into the observed discrepancies. We observe that for the parameter $\Omega_{m0}h^2$ , the combination of BAO 3D data with SNe calibrations derived from high-redshift experiments demonstrates excellent agreement with the Planck estimated value. However, the combination previously referred to as the \textbf{\textit{baseline analysis}} yields a value for $\Omega_{m0}h^2$ that deviates by more than $3.5\sigma$ from the Planck-estimated value.\footnote{Since $r_d$ is a function of $\Omega_{m0}h^2$ and $\Omega_{b0}h^2$ (Equation \ref{eq:rd}), maintaining consistency between $r_d$ and $\Omega_{m0}h^2$ may require adjustments to the parameter $\Omega_{b0}h^2$, which is very tightly constrained by Big Bang Nucleosynthesis (BBN). We plan to investigate this aspect comprehensively in future work.}

This discrepancy highlights the importance of carefully assessing the impact of model assumptions and systematic effects when using 2D BAO data to address the Hubble tension. Understanding the source of this 2D-3D inconsistency should be a priority for future investigations, as it may have profound implications for the interpretation of cosmological parameters. A detailed investigation into the origin of the differences between 2D and 3D BAO measurements is currently underway and will aim to provide clarity on this issue \cite{Ruchika_bao}.

\section{Analysis with DESI data}\label{sec6:desi}

\subsection{Model, Methodology, and Dataset}

In our analysis with DESI data, we adhered to the standard cosmological model, $\Lambda$CDM, to compare with the results obtained in the previous sections. However, to highlight DESI's main findings, we extended our analysis to include the Chevallier-Polarski-Linder (CPL) model \citep{Polarski}, which introduces a redshift-dependent equation of state parameter $w(z) = w_0 + w_a(1-a) = w_0 + w_a\frac{z}{1+z}$, providing a more nuanced characterization of cosmic dynamics.

In the CPL model, the Hubble parameter $H(z)$ is expressed as:
\begin{equation}
\frac{H^2(z)}{H_0^2} = \Omega_{m0}(1+z)^3 + (1-\Omega_{m0})f(z),
\end{equation}
where $f(z) = \exp\left(3\int^z\frac{(1+w(x))}{1+x}dx\right)$. 

\begin{table}
  \centering
    \begin{tabular}{|c|c|c|c|c|c|}
    \toprule
    Tracer & $z$ & $D_M/r_d$ & $D_H/r_d$ & $D_V/r_d$ \\
    \hline
    BGS   & 0.30  & -     & -     & $7.93 \pm 0.15$ \\
    LRG   & 0.51  & $13.62 \pm 0.25$ & $20.98 \pm 0.61$ & - \\
    LRG   & 0.71  & $16.85 \pm 0.32$ & $20.08 \pm 0.60$ & - \\
    LRG+ELG & 0.93  & $21.71 \pm 0.28$ & $17.88 \pm 0.35$ & - \\
    ELG   & 1.32  & $27.79 \pm 0.69$ & $13.82 \pm 0.42$ & - \\
    QSO   & 1.49  & -     & -     & $26.07 \pm 0.67$ \\
    Lya QSO & 2.33  & $39.71 \pm 0.94$ & $8.52 \pm 0.17$ & - \\
    \hline
    \end{tabular}%
      \caption{DESI Data Release 1 BAO Measurement used}
  \label{tab:desi}%
\end{table}%

\begin{figure*}
\resizebox{490pt}{120pt}{\includegraphics{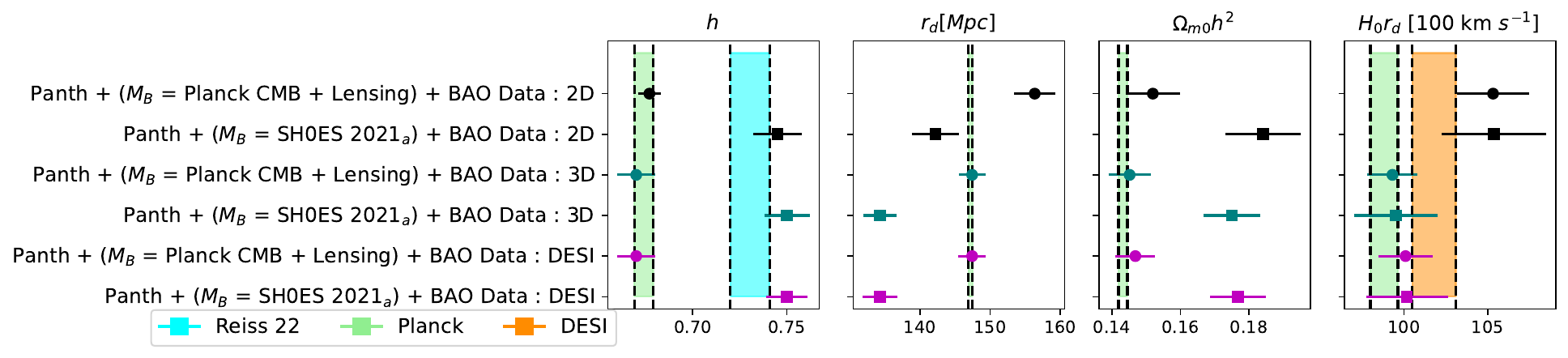}}
\vspace{-0.1cm}
\caption{\label{fig:comparefourparams}This plot shows cosmological parameters $h$, $r_d$ along with the derived parameter $\Omega_{m0}h^2$ and product of Hubble parameter and sound horizon at the drag epoch $H_0r_d$ obtained using various calibration for Pantheon Plus sample combined with two different BAO datasets. In the $h$ subplot, we show green and cyan bands from Planck 2018 and Reiss '22 for comparison. In the $r_d$ and $\Omega_{m0}h^2$ subplots, we show Planck corresponding values. Further in the extreme right subplot of $H_0r_d$, the orange and light green bands used for comparison are the results from DESI ($r_dh = 101.8 \pm 1.3$ Mpc) and CMB temperature, polarization and lensing ($r_dh = 98.82 \pm 0.82$ Mpc). We used the notation $r_dh$ $\equiv$ $H_0r_d$/(100 \Hunit).  }
\end{figure*}

We applied the same methodology and definitions used for conducting the Chisq analysis of the 3D BAO dataset to analyze the DESI data. This includes the Equation (\ref{eq:chiPCF-aniso}) and Equation (\ref{eq:chiPCF-iso}) used for calculating the Chisq values for the DESI dataset. The results obtained for the $\Lambda$CDM model are presented in Table \ref{Tab:resdesi}. Our study involved a comparison of data while maintaining the same model. Table \ref{Tab:res}, Table \ref{Tab:resdesi}, and Figure \ref{fig:comparefourparams} depict the comparison of three BAO datasets and their impact on cosmological parameters under the assumption of cosmology as standard $\Lambda$CDM. 

\begin{table*}
\centering
\begin{tabular}{|c|c|c|c|}
\hline
& \multicolumn{3}{c|}{Panth  Plus + BAO Data : DESI}  \\
\hline \hline
Measurements & $r_d$ (Mpc) & $h$ & $\Omega_{m0}$  \\
\hline
SH0ES 2021$_a$ & $134.33 \pm 2.46$ & $0.75 \pm 0.011$ & $0.318 \pm 0.0124$  \\ \hline
SH0ES 2021 & $135.83 \pm 2.74$ & $0.74 \pm 0.013$ & $0.318 \pm 0.0125$  \\ \hline
Masers & $136.56 \pm 4.44$ & $0.73 \pm 0.02$ & $0.317 \pm 0.0126$  \\ \hline
SH0ES 2019 & $134.68 \pm 2.67$ & $0.74 \pm 0.01$ & $0.318 \pm 0.0123$  \\ \hline
Tully Fisher & $134.26 \pm 3.12$ & $0.75 \pm 0.02$ & $0.311 \pm 0.0120$  \\
\hline \hline
Planck CMB + Lensing & $147.44 \pm 1.95$ & $0.68 \pm 0.01$ & $0.319 \pm 0.0126$   \\ \hline
BOSS DR12 + BBN & $145.10 \pm 4.87$ & $0.69 \pm 0.022$ & $0.318 \pm 0.0127$   \\ \hline
ACT + WMAP CMB & $146.95 \pm 2.82$ & $0.68 \pm 0.01$ & $0.318 \pm 0.0127$  \\
\hline
\end{tabular}
\caption{Constraints on parameters for $\Lambda$CDM for BAO Data DESI data. The error bars quoted are at $1\sigma$ confidence interval.}
\label{Tab:resdesi}
\end{table*}

\section{Discussion of 2D, 3D and DESI BAO Results}\label{sec7:disussionallbao}
Here, we delve into the results and comparison between the 2D and 3D BAO datasets, alongside the new DESI data release. Figure \ref{fig:comparefourparams} shows the comparison of two cosmological parameters $h$, $r_d$ along with the product of Hubble parameter and sound horizon at the drag epoch $H_0r_d$ and derived parameter $\Omega_{m0}h^2$ obtained while using various calibrations for Pantheon Plus sample and combined with three BAO datasets 2D, 3D and BAO. In the $h$ subplot, we show green and cyan bands from Planck 2018 and Reiss '22 \cite{R22} for comparison. In the $r_d$ and $\Omega_{m0}h^2$ subplots, we show Planck corresponding values. Further in the extreme right subplot of $H_0r_d$, the orange and light green bands used for comparison are the results from DESI ($r_dh = 101.8 \pm 1.3$ Mpc) and CMB temperature, polarization and lensing ($r_dh = 98.82 \pm 0.82$ Mpc). We used the notation $r_dh$ $\equiv$ $H_0r_d$/(100 \Hunit). To distinguish between the BAO datasets and draw conclusions independent of the Pantheon calibration, it is important to focus on the quantity that BAO measurements constrain: the product $h r_d$. As shown here, both 2D and 3D BAO are consistent with the $h r_d$ values obtained from DESI and the CMB within 2$\sigma$. Several other studies \citep{ Nunes1, Nunes2,Gomez-Valent:2023uof, lemos} have reported similar cosmological constraints while utilising 2D BAO dataset. \citep{favale} also clearly presents the discrepancy between 2D and 3D BAO datasets.

\section{Conclusions}
Through a critical examination and detailed analysis of 2D and 3D BAO datasets within the theoretical framework of the standard model of cosmology, we assessed the consistency between these datasets. Our analysis reveals that the product $h r_d$ obtained from both 2D and 3D BAO datasets is consistent within 2$\sigma$. However, the $h r_d$ value derived from the 2D BAO analysis is systematically higher, which can lead to a higher $h$ (comparable to \citep{R22}) as well as a higher $r_d$ (closely matching the Planck-estimated value). This suggests that while the 2D BAO dataset holds significant potential, it should be used with caution when addressing the Hubble tension.

This systematic difference has interesting implications for theoretical models addressing the Hubble tension. When using 2D BAO measurements, the naturally higher $h r_d$ values inherent to this methodology might influence the derived cosmological parameters. In particular, theoretical models achieving higher $H_0$ values while maintaining Planck-compatible $r_d$ values and proposing a resolution to the Hubble tension would benefit from additional validation using multiple BAO methodologies. Such cross-validation would help distinguish whether the resolution stems from the physical mechanisms proposed by the models or reflects the systematic properties of the 2D BAO measurements.

Additionally, focusing on $\Omega_{m0}h^2$, a parameter directly constrained by Planck, provides valuable insights into the observed discrepancies. Our results show that $\Omega_{m0}$ remains consistent with Planck constraints, regardless of the combination of data (Pantheon Plus + BAO 2D or 3D) used. However, the combination previously referred to as the baseline analysis yields a value for $\Omega_{m0}h^2$ that deviates by more than $3.5\sigma$ from Planck’s estimate. This highlights the importance of considering $\Omega_{m0}h^2$ for direct comparisons with Planck and understanding how its correlation with $r_d$ (as shown in Figure \ref{fig:rdom0h2countour}) contributes to these discrepancies.

As shown in the $\Omega_{m0} - H_0 r_d$ plane (Figure \ref{fig:resrdhcompareandom0countour} and Figure \ref{fig:rdom0h2countour}), resolving the tension between the 2D and 3D BAO contours may require identifying potential systematic uncertainties within the datasets or reconsidering key assumptions in their analysis. These systematics or assumptions could impact how cosmological parameters are derived, and addressing them is crucial for improving the reliability of results and ensuring consistency between observations.

Our future work will focus on exploring additional 2D and 3D BAO datasets to further investigate these discrepancies. This effort will aim to pinpoint the source of the differences and refine the precision of cosmological parameter estimates. Understanding these systematic differences is essential not only for the proper interpretation of cosmological measurements but also for evaluating proposed solutions to the Hubble tension.

\section*{Acknowledgements}
The author Ruchika would like to thank Alessandro Melchiorri, Florian Beutler, Ravi Seth, M. M. Sheikh Jabbari, Nils Schöneberg, Jalison Alcaniz, Thais Lemos, Giacomo Gradenigo and Anjan Ananda Sen for useful discussions. We acknowledge IUCAA, Pune, India, for providing access to their computational facilities. We also acknowledge financial support from TASP, iniziativa specifica INFN. This work was partially supported by Project SA097P24, funded by Junta de Castilla y León. Finally, we sincerely thank the anonymous referee for their detailed and constructive feedback, which has significantly improved the quality and clarity of our manuscript.

\appendix

\section{Appendix A}
\subsection{Covariance matrix for 3D BAO}
The full correlation matrix corresponding to anisotropic constraints corresponding to elements in Table \ref{tabledata:PCF} is given in Equation \eqref{eq:covmatrix}.

\begin{figure*}
\centering
\begin{equation}\label{eq:covmatrix}\tag{A.1}
\mathbf{C} = 
\normalsize
\begin{pmatrix}
0.0150 & -0.0358 & 0.0071 & -0.0100 & 0.0032 & -0.0036 & 0 & 0 \\
-0.0357 & 0.5304 & -0.0160 & 0.1766 & -0.0083 & 0.0616 & 0 & 0 \\
0.0071 & -0.0160 & 0.0182 & -0.0323 & 0.0097 & -0.0131 & 0 & 0 \\
-0.0100 & 0.1766 & -0.0323 & 0.3267 & -0.0167 & 0.1450 & 0 & 0 \\
0.0032 & -0.0083 & 0.0097 & -0.0167 & 0.0243 & -0.0352 & 0 & 0 \\
-0.0036 & 0.0616 & -0.0131 & 0.1450 & -0.0352 & 0.2684 & 0 & 0 \\
0 & 0 & 0 & 0 & 0 & 0 & 0.1358 & -0.0296 \\
0 & 0 & 0 & 0 & 0 & 0 & -0.0296 & 0.0492
\end{pmatrix}
\end{equation}
\end{figure*}

\section{ BAO Data for redshift less than 1}

\begin{figure}
\resizebox{240pt}{180pt}{\includegraphics{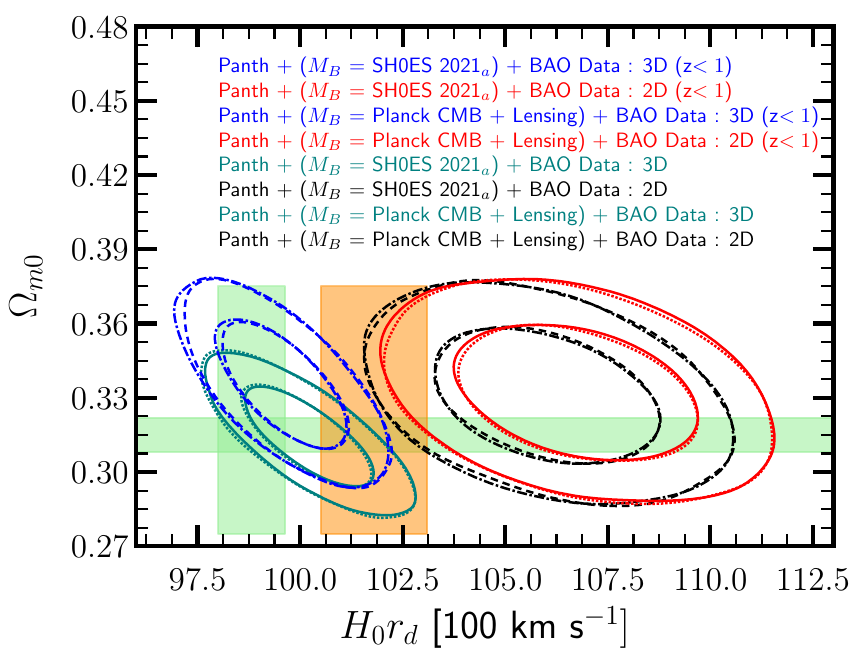}}
\vspace{0.1cm}
\caption{\label{fig:appendixzlessthan1z} This plot shows the contours of $H_0r_d$ and $\Omega_{m0}$ matter density at present obtained for various calibrations for the Pantheon Plus sample when combined with two different BAO datasets (Teal and Black contours are for full BAO dataset whereas blue and red contours are for BAO Dataset (z$<1$)). The orange and light green bands used for comparison are the results from DESI ($r_dh = 101.8 \pm 1.3$ Mpc) and CMB temperature, polarization and lensing ($r_dh = 98.82 \pm 0.82$ Mpc and $\Omega_{m0} = 0.315 \pm 0.007$) respectively. We used the notation $r_dh$ $\equiv$ $H_0r_d$/(100 \Hunit) in the plot. }
\end{figure}

To assess the potential bias introduced by BAO data at redshifts greater than one, we removed BAO data for redshift $z>1$ and performed the analysis again. The results presented in Figure \ref{fig:appendixzlessthan1z} are similar to those obtained when all BAO redshift measurements were included (Figure \ref{fig:resrdhcompareandom0countour}-right). However, the contours obtained while using 3D BAO data shift toward a higher value of $\Omega_{m0}$. Despite this shift, our conclusion that there is more than two-sigma tension in $H_0r_d$ in $\Omega_{m0}-H_0r_d$ plane remains unchanged.

\subsection{Redshift Dependence of Sound Horizon Measurements}

\begin{figure*}
\vspace{0.1cm}
\resizebox{240pt}{180pt}{\includegraphics{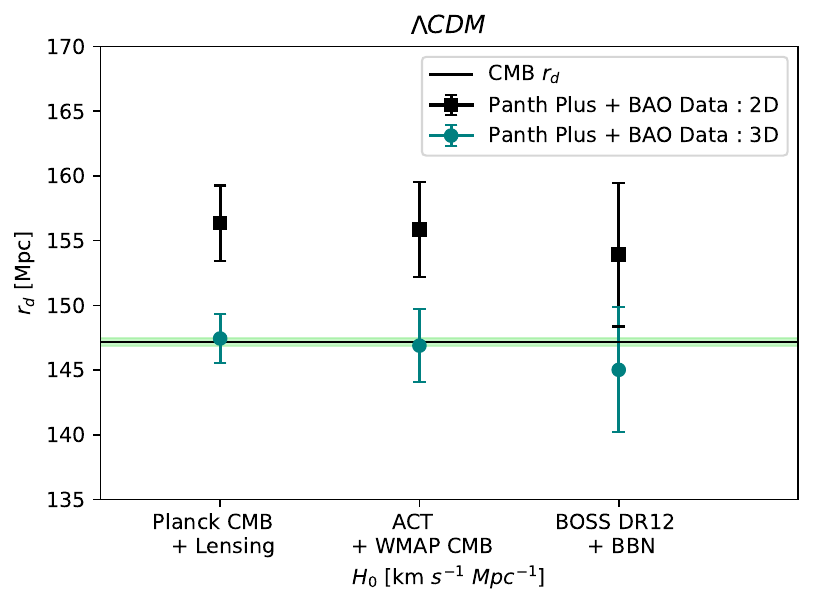}}
\vspace{0.1cm}
\resizebox{240pt}{180pt}{\includegraphics{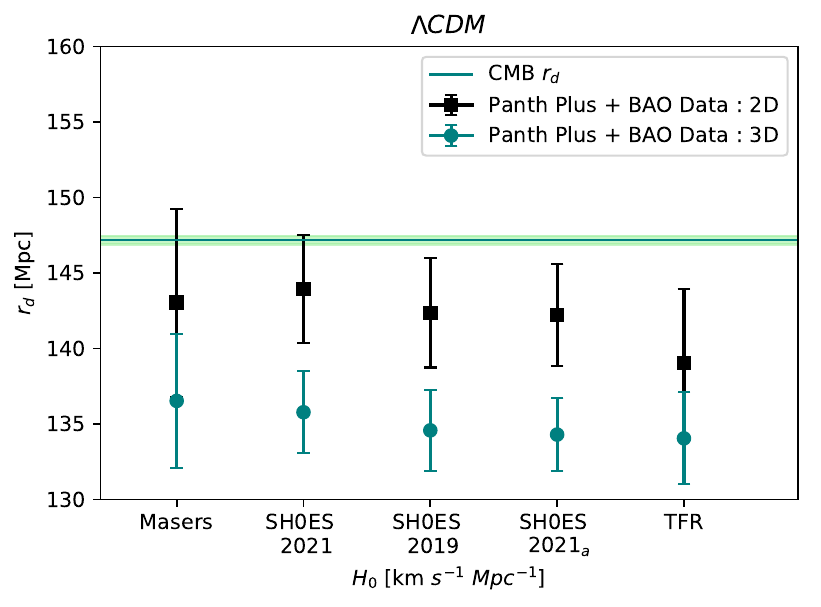}}
\caption{\label{fig:comaparerdlowandhighz} This plot illustrates the constraints obtained on $r_d$ given $M_B$ or $H_0$ values derived from various experiments. A green horizontal band across both panels represents the $r_d$ constraints from the Planck CMB observations, serving as a reference scale.}

\end{figure*}

Figure \ref{fig:comaparerdlowandhighz} examines $r_d$ values derived from various experimental measurements, divided into high-redshift (left panel) and low-redshift (right panel) surveys. Using 3D BAO with SNe data, we find that experiments like Planck CMB, BOSS+BBN, and ACT+WMAP (left panel) yield $r_d$ values clustering around Planck's CMB constraint (green band). In contrast, when using measurements from low-redshift experiments like SH0ES, Masers, and Tully-Fisher (right panel), we obtain systematically lower $r_d$ values compared to Planck's measurement. This pattern, well recognized in the literature, persists across different experimental calibrations, reinforcing the known tension between high and low redshift measurements.

Notably, while 3D BAO measurements calibrated with low-redshift experiments yield $r_d \sim 135$ Mpc ( $> 2 \sigma$ tension with Planck), 2D BAO measurements with identical experimental calibrations give $r_d \sim 142$ Mpc, consistent with Planck within $1 \sigma$. This systematic difference between 2D and 3D BAO methodologies suggests that BAO analysis choice significantly impacts cosmological inferences. This figure complements Figure \ref{fig:resrdhcompareandom0countour}-right by focusing on individual $r_d$ values rather than the combined parameter $H_0r_d$.

\subsection{Additional analysis for DESI results}
\begin{figure*}
\vspace{0.1cm}
\resizebox{240pt}{180pt}{\includegraphics{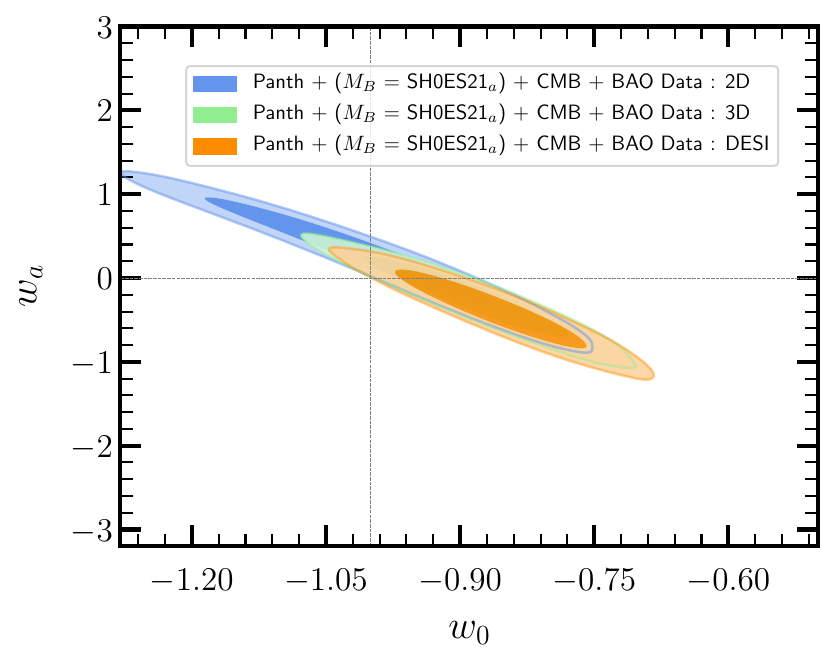}}
\vspace{0.1cm}
\resizebox{240pt}{180pt}{\includegraphics{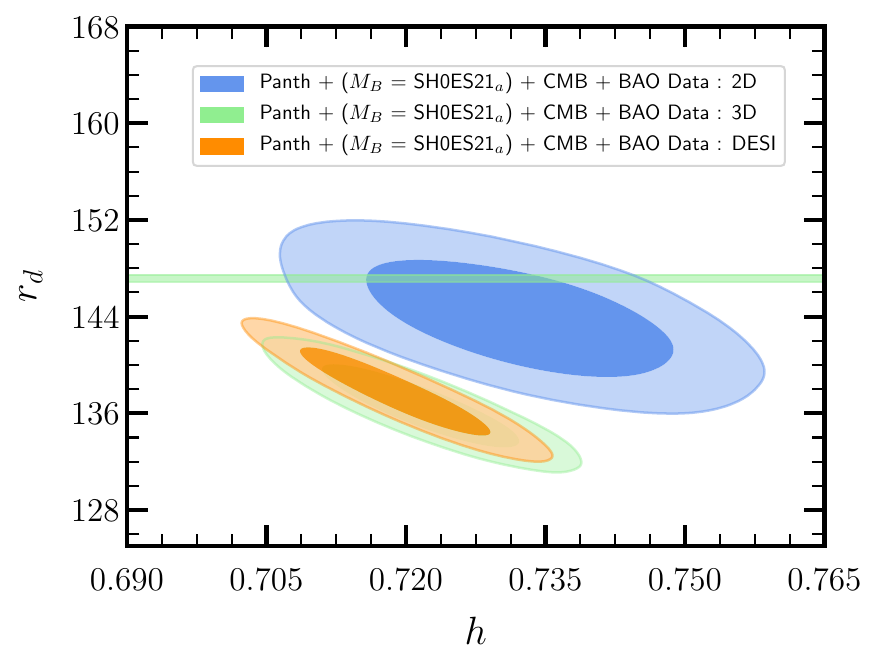}}
\caption{\label{fig:cmb+data} Left : This plot shows how the $w_0-w_a$ contour plot shifts when we replace 3D dataset to 2D dataset. It also shows that the BAO 3D and DESI BAO datasets have contours that are nearly overlapping. Right: This plot shows three $h-r_d$ contours for the combination of Pantheon Plus + CMB + BAO dataset. The three contours are for three different data combinations. It clearly agrees with the fact that the product $hr_d$ for BAO: 2D is greater than that for BAO : 3D or BAO DESI. Keeping $h$ constant, $r_d$ estimated values from Panth Plus + BAO: 2D data combination agrees with Planck $r_d$ within one sigma while Planck $r_d$ is more than 2~$\sigma$ away from Panth Plus + BAO 3D estimate. In both the plots, it is evident that BAO 2D has less constraining power hence giving bigger contours than the BAO 3D dataset.}
\end{figure*}

Other than the Pantheon Plus dataset and DESI BAO dataset as provided in Table \ref{tab:desi}, we also incorporated CMB data.
In particular, we utilised the \textit{Planck} 2018 compressed likelihood for TT, TE, EE + lowE as obtained by \citep{Chen:2018dbv} (for the detailed method for obtaining the compressed likelihood (see  \citep{ade2}).
To study these data combinations, we also utilize CPL model \citep{Polarski}. In Figure \ref{fig:cmb+data}, we present the $w_0-w_a$ and $h-r_d$ planes. Both plots feature the Pantheon Plus sample calibrated with SH0ES21$_a$ along with CMB data, and three BAO datasets: 2D, 3D, and DESI Release.

In the left contour plot, showcasing the $w_0-w_a$ plane, it's notable that the (-1.0,0) point lies on the edge of the two-sigma contour across all three datasets. The DESI and 3D BAO datasets exhibit comparable constraining power, while the 2D dataset displays larger contours. Moreover, the contour in the one-sigma region extends towards lower values of $w_0$ and higher values of $w_a$ for the 2D dataset, while maintaining the correlation. In the right plot, a particularly noteworthy observation emerges. By maintaining consistency with the other two datasets (CMB and Sne Ia) and solely altering the BAO datasets, a significant finding surfaces: when utilizing the BAO 2D dataset, our estimated $r_d$ aligns with Planck's $r_d$ within a one-sigma region. However, for 3D BAO and DESI Release, the constraints on $r_d$ deviate from Planck's $r_d$ by more than two sigma.

\begin{figure*}
\vspace{0.1cm}
\resizebox{240pt}{180pt}{\includegraphics{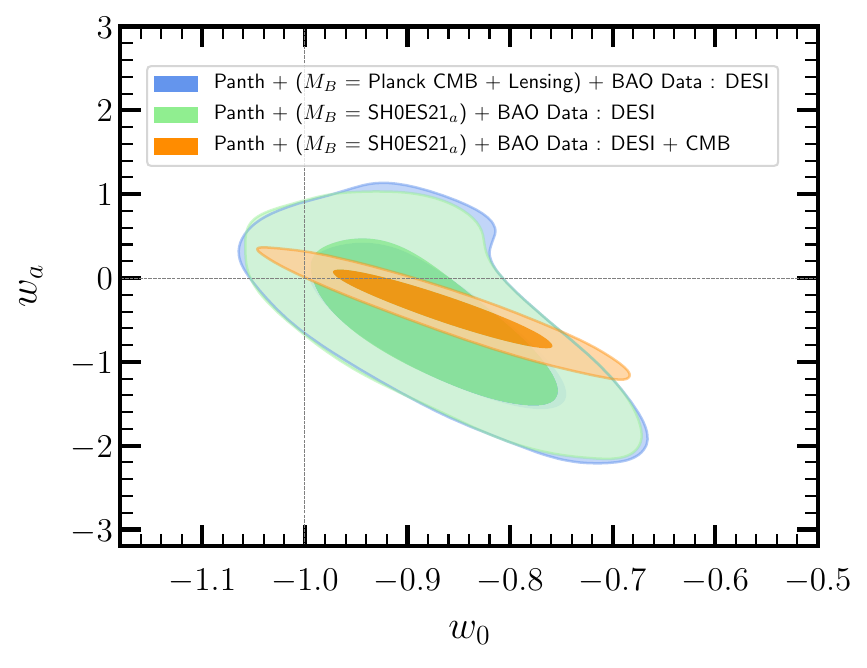}}
\vspace{0.1cm}
\resizebox{240pt}{180pt}{\includegraphics{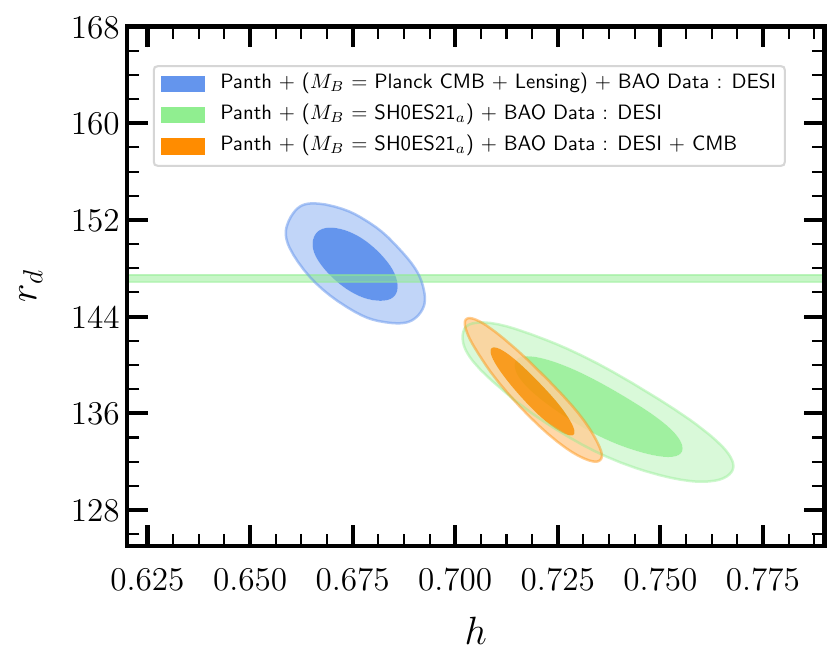}}
\caption{\label{fig:withandwithoutcmb} The left plot shows the $w_0-w_a$ plane when we change the calibration of Pantheon Plus. It also shows how the constraints get better as we add CMB data. In the right plot, we can see the $h-r_d$ contours for different cosmological dataset combinations.}
\end{figure*}

In Figure \ref{fig:withandwithoutcmb}, the orange contour is the same as in Figure \ref{fig:cmb+data} and is provided for comparison and reference. In the left plot, we observe that the contours expand notably when the CMB dataset is removed. This behavior is a direct consequence of CMB's ability to place strong constraints on both $\Omega_M h^2$ and $\Omega_b h^2$ parameters, which together determine the sound horizon scale $r_d$. The removal of CMB data eliminates these tight parameter constraints, resulting in significantly broader contours that reflect the reduced precision in our $r_d$ determination when relying solely on non-CMB measurements. Interestingly, altering the calibration of the Pantheon Plus sample does not induce significant changes in the behaviour of the $w_0-w_a$ contours. However, it does lead to substantial shifts in the values of both parameters $h$ and $r_d$ in the right plot.

\bibliographystyle{apsrev4-2} 
\bibliography{main}

\begin{thebibliography}{133}%
\makeatletter
\providecommand \@ifxundefined [1]{%
 \@ifx{#1\undefined}
}%
\providecommand \@ifnum [1]{%
 \ifnum #1\expandafter \@firstoftwo
 \else \expandafter \@secondoftwo
 \fi
}%
\providecommand \@ifx [1]{%
 \ifx #1\expandafter \@firstoftwo
 \else \expandafter \@secondoftwo
 \fi
}%
\providecommand \natexlab [1]{#1}%
\providecommand \enquote  [1]{``#1''}%
\providecommand \bibnamefont  [1]{#1}%
\providecommand \bibfnamefont [1]{#1}%
\providecommand \citenamefont [1]{#1}%
\providecommand \href@noop [0]{\@secondoftwo}%
\providecommand \href [0]{\begingroup \@sanitize@url \@href}%
\providecommand \@href[1]{\@@startlink{#1}\@@href}%
\providecommand \@@href[1]{\endgroup#1\@@endlink}%
\providecommand \@sanitize@url [0]{\catcode `\\12\catcode `\$12\catcode `\&12\catcode `\#12\catcode `\^12\catcode `\_12\catcode `\%12\relax}%
\providecommand \@@startlink[1]{}%
\providecommand \@@endlink[0]{}%
\providecommand \url  [0]{\begingroup\@sanitize@url \@url }%
\providecommand \@url [1]{\endgroup\@href {#1}{\urlprefix }}%
\providecommand \urlprefix  [0]{URL }%
\providecommand \Eprint [0]{\href }%
\providecommand \doibase [0]{https://doi.org/}%
\providecommand \selectlanguage [0]{\@gobble}%
\providecommand \bibinfo  [0]{\@secondoftwo}%
\providecommand \bibfield  [0]{\@secondoftwo}%
\providecommand \translation [1]{[#1]}%
\providecommand \BibitemOpen [0]{}%
\providecommand \bibitemStop [0]{}%
\providecommand \bibitemNoStop [0]{.\EOS\space}%
\providecommand \EOS [0]{\spacefactor3000\relax}%
\providecommand \BibitemShut  [1]{\csname bibitem#1\endcsname}%
\let\auto@bib@innerbib\@empty
\bibitem [{\citenamefont {Ade}\ \emph {et~al.}(2016{\natexlab{a}})\citenamefont {Ade} \emph {et~al.}}]{ade1}%
  \BibitemOpen
  \bibfield  {author} {\bibinfo {author} {\bibfnamefont {P.~A.~R.}\ \bibnamefont {Ade}} \emph {et~al.} (\bibinfo {collaboration} {Planck}),\ }\href {https://doi.org/10.1051/0004-6361/201525830} {\bibfield  {journal} {\bibinfo  {journal} {Astron. Astrophys.}\ }\textbf {\bibinfo {volume} {594}},\ \bibinfo {pages} {A13} (\bibinfo {year} {2016}{\natexlab{a}})},\ \Eprint {https://arxiv.org/abs/1502.01589} {arXiv:1502.01589 [astro-ph.CO]} \BibitemShut {NoStop}%
\bibitem [{\citenamefont {Ade}\ \emph {et~al.}(2016{\natexlab{b}})\citenamefont {Ade} \emph {et~al.}}]{ade2}%
  \BibitemOpen
  \bibfield  {author} {\bibinfo {author} {\bibfnamefont {P.~A.~R.}\ \bibnamefont {Ade}} \emph {et~al.} (\bibinfo {collaboration} {Planck}),\ }\href {https://doi.org/10.1051/0004-6361/201525814} {\bibfield  {journal} {\bibinfo  {journal} {Astron. Astrophys.}\ }\textbf {\bibinfo {volume} {594}},\ \bibinfo {pages} {A14} (\bibinfo {year} {2016}{\natexlab{b}})},\ \Eprint {https://arxiv.org/abs/1502.01590} {arXiv:1502.01590 [astro-ph.CO]} \BibitemShut {NoStop}%
\bibitem [{\citenamefont {Aghanim}\ \emph {et~al.}(2020)\citenamefont {Aghanim} \emph {et~al.}}]{aghanim}%
  \BibitemOpen
  \bibfield  {author} {\bibinfo {author} {\bibfnamefont {N.}~\bibnamefont {Aghanim}} \emph {et~al.} (\bibinfo {collaboration} {Planck}),\ }\href {https://doi.org/10.1051/0004-6361/201833910} {\bibfield  {journal} {\bibinfo  {journal} {Astron. Astrophys.}\ }\textbf {\bibinfo {volume} {641}},\ \bibinfo {pages} {A6} (\bibinfo {year} {2020})},\ \bibinfo {note} {[Erratum: Astron. Astrophys. 652, C4 (2021)]},\ \Eprint {https://arxiv.org/abs/1807.06209} {arXiv:1807.06209 [astro-ph.CO]} \BibitemShut {NoStop}%
\bibitem [{\citenamefont {Betoule}\ \emph {et~al.}(2014)\citenamefont {Betoule} \emph {et~al.}}]{sn1}%
  \BibitemOpen
  \bibfield  {author} {\bibinfo {author} {\bibfnamefont {M.}~\bibnamefont {Betoule}} \emph {et~al.} (\bibinfo {collaboration} {SDSS}),\ }\href {https://doi.org/10.1051/0004-6361/201423413} {\bibfield  {journal} {\bibinfo  {journal} {Astron. Astrophys.}\ }\textbf {\bibinfo {volume} {568}},\ \bibinfo {pages} {A22} (\bibinfo {year} {2014})},\ \Eprint {https://arxiv.org/abs/1401.4064} {arXiv:1401.4064 [astro-ph.CO]} \BibitemShut {NoStop}%
\bibitem [{\citenamefont {Perlmutter}\ \emph {et~al.}(1997)\citenamefont {Perlmutter} \emph {et~al.}}]{sn2}%
  \BibitemOpen
  \bibfield  {author} {\bibinfo {author} {\bibfnamefont {S.}~\bibnamefont {Perlmutter}} \emph {et~al.},\ }\href {https://doi.org/10.1086/304265} {\bibfield  {journal} {\bibinfo  {journal} {Astrophys. J.}\ }\textbf {\bibinfo {volume} {483}},\ \bibinfo {pages} {565} (\bibinfo {year} {1997})},\ \Eprint {https://arxiv.org/abs/astro-ph/9608192} {arXiv:astro-ph/9608192 [astro-ph]} \BibitemShut {NoStop}%
\bibitem [{\citenamefont {Riess}\ \emph {et~al.}(1998)\citenamefont {Riess} \emph {et~al.}}]{sn3}%
  \BibitemOpen
  \bibfield  {author} {\bibinfo {author} {\bibfnamefont {A.~G.}\ \bibnamefont {Riess}} \emph {et~al.},\ }\href {https://doi.org/10.1086/300499} {\bibfield  {journal} {\bibinfo  {journal} {Astron. J.}\ }\textbf {\bibinfo {volume} {116}},\ \bibinfo {pages} {1009} (\bibinfo {year} {1998})},\ \Eprint {https://arxiv.org/abs/astro-ph/9805201} {arXiv:astro-ph/9805201 [astro-ph]} \BibitemShut {NoStop}%
\bibitem [{\citenamefont {Beutler}\ \emph {et~al.}(2011{\natexlab{a}})\citenamefont {Beutler}, \citenamefont {Blake}, \citenamefont {Colless}, \citenamefont {Jones}, \citenamefont {Staveley-Smith}, \citenamefont {Campbell}, \citenamefont {Parker}, \citenamefont {Saunders},\ and\ \citenamefont {Watson}}]{bao1}%
  \BibitemOpen
  \bibfield  {author} {\bibinfo {author} {\bibfnamefont {F.}~\bibnamefont {Beutler}}, \bibinfo {author} {\bibfnamefont {C.}~\bibnamefont {Blake}}, \bibinfo {author} {\bibfnamefont {M.}~\bibnamefont {Colless}}, \bibinfo {author} {\bibfnamefont {D.~H.}\ \bibnamefont {Jones}}, \bibinfo {author} {\bibfnamefont {L.}~\bibnamefont {Staveley-Smith}}, \bibinfo {author} {\bibfnamefont {L.}~\bibnamefont {Campbell}}, \bibinfo {author} {\bibfnamefont {Q.}~\bibnamefont {Parker}}, \bibinfo {author} {\bibfnamefont {W.}~\bibnamefont {Saunders}},\ and\ \bibinfo {author} {\bibfnamefont {F.}~\bibnamefont {Watson}},\ }\href {https://doi.org/10.1111/j.1365-2966.2011.19250.x} {\bibfield  {journal} {\bibinfo  {journal} {Mon. Not. Roy. Astron. Soc.}\ }\textbf {\bibinfo {volume} {416}},\ \bibinfo {pages} {3017} (\bibinfo {year} {2011}{\natexlab{a}})},\ \Eprint {https://arxiv.org/abs/1106.3366} {arXiv:1106.3366 [astro-ph.CO]} \BibitemShut {NoStop}%
\bibitem [{\citenamefont {Beutler}\ \emph {et~al.}(2012)\citenamefont {Beutler} \emph {et~al.}}]{bao2}%
  \BibitemOpen
  \bibfield  {author} {\bibinfo {author} {\bibfnamefont {F.}~\bibnamefont {Beutler}} \emph {et~al.},\ }\href {https://doi.org/10.1111/j.1365-2966.2012.21136.x} {\bibfield  {journal} {\bibinfo  {journal} {Mon. Not. Roy. Astron. Soc.}\ }\textbf {\bibinfo {volume} {423}},\ \bibinfo {pages} {3430} (\bibinfo {year} {2012})},\ \Eprint {https://arxiv.org/abs/1204.4725} {arXiv:1204.4725 [astro-ph.CO]} \BibitemShut {NoStop}%
\bibitem [{\citenamefont {Blake}\ \emph {et~al.}(2012)\citenamefont {Blake} \emph {et~al.}}]{bao3}%
  \BibitemOpen
  \bibfield  {author} {\bibinfo {author} {\bibfnamefont {C.}~\bibnamefont {Blake}} \emph {et~al.},\ }\href {https://doi.org/10.1111/j.1365-2966.2012.21473.x} {\bibfield  {journal} {\bibinfo  {journal} {Mon. Not. Roy. Astron. Soc.}\ }\textbf {\bibinfo {volume} {425}},\ \bibinfo {pages} {405} (\bibinfo {year} {2012})},\ \Eprint {https://arxiv.org/abs/1204.3674} {arXiv:1204.3674 [astro-ph.CO]} \BibitemShut {NoStop}%
\bibitem [{\citenamefont {Anderson}\ \emph {et~al.}(2013)\citenamefont {Anderson} \emph {et~al.}}]{bao4}%
  \BibitemOpen
  \bibfield  {author} {\bibinfo {author} {\bibfnamefont {L.}~\bibnamefont {Anderson}} \emph {et~al.},\ }\href {https://doi.org/10.1111/j.1365-2966.2012.22066.x} {\bibfield  {journal} {\bibinfo  {journal} {Mon. Not. Roy. Astron. Soc.}\ }\textbf {\bibinfo {volume} {427}},\ \bibinfo {pages} {3435} (\bibinfo {year} {2013})},\ \Eprint {https://arxiv.org/abs/1203.6594} {arXiv:1203.6594 [astro-ph.CO]} \BibitemShut {NoStop}%
\bibitem [{\citenamefont {Anderson}\ \emph {et~al.}(2014)\citenamefont {Anderson} \emph {et~al.}}]{bao5}%
  \BibitemOpen
  \bibfield  {author} {\bibinfo {author} {\bibfnamefont {L.}~\bibnamefont {Anderson}} \emph {et~al.},\ }\href {https://doi.org/10.1093/mnras/stu523} {\bibfield  {journal} {\bibinfo  {journal} {Mon. Not. Roy. Astron. Soc.}\ }\textbf {\bibinfo {volume} {441}},\ \bibinfo {pages} {24} (\bibinfo {year} {2014})},\ \Eprint {https://arxiv.org/abs/1312.4877} {arXiv:1312.4877 [astro-ph.CO]} \BibitemShut {NoStop}%
\bibitem [{\citenamefont {Riess}\ \emph {et~al.}(2022)\citenamefont {Riess} \emph {et~al.}}]{R22}%
  \BibitemOpen
  \bibfield  {author} {\bibinfo {author} {\bibfnamefont {A.~G.}\ \bibnamefont {Riess}} \emph {et~al.} (\bibinfo {collaboration} {SH0ES}),\ }\href {https://doi.org/10.3847/2041-8213/ac5c5b} {\bibfield  {journal} {\bibinfo  {journal} {Astrophys. J. Lett.}\ }\textbf {\bibinfo {volume} {934}},\ \bibinfo {pages} {L7} (\bibinfo {year} {2022})},\ \Eprint {https://arxiv.org/abs/2112.04510} {arXiv:2112.04510 [astro-ph.CO]} \BibitemShut {NoStop}%
\bibitem [{\citenamefont {Knox}\ and\ \citenamefont {Millea}(2020{\natexlab{a}})}]{Hhg}%
  \BibitemOpen
  \bibfield  {author} {\bibinfo {author} {\bibfnamefont {L.}~\bibnamefont {Knox}}\ and\ \bibinfo {author} {\bibfnamefont {M.}~\bibnamefont {Millea}},\ }\href {https://doi.org/10.1103/PhysRevD.101.043533} {\bibfield  {journal} {\bibinfo  {journal} {Phys. Rev. D}\ }\textbf {\bibinfo {volume} {101}},\ \bibinfo {pages} {043533} (\bibinfo {year} {2020}{\natexlab{a}})},\ \Eprint {https://arxiv.org/abs/1908.03663} {arXiv:1908.03663 [astro-ph.CO]} \BibitemShut {NoStop}%
\bibitem [{\citenamefont {Forconi}\ \emph {et~al.}(2023)\citenamefont {Forconi}, \citenamefont {Ruchika}, \citenamefont {Melchiorri}, \citenamefont {Mena},\ and\ \citenamefont {Menci}}]{Matteo}%
  \BibitemOpen
  \bibfield  {author} {\bibinfo {author} {\bibfnamefont {M.}~\bibnamefont {Forconi}}, \bibinfo {author} {\bibnamefont {Ruchika}}, \bibinfo {author} {\bibfnamefont {A.}~\bibnamefont {Melchiorri}}, \bibinfo {author} {\bibfnamefont {O.}~\bibnamefont {Mena}},\ and\ \bibinfo {author} {\bibfnamefont {N.}~\bibnamefont {Menci}},\ }\href {https://doi.org/10.1088/1475-7516/2023/10/012} {\bibfield  {journal} {\bibinfo  {journal} {JCAP}\ }\textbf {\bibinfo {volume} {10}},\ \bibinfo {pages} {012}},\ \Eprint {https://arxiv.org/abs/2306.07781} {arXiv:2306.07781 [astro-ph.CO]} \BibitemShut {NoStop}%
\bibitem [{\citenamefont {Ruchika}\ \emph {et~al.}(2024)\citenamefont {Ruchika}, \citenamefont {Rathore}, \citenamefont {Roy~Choudhury},\ and\ \citenamefont {Rentala}}]{Ruchika:2023ugh}%
  \BibitemOpen
  \bibfield  {author} {\bibinfo {author} {\bibnamefont {Ruchika}}, \bibinfo {author} {\bibfnamefont {H.}~\bibnamefont {Rathore}}, \bibinfo {author} {\bibfnamefont {S.}~\bibnamefont {Roy~Choudhury}},\ and\ \bibinfo {author} {\bibfnamefont {V.}~\bibnamefont {Rentala}},\ }\href {https://doi.org/10.1088/1475-7516/2024/06/056} {\bibfield  {journal} {\bibinfo  {journal} {JCAP}\ }\textbf {\bibinfo {volume} {06}},\ \bibinfo {pages} {056}},\ \Eprint {https://arxiv.org/abs/2306.05450} {arXiv:2306.05450 [astro-ph.CO]} \BibitemShut {NoStop}%
\bibitem [{\citenamefont {Kable}\ \emph {et~al.}(2023)\citenamefont {Kable}, \citenamefont {Benevento}, \citenamefont {Addison},\ and\ \citenamefont {Bennett}}]{Giampaolo}%
  \BibitemOpen
  \bibfield  {author} {\bibinfo {author} {\bibfnamefont {J.~A.}\ \bibnamefont {Kable}}, \bibinfo {author} {\bibfnamefont {G.}~\bibnamefont {Benevento}}, \bibinfo {author} {\bibfnamefont {G.~E.}\ \bibnamefont {Addison}},\ and\ \bibinfo {author} {\bibfnamefont {C.~L.}\ \bibnamefont {Bennett}},\ }\href {https://doi.org/10.3847/1538-4357/acfed0} {\bibfield  {journal} {\bibinfo  {journal} {Astrophys. J.}\ }\textbf {\bibinfo {volume} {959}},\ \bibinfo {pages} {143} (\bibinfo {year} {2023})},\ \Eprint {https://arxiv.org/abs/2307.12174} {arXiv:2307.12174 [astro-ph.CO]} \BibitemShut {NoStop}%
\bibitem [{\citenamefont {Vagnozzi}(2023)}]{Vagnozzi:2023nrq}%
  \BibitemOpen
  \bibfield  {author} {\bibinfo {author} {\bibfnamefont {S.}~\bibnamefont {Vagnozzi}},\ }\href {https://doi.org/10.3390/universe9090393} {\bibfield  {journal} {\bibinfo  {journal} {Universe}\ }\textbf {\bibinfo {volume} {9}},\ \bibinfo {pages} {393} (\bibinfo {year} {2023})},\ \Eprint {https://arxiv.org/abs/2308.16628} {arXiv:2308.16628 [astro-ph.CO]} \BibitemShut {NoStop}%
\bibitem [{\citenamefont {Anchordoqui}\ \emph {et~al.}(2015)\citenamefont {Anchordoqui}, \citenamefont {Barger}, \citenamefont {Goldberg}, \citenamefont {Huang}, \citenamefont {Marfatia}, \citenamefont {da~Silva},\ and\ \citenamefont {Weiler}}]{Anchordoqui:2015lqa}%
  \BibitemOpen
  \bibfield  {author} {\bibinfo {author} {\bibfnamefont {L.~A.}\ \bibnamefont {Anchordoqui}}, \bibinfo {author} {\bibfnamefont {V.}~\bibnamefont {Barger}}, \bibinfo {author} {\bibfnamefont {H.}~\bibnamefont {Goldberg}}, \bibinfo {author} {\bibfnamefont {X.}~\bibnamefont {Huang}}, \bibinfo {author} {\bibfnamefont {D.}~\bibnamefont {Marfatia}}, \bibinfo {author} {\bibfnamefont {L.~H.~M.}\ \bibnamefont {da~Silva}},\ and\ \bibinfo {author} {\bibfnamefont {T.~J.}\ \bibnamefont {Weiler}},\ }\href {https://doi.org/10.1103/PhysRevD.94.069901} {\bibfield  {journal} {\bibinfo  {journal} {Phys. Rev. D}\ }\textbf {\bibinfo {volume} {92}},\ \bibinfo {pages} {061301} (\bibinfo {year} {2015})},\ \bibinfo {note} {[Erratum: Phys.Rev.D 94, 069901 (2016)]},\ \Eprint {https://arxiv.org/abs/1506.08788} {arXiv:1506.08788 [hep-ph]} \BibitemShut {NoStop}%
\bibitem [{\citenamefont {Karwal}\ and\ \citenamefont {Kamionkowski}(2016)}]{Karwal:2016vyq}%
  \BibitemOpen
  \bibfield  {author} {\bibinfo {author} {\bibfnamefont {T.}~\bibnamefont {Karwal}}\ and\ \bibinfo {author} {\bibfnamefont {M.}~\bibnamefont {Kamionkowski}},\ }\href {https://doi.org/10.1103/PhysRevD.94.103523} {\bibfield  {journal} {\bibinfo  {journal} {Phys. Rev. D}\ }\textbf {\bibinfo {volume} {94}},\ \bibinfo {pages} {103523} (\bibinfo {year} {2016})},\ \Eprint {https://arxiv.org/abs/1608.01309} {arXiv:1608.01309 [astro-ph.CO]} \BibitemShut {NoStop}%
\bibitem [{\citenamefont {Benetti}\ \emph {et~al.}(2018)\citenamefont {Benetti}, \citenamefont {Graef},\ and\ \citenamefont {Alcaniz}}]{Benetti:2017juy}%
  \BibitemOpen
  \bibfield  {author} {\bibinfo {author} {\bibfnamefont {M.}~\bibnamefont {Benetti}}, \bibinfo {author} {\bibfnamefont {L.~L.}\ \bibnamefont {Graef}},\ and\ \bibinfo {author} {\bibfnamefont {J.~S.}\ \bibnamefont {Alcaniz}},\ }\href {https://doi.org/10.1088/1475-7516/2018/07/066} {\bibfield  {journal} {\bibinfo  {journal} {JCAP}\ }\textbf {\bibinfo {volume} {07}},\ \bibinfo {pages} {066}},\ \Eprint {https://arxiv.org/abs/1712.00677} {arXiv:1712.00677 [astro-ph.CO]} \BibitemShut {NoStop}%
\bibitem [{\citenamefont {M\"ortsell}\ and\ \citenamefont {Dhawan}(2018)}]{Mortsell:2018mfj}%
  \BibitemOpen
  \bibfield  {author} {\bibinfo {author} {\bibfnamefont {E.}~\bibnamefont {M\"ortsell}}\ and\ \bibinfo {author} {\bibfnamefont {S.}~\bibnamefont {Dhawan}},\ }\href {https://doi.org/10.1088/1475-7516/2018/09/025} {\bibfield  {journal} {\bibinfo  {journal} {JCAP}\ }\textbf {\bibinfo {volume} {09}},\ \bibinfo {pages} {025}},\ \Eprint {https://arxiv.org/abs/1801.07260} {arXiv:1801.07260 [astro-ph.CO]} \BibitemShut {NoStop}%
\bibitem [{\citenamefont {Kumar}\ \emph {et~al.}(2018)\citenamefont {Kumar}, \citenamefont {Nunes},\ and\ \citenamefont {Yadav}}]{Kumar:2018yhh}%
  \BibitemOpen
  \bibfield  {author} {\bibinfo {author} {\bibfnamefont {S.}~\bibnamefont {Kumar}}, \bibinfo {author} {\bibfnamefont {R.~C.}\ \bibnamefont {Nunes}},\ and\ \bibinfo {author} {\bibfnamefont {S.~K.}\ \bibnamefont {Yadav}},\ }\href {https://doi.org/10.1103/PhysRevD.98.043521} {\bibfield  {journal} {\bibinfo  {journal} {Phys. Rev. D}\ }\textbf {\bibinfo {volume} {98}},\ \bibinfo {pages} {043521} (\bibinfo {year} {2018})},\ \Eprint {https://arxiv.org/abs/1803.10229} {arXiv:1803.10229 [astro-ph.CO]} \BibitemShut {NoStop}%
\bibitem [{\citenamefont {Guo}\ \emph {et~al.}(2019)\citenamefont {Guo}, \citenamefont {Zhang},\ and\ \citenamefont {Zhang}}]{Guo:2018ans}%
  \BibitemOpen
  \bibfield  {author} {\bibinfo {author} {\bibfnamefont {R.-Y.}\ \bibnamefont {Guo}}, \bibinfo {author} {\bibfnamefont {J.-F.}\ \bibnamefont {Zhang}},\ and\ \bibinfo {author} {\bibfnamefont {X.}~\bibnamefont {Zhang}},\ }\href {https://doi.org/10.1088/1475-7516/2019/02/054} {\bibfield  {journal} {\bibinfo  {journal} {JCAP}\ }\textbf {\bibinfo {volume} {02}},\ \bibinfo {pages} {054}},\ \Eprint {https://arxiv.org/abs/1809.02340} {arXiv:1809.02340 [astro-ph.CO]} \BibitemShut {NoStop}%
\bibitem [{\citenamefont {Poulin}\ \emph {et~al.}(2019)\citenamefont {Poulin}, \citenamefont {Smith}, \citenamefont {Karwal},\ and\ \citenamefont {Kamionkowski}}]{Poulin:2018cxd}%
  \BibitemOpen
  \bibfield  {author} {\bibinfo {author} {\bibfnamefont {V.}~\bibnamefont {Poulin}}, \bibinfo {author} {\bibfnamefont {T.~L.}\ \bibnamefont {Smith}}, \bibinfo {author} {\bibfnamefont {T.}~\bibnamefont {Karwal}},\ and\ \bibinfo {author} {\bibfnamefont {M.}~\bibnamefont {Kamionkowski}},\ }\href {https://doi.org/10.1103/PhysRevLett.122.221301} {\bibfield  {journal} {\bibinfo  {journal} {Phys. Rev. Lett.}\ }\textbf {\bibinfo {volume} {122}},\ \bibinfo {pages} {221301} (\bibinfo {year} {2019})},\ \Eprint {https://arxiv.org/abs/1811.04083} {arXiv:1811.04083 [astro-ph.CO]} \BibitemShut {NoStop}%
\bibitem [{\citenamefont {Graef}\ \emph {et~al.}(2019)\citenamefont {Graef}, \citenamefont {Benetti},\ and\ \citenamefont {Alcaniz}}]{Graef:2018fzu}%
  \BibitemOpen
  \bibfield  {author} {\bibinfo {author} {\bibfnamefont {L.~L.}\ \bibnamefont {Graef}}, \bibinfo {author} {\bibfnamefont {M.}~\bibnamefont {Benetti}},\ and\ \bibinfo {author} {\bibfnamefont {J.~S.}\ \bibnamefont {Alcaniz}},\ }\href {https://doi.org/10.1103/PhysRevD.99.043519} {\bibfield  {journal} {\bibinfo  {journal} {Phys. Rev. D}\ }\textbf {\bibinfo {volume} {99}},\ \bibinfo {pages} {043519} (\bibinfo {year} {2019})},\ \Eprint {https://arxiv.org/abs/1809.04501} {arXiv:1809.04501 [astro-ph.CO]} \BibitemShut {NoStop}%
\bibitem [{\citenamefont {Agrawal}\ \emph {et~al.}(2023)\citenamefont {Agrawal}, \citenamefont {Cyr-Racine}, \citenamefont {Pinner},\ and\ \citenamefont {Randall}}]{Agrawal:2019lmo}%
  \BibitemOpen
  \bibfield  {author} {\bibinfo {author} {\bibfnamefont {P.}~\bibnamefont {Agrawal}}, \bibinfo {author} {\bibfnamefont {F.-Y.}\ \bibnamefont {Cyr-Racine}}, \bibinfo {author} {\bibfnamefont {D.}~\bibnamefont {Pinner}},\ and\ \bibinfo {author} {\bibfnamefont {L.}~\bibnamefont {Randall}},\ }\href {https://doi.org/10.1016/j.dark.2023.101347} {\bibfield  {journal} {\bibinfo  {journal} {Phys. Dark Univ.}\ }\textbf {\bibinfo {volume} {42}},\ \bibinfo {pages} {101347} (\bibinfo {year} {2023})},\ \Eprint {https://arxiv.org/abs/1904.01016} {arXiv:1904.01016 [astro-ph.CO]} \BibitemShut {NoStop}%
\bibitem [{\citenamefont {Escudero}\ and\ \citenamefont {Witte}(2020)}]{Escudero:2019gvw}%
  \BibitemOpen
  \bibfield  {author} {\bibinfo {author} {\bibfnamefont {M.}~\bibnamefont {Escudero}}\ and\ \bibinfo {author} {\bibfnamefont {S.~J.}\ \bibnamefont {Witte}},\ }\href {https://doi.org/10.1140/epjc/s10052-020-7854-5} {\bibfield  {journal} {\bibinfo  {journal} {Eur. Phys. J. C}\ }\textbf {\bibinfo {volume} {80}},\ \bibinfo {pages} {294} (\bibinfo {year} {2020})},\ \Eprint {https://arxiv.org/abs/1909.04044} {arXiv:1909.04044 [astro-ph.CO]} \BibitemShut {NoStop}%
\bibitem [{\citenamefont {Niedermann}\ and\ \citenamefont {Sloth}(2021)}]{Niedermann:2019olb}%
  \BibitemOpen
  \bibfield  {author} {\bibinfo {author} {\bibfnamefont {F.}~\bibnamefont {Niedermann}}\ and\ \bibinfo {author} {\bibfnamefont {M.~S.}\ \bibnamefont {Sloth}},\ }\href {https://doi.org/10.1103/PhysRevD.103.L041303} {\bibfield  {journal} {\bibinfo  {journal} {Phys. Rev. D}\ }\textbf {\bibinfo {volume} {103}},\ \bibinfo {pages} {L041303} (\bibinfo {year} {2021})},\ \Eprint {https://arxiv.org/abs/1910.10739} {arXiv:1910.10739 [astro-ph.CO]} \BibitemShut {NoStop}%
\bibitem [{\citenamefont {Sakstein}\ and\ \citenamefont {Trodden}(2020)}]{Sakstein:2019fmf}%
  \BibitemOpen
  \bibfield  {author} {\bibinfo {author} {\bibfnamefont {J.}~\bibnamefont {Sakstein}}\ and\ \bibinfo {author} {\bibfnamefont {M.}~\bibnamefont {Trodden}},\ }\href {https://doi.org/10.1103/PhysRevLett.124.161301} {\bibfield  {journal} {\bibinfo  {journal} {Phys. Rev. Lett.}\ }\textbf {\bibinfo {volume} {124}},\ \bibinfo {pages} {161301} (\bibinfo {year} {2020})},\ \Eprint {https://arxiv.org/abs/1911.11760} {arXiv:1911.11760 [astro-ph.CO]} \BibitemShut {NoStop}%
\bibitem [{\citenamefont {Knox}\ and\ \citenamefont {Millea}(2020{\natexlab{b}})}]{Knox:2019rjx}%
  \BibitemOpen
  \bibfield  {author} {\bibinfo {author} {\bibfnamefont {L.}~\bibnamefont {Knox}}\ and\ \bibinfo {author} {\bibfnamefont {M.}~\bibnamefont {Millea}},\ }\href {https://doi.org/10.1103/PhysRevD.101.043533} {\bibfield  {journal} {\bibinfo  {journal} {Phys. Rev. D}\ }\textbf {\bibinfo {volume} {101}},\ \bibinfo {pages} {043533} (\bibinfo {year} {2020}{\natexlab{b}})},\ \Eprint {https://arxiv.org/abs/1908.03663} {arXiv:1908.03663 [astro-ph.CO]} \BibitemShut {NoStop}%
\bibitem [{\citenamefont {Hart}\ and\ \citenamefont {Chluba}(2020)}]{Hart:2019dxi}%
  \BibitemOpen
  \bibfield  {author} {\bibinfo {author} {\bibfnamefont {L.}~\bibnamefont {Hart}}\ and\ \bibinfo {author} {\bibfnamefont {J.}~\bibnamefont {Chluba}},\ }\href {https://doi.org/10.1093/mnras/staa412} {\bibfield  {journal} {\bibinfo  {journal} {Mon. Not. Roy. Astron. Soc.}\ }\textbf {\bibinfo {volume} {493}},\ \bibinfo {pages} {3255} (\bibinfo {year} {2020})},\ \Eprint {https://arxiv.org/abs/1912.03986} {arXiv:1912.03986 [astro-ph.CO]} \BibitemShut {NoStop}%
\bibitem [{\citenamefont {Ballesteros}\ \emph {et~al.}(2020)\citenamefont {Ballesteros}, \citenamefont {Notari},\ and\ \citenamefont {Rompineve}}]{Ballesteros:2020sik}%
  \BibitemOpen
  \bibfield  {author} {\bibinfo {author} {\bibfnamefont {G.}~\bibnamefont {Ballesteros}}, \bibinfo {author} {\bibfnamefont {A.}~\bibnamefont {Notari}},\ and\ \bibinfo {author} {\bibfnamefont {F.}~\bibnamefont {Rompineve}},\ }\href {https://doi.org/10.1088/1475-7516/2020/11/024} {\bibfield  {journal} {\bibinfo  {journal} {JCAP}\ }\textbf {\bibinfo {volume} {11}},\ \bibinfo {pages} {024}},\ \Eprint {https://arxiv.org/abs/2004.05049} {arXiv:2004.05049 [astro-ph.CO]} \BibitemShut {NoStop}%
\bibitem [{\citenamefont {Jedamzik}\ and\ \citenamefont {Pogosian}(2020)}]{Jedamzik:2020krr}%
  \BibitemOpen
  \bibfield  {author} {\bibinfo {author} {\bibfnamefont {K.}~\bibnamefont {Jedamzik}}\ and\ \bibinfo {author} {\bibfnamefont {L.}~\bibnamefont {Pogosian}},\ }\href {https://doi.org/10.1103/PhysRevLett.125.181302} {\bibfield  {journal} {\bibinfo  {journal} {Phys. Rev. Lett.}\ }\textbf {\bibinfo {volume} {125}},\ \bibinfo {pages} {181302} (\bibinfo {year} {2020})},\ \Eprint {https://arxiv.org/abs/2004.09487} {arXiv:2004.09487 [astro-ph.CO]} \BibitemShut {NoStop}%
\bibitem [{\citenamefont {Ballardini}\ \emph {et~al.}(2020)\citenamefont {Ballardini}, \citenamefont {Braglia}, \citenamefont {Finelli}, \citenamefont {Paoletti}, \citenamefont {Starobinsky},\ and\ \citenamefont {Umilt\`a}}]{Ballardini:2020iws}%
  \BibitemOpen
  \bibfield  {author} {\bibinfo {author} {\bibfnamefont {M.}~\bibnamefont {Ballardini}}, \bibinfo {author} {\bibfnamefont {M.}~\bibnamefont {Braglia}}, \bibinfo {author} {\bibfnamefont {F.}~\bibnamefont {Finelli}}, \bibinfo {author} {\bibfnamefont {D.}~\bibnamefont {Paoletti}}, \bibinfo {author} {\bibfnamefont {A.~A.}\ \bibnamefont {Starobinsky}},\ and\ \bibinfo {author} {\bibfnamefont {C.}~\bibnamefont {Umilt\`a}},\ }\href {https://doi.org/10.1088/1475-7516/2020/10/044} {\bibfield  {journal} {\bibinfo  {journal} {JCAP}\ }\textbf {\bibinfo {volume} {10}},\ \bibinfo {pages} {044}},\ \Eprint {https://arxiv.org/abs/2004.14349} {arXiv:2004.14349 [astro-ph.CO]} \BibitemShut {NoStop}%
\bibitem [{\citenamefont {Di~Valentino}\ \emph {et~al.}(2020)\citenamefont {Di~Valentino}, \citenamefont {Gariazzo}, \citenamefont {Mena},\ and\ \citenamefont {Vagnozzi}}]{DiValentino:2020evt}%
  \BibitemOpen
  \bibfield  {author} {\bibinfo {author} {\bibfnamefont {E.}~\bibnamefont {Di~Valentino}}, \bibinfo {author} {\bibfnamefont {S.}~\bibnamefont {Gariazzo}}, \bibinfo {author} {\bibfnamefont {O.}~\bibnamefont {Mena}},\ and\ \bibinfo {author} {\bibfnamefont {S.}~\bibnamefont {Vagnozzi}},\ }\href {https://doi.org/10.1088/1475-7516/2020/07/045} {\bibfield  {journal} {\bibinfo  {journal} {JCAP}\ }\textbf {\bibinfo {volume} {07}}\bibfield  {number} {\bibinfo  {number} { (07)},\ \bibinfo {pages} {045}},\ }\Eprint {https://arxiv.org/abs/2005.02062} {arXiv:2005.02062 [astro-ph.CO]} \BibitemShut {NoStop}%
\bibitem [{\citenamefont {Niedermann}\ and\ \citenamefont {Sloth}(2020)}]{Niedermann:2020dwg}%
  \BibitemOpen
  \bibfield  {author} {\bibinfo {author} {\bibfnamefont {F.}~\bibnamefont {Niedermann}}\ and\ \bibinfo {author} {\bibfnamefont {M.~S.}\ \bibnamefont {Sloth}},\ }\href {https://doi.org/10.1103/PhysRevD.102.063527} {\bibfield  {journal} {\bibinfo  {journal} {Phys. Rev. D}\ }\textbf {\bibinfo {volume} {102}},\ \bibinfo {pages} {063527} (\bibinfo {year} {2020})},\ \Eprint {https://arxiv.org/abs/2006.06686} {arXiv:2006.06686 [astro-ph.CO]} \BibitemShut {NoStop}%
\bibitem [{\citenamefont {Gonzalez}\ \emph {et~al.}(2020)\citenamefont {Gonzalez}, \citenamefont {Hertzberg},\ and\ \citenamefont {Rompineve}}]{Gonzalez:2020fdy}%
  \BibitemOpen
  \bibfield  {author} {\bibinfo {author} {\bibfnamefont {M.}~\bibnamefont {Gonzalez}}, \bibinfo {author} {\bibfnamefont {M.~P.}\ \bibnamefont {Hertzberg}},\ and\ \bibinfo {author} {\bibfnamefont {F.}~\bibnamefont {Rompineve}},\ }\href {https://doi.org/10.1088/1475-7516/2020/10/028} {\bibfield  {journal} {\bibinfo  {journal} {JCAP}\ }\textbf {\bibinfo {volume} {10}},\ \bibinfo {pages} {028}},\ \Eprint {https://arxiv.org/abs/2006.13959} {arXiv:2006.13959 [astro-ph.CO]} \BibitemShut {NoStop}%
\bibitem [{\citenamefont {Braglia}\ \emph {et~al.}(2021)\citenamefont {Braglia}, \citenamefont {Ballardini}, \citenamefont {Finelli},\ and\ \citenamefont {Koyama}}]{Braglia:2020auw}%
  \BibitemOpen
  \bibfield  {author} {\bibinfo {author} {\bibfnamefont {M.}~\bibnamefont {Braglia}}, \bibinfo {author} {\bibfnamefont {M.}~\bibnamefont {Ballardini}}, \bibinfo {author} {\bibfnamefont {F.}~\bibnamefont {Finelli}},\ and\ \bibinfo {author} {\bibfnamefont {K.}~\bibnamefont {Koyama}},\ }\href {https://doi.org/10.1103/PhysRevD.103.043528} {\bibfield  {journal} {\bibinfo  {journal} {Phys. Rev. D}\ }\textbf {\bibinfo {volume} {103}},\ \bibinfo {pages} {043528} (\bibinfo {year} {2021})},\ \Eprint {https://arxiv.org/abs/2011.12934} {arXiv:2011.12934 [astro-ph.CO]} \BibitemShut {NoStop}%
\bibitem [{\citenamefont {Roy~Choudhury}\ \emph {et~al.}(2021)\citenamefont {Roy~Choudhury}, \citenamefont {Hannestad},\ and\ \citenamefont {Tram}}]{RoyChoudhury:2020dmd}%
  \BibitemOpen
  \bibfield  {author} {\bibinfo {author} {\bibfnamefont {S.}~\bibnamefont {Roy~Choudhury}}, \bibinfo {author} {\bibfnamefont {S.}~\bibnamefont {Hannestad}},\ and\ \bibinfo {author} {\bibfnamefont {T.}~\bibnamefont {Tram}},\ }\href {https://doi.org/10.1088/1475-7516/2021/03/084} {\bibfield  {journal} {\bibinfo  {journal} {JCAP}\ }\textbf {\bibinfo {volume} {03}},\ \bibinfo {pages} {084}},\ \Eprint {https://arxiv.org/abs/2012.07519} {arXiv:2012.07519 [astro-ph.CO]} \BibitemShut {NoStop}%
\bibitem [{\citenamefont {Brinckmann}\ \emph {et~al.}(2021)\citenamefont {Brinckmann}, \citenamefont {Chang},\ and\ \citenamefont {LoVerde}}]{Brinckmann:2020bcn}%
  \BibitemOpen
  \bibfield  {author} {\bibinfo {author} {\bibfnamefont {T.}~\bibnamefont {Brinckmann}}, \bibinfo {author} {\bibfnamefont {J.~H.}\ \bibnamefont {Chang}},\ and\ \bibinfo {author} {\bibfnamefont {M.}~\bibnamefont {LoVerde}},\ }\href {https://doi.org/10.1103/PhysRevD.104.063523} {\bibfield  {journal} {\bibinfo  {journal} {Phys. Rev. D}\ }\textbf {\bibinfo {volume} {104}},\ \bibinfo {pages} {063523} (\bibinfo {year} {2021})},\ \Eprint {https://arxiv.org/abs/2012.11830} {arXiv:2012.11830 [astro-ph.CO]} \BibitemShut {NoStop}%
\bibitem [{\citenamefont {Karwal}\ \emph {et~al.}(2022)\citenamefont {Karwal}, \citenamefont {Raveri}, \citenamefont {Jain}, \citenamefont {Khoury},\ and\ \citenamefont {Trodden}}]{Karwal:2021vpk}%
  \BibitemOpen
  \bibfield  {author} {\bibinfo {author} {\bibfnamefont {T.}~\bibnamefont {Karwal}}, \bibinfo {author} {\bibfnamefont {M.}~\bibnamefont {Raveri}}, \bibinfo {author} {\bibfnamefont {B.}~\bibnamefont {Jain}}, \bibinfo {author} {\bibfnamefont {J.}~\bibnamefont {Khoury}},\ and\ \bibinfo {author} {\bibfnamefont {M.}~\bibnamefont {Trodden}},\ }\href {https://doi.org/10.1103/PhysRevD.105.063535} {\bibfield  {journal} {\bibinfo  {journal} {Phys. Rev. D}\ }\textbf {\bibinfo {volume} {105}},\ \bibinfo {pages} {063535} (\bibinfo {year} {2022})},\ \Eprint {https://arxiv.org/abs/2106.13290} {arXiv:2106.13290 [astro-ph.CO]} \BibitemShut {NoStop}%
\bibitem [{\citenamefont {Herold}\ and\ \citenamefont {Ferreira}(2023)}]{Herold:2022iib}%
  \BibitemOpen
  \bibfield  {author} {\bibinfo {author} {\bibfnamefont {L.}~\bibnamefont {Herold}}\ and\ \bibinfo {author} {\bibfnamefont {E.~G.~M.}\ \bibnamefont {Ferreira}},\ }\href {https://doi.org/10.1103/PhysRevD.108.043513} {\bibfield  {journal} {\bibinfo  {journal} {Phys. Rev. D}\ }\textbf {\bibinfo {volume} {108}},\ \bibinfo {pages} {043513} (\bibinfo {year} {2023})},\ \Eprint {https://arxiv.org/abs/2210.16296} {arXiv:2210.16296 [astro-ph.CO]} \BibitemShut {NoStop}%
\bibitem [{\citenamefont {G\'omez-Valent}\ \emph {et~al.}(2021)\citenamefont {G\'omez-Valent}, \citenamefont {Zheng}, \citenamefont {Amendola}, \citenamefont {Pettorino},\ and\ \citenamefont {Wetterich}}]{Gomez-Valent:2021cbe}%
  \BibitemOpen
  \bibfield  {author} {\bibinfo {author} {\bibfnamefont {A.}~\bibnamefont {G\'omez-Valent}}, \bibinfo {author} {\bibfnamefont {Z.}~\bibnamefont {Zheng}}, \bibinfo {author} {\bibfnamefont {L.}~\bibnamefont {Amendola}}, \bibinfo {author} {\bibfnamefont {V.}~\bibnamefont {Pettorino}},\ and\ \bibinfo {author} {\bibfnamefont {C.}~\bibnamefont {Wetterich}},\ }\href {https://doi.org/10.1103/PhysRevD.104.083536} {\bibfield  {journal} {\bibinfo  {journal} {Phys. Rev. D}\ }\textbf {\bibinfo {volume} {104}},\ \bibinfo {pages} {083536} (\bibinfo {year} {2021})},\ \Eprint {https://arxiv.org/abs/2107.11065} {arXiv:2107.11065 [astro-ph.CO]} \BibitemShut {NoStop}%
\bibitem [{\citenamefont {Cyr-Racine}\ \emph {et~al.}(2022)\citenamefont {Cyr-Racine}, \citenamefont {Ge},\ and\ \citenamefont {Knox}}]{Cyr-Racine:2021oal}%
  \BibitemOpen
  \bibfield  {author} {\bibinfo {author} {\bibfnamefont {F.-Y.}\ \bibnamefont {Cyr-Racine}}, \bibinfo {author} {\bibfnamefont {F.}~\bibnamefont {Ge}},\ and\ \bibinfo {author} {\bibfnamefont {L.}~\bibnamefont {Knox}},\ }\href {https://doi.org/10.1103/PhysRevLett.128.201301} {\bibfield  {journal} {\bibinfo  {journal} {Phys. Rev. Lett.}\ }\textbf {\bibinfo {volume} {128}},\ \bibinfo {pages} {201301} (\bibinfo {year} {2022})},\ \Eprint {https://arxiv.org/abs/2107.13000} {arXiv:2107.13000 [astro-ph.CO]} \BibitemShut {NoStop}%
\bibitem [{\citenamefont {Niedermann}\ and\ \citenamefont {Sloth}(2022)}]{Niedermann:2021ijp}%
  \BibitemOpen
  \bibfield  {author} {\bibinfo {author} {\bibfnamefont {F.}~\bibnamefont {Niedermann}}\ and\ \bibinfo {author} {\bibfnamefont {M.~S.}\ \bibnamefont {Sloth}},\ }\href {https://doi.org/10.1016/j.physletb.2022.137555} {\bibfield  {journal} {\bibinfo  {journal} {Phys. Lett. B}\ }\textbf {\bibinfo {volume} {835}},\ \bibinfo {pages} {137555} (\bibinfo {year} {2022})},\ \Eprint {https://arxiv.org/abs/2112.00759} {arXiv:2112.00759 [hep-ph]} \BibitemShut {NoStop}%
\bibitem [{\citenamefont {Saridakis}\ \emph {et~al.}(2023)\citenamefont {Saridakis}, \citenamefont {Yang}, \citenamefont {Pan}, \citenamefont {Anagnostopoulos},\ and\ \citenamefont {Basilakos}}]{Saridakis:2021xqy}%
  \BibitemOpen
  \bibfield  {author} {\bibinfo {author} {\bibfnamefont {E.~N.}\ \bibnamefont {Saridakis}}, \bibinfo {author} {\bibfnamefont {W.}~\bibnamefont {Yang}}, \bibinfo {author} {\bibfnamefont {S.}~\bibnamefont {Pan}}, \bibinfo {author} {\bibfnamefont {F.~K.}\ \bibnamefont {Anagnostopoulos}},\ and\ \bibinfo {author} {\bibfnamefont {S.}~\bibnamefont {Basilakos}},\ }\href {https://doi.org/10.1016/j.nuclphysb.2022.116042} {\bibfield  {journal} {\bibinfo  {journal} {Nucl. Phys. B}\ }\textbf {\bibinfo {volume} {986}},\ \bibinfo {pages} {116042} (\bibinfo {year} {2023})},\ \Eprint {https://arxiv.org/abs/2112.08330} {arXiv:2112.08330 [astro-ph.CO]} \BibitemShut {NoStop}%
\bibitem [{\citenamefont {Herold}\ \emph {et~al.}(2022)\citenamefont {Herold}, \citenamefont {Ferreira},\ and\ \citenamefont {Komatsu}}]{Herold:2021ksg}%
  \BibitemOpen
  \bibfield  {author} {\bibinfo {author} {\bibfnamefont {L.}~\bibnamefont {Herold}}, \bibinfo {author} {\bibfnamefont {E.~G.~M.}\ \bibnamefont {Ferreira}},\ and\ \bibinfo {author} {\bibfnamefont {E.}~\bibnamefont {Komatsu}},\ }\href {https://doi.org/10.3847/2041-8213/ac63a3} {\bibfield  {journal} {\bibinfo  {journal} {Astrophys. J. Lett.}\ }\textbf {\bibinfo {volume} {929}},\ \bibinfo {pages} {L16} (\bibinfo {year} {2022})},\ \Eprint {https://arxiv.org/abs/2112.12140} {arXiv:2112.12140 [astro-ph.CO]} \BibitemShut {NoStop}%
\bibitem [{\citenamefont {Odintsov}\ and\ \citenamefont {Oikonomou}(2022{\natexlab{a}})}]{Odintsov:2022eqm}%
  \BibitemOpen
  \bibfield  {author} {\bibinfo {author} {\bibfnamefont {S.~D.}\ \bibnamefont {Odintsov}}\ and\ \bibinfo {author} {\bibfnamefont {V.~K.}\ \bibnamefont {Oikonomou}},\ }\href {https://doi.org/10.1209/0295-5075/ac52dc} {\bibfield  {journal} {\bibinfo  {journal} {EPL}\ }\textbf {\bibinfo {volume} {137}},\ \bibinfo {pages} {39001} (\bibinfo {year} {2022}{\natexlab{a}})},\ \Eprint {https://arxiv.org/abs/2201.07647} {arXiv:2201.07647 [gr-qc]} \BibitemShut {NoStop}%
\bibitem [{\citenamefont {Aboubrahim}\ \emph {et~al.}(2022)\citenamefont {Aboubrahim}, \citenamefont {Klasen},\ and\ \citenamefont {Nath}}]{Aboubrahim:2022gjb}%
  \BibitemOpen
  \bibfield  {author} {\bibinfo {author} {\bibfnamefont {A.}~\bibnamefont {Aboubrahim}}, \bibinfo {author} {\bibfnamefont {M.}~\bibnamefont {Klasen}},\ and\ \bibinfo {author} {\bibfnamefont {P.}~\bibnamefont {Nath}},\ }\href {https://doi.org/10.1088/1475-7516/2022/04/042} {\bibfield  {journal} {\bibinfo  {journal} {JCAP}\ }\textbf {\bibinfo {volume} {04}}\bibfield  {number} {\bibinfo  {number} { (04)},\ \bibinfo {pages} {042}},\ }\Eprint {https://arxiv.org/abs/2202.04453} {arXiv:2202.04453 [astro-ph.CO]} \BibitemShut {NoStop}%
\bibitem [{\citenamefont {Ren}\ \emph {et~al.}(2022)\citenamefont {Ren}, \citenamefont {Yan}, \citenamefont {Zhao}, \citenamefont {Cai},\ and\ \citenamefont {Saridakis}}]{Ren:2022aeo}%
  \BibitemOpen
  \bibfield  {author} {\bibinfo {author} {\bibfnamefont {X.}~\bibnamefont {Ren}}, \bibinfo {author} {\bibfnamefont {S.-F.}\ \bibnamefont {Yan}}, \bibinfo {author} {\bibfnamefont {Y.}~\bibnamefont {Zhao}}, \bibinfo {author} {\bibfnamefont {Y.-F.}\ \bibnamefont {Cai}},\ and\ \bibinfo {author} {\bibfnamefont {E.~N.}\ \bibnamefont {Saridakis}},\ }\href {https://doi.org/10.3847/1538-4357/ac6ba5} {\bibfield  {journal} {\bibinfo  {journal} {Astrophys. J.}\ }\textbf {\bibinfo {volume} {932}},\ \bibinfo {pages} {131} (\bibinfo {year} {2022})},\ \Eprint {https://arxiv.org/abs/2203.01926} {arXiv:2203.01926 [astro-ph.CO]} \BibitemShut {NoStop}%
\bibitem [{\citenamefont {Adhikari}(2022)}]{Adhikari:2022moo}%
  \BibitemOpen
  \bibfield  {author} {\bibinfo {author} {\bibfnamefont {S.}~\bibnamefont {Adhikari}},\ }\href {https://doi.org/10.1016/j.dark.2022.101005} {\bibfield  {journal} {\bibinfo  {journal} {Phys. Dark Univ.}\ }\textbf {\bibinfo {volume} {36}},\ \bibinfo {pages} {101005} (\bibinfo {year} {2022})},\ \Eprint {https://arxiv.org/abs/2203.04835} {arXiv:2203.04835 [astro-ph.CO]} \BibitemShut {NoStop}%
\bibitem [{\citenamefont {Nojiri}\ \emph {et~al.}(2022)\citenamefont {Nojiri}, \citenamefont {Odintsov},\ and\ \citenamefont {Oikonomou}}]{Nojiri:2022ski}%
  \BibitemOpen
  \bibfield  {author} {\bibinfo {author} {\bibfnamefont {S.}~\bibnamefont {Nojiri}}, \bibinfo {author} {\bibfnamefont {S.~D.}\ \bibnamefont {Odintsov}},\ and\ \bibinfo {author} {\bibfnamefont {V.~K.}\ \bibnamefont {Oikonomou}},\ }\href {https://doi.org/10.1016/j.nuclphysb.2022.115850} {\bibfield  {journal} {\bibinfo  {journal} {Nucl. Phys. B}\ }\textbf {\bibinfo {volume} {980}},\ \bibinfo {pages} {115850} (\bibinfo {year} {2022})},\ \Eprint {https://arxiv.org/abs/2205.11681} {arXiv:2205.11681 [gr-qc]} \BibitemShut {NoStop}%
\bibitem [{\citenamefont {Sch\"oneberg}\ and\ \citenamefont {Franco~Abell\'an}(2022)}]{Schoneberg:2022grr}%
  \BibitemOpen
  \bibfield  {author} {\bibinfo {author} {\bibfnamefont {N.}~\bibnamefont {Sch\"oneberg}}\ and\ \bibinfo {author} {\bibfnamefont {G.}~\bibnamefont {Franco~Abell\'an}},\ }\href {https://doi.org/10.1088/1475-7516/2022/12/001} {\bibfield  {journal} {\bibinfo  {journal} {JCAP}\ }\textbf {\bibinfo {volume} {12}},\ \bibinfo {pages} {001}},\ \Eprint {https://arxiv.org/abs/2206.11276} {arXiv:2206.11276 [astro-ph.CO]} \BibitemShut {NoStop}%
\bibitem [{\citenamefont {Joseph}\ \emph {et~al.}(2023)\citenamefont {Joseph}, \citenamefont {Aloni}, \citenamefont {Schmaltz}, \citenamefont {Sivarajan},\ and\ \citenamefont {Weiner}}]{Joseph:2022jsf}%
  \BibitemOpen
  \bibfield  {author} {\bibinfo {author} {\bibfnamefont {M.}~\bibnamefont {Joseph}}, \bibinfo {author} {\bibfnamefont {D.}~\bibnamefont {Aloni}}, \bibinfo {author} {\bibfnamefont {M.}~\bibnamefont {Schmaltz}}, \bibinfo {author} {\bibfnamefont {E.~N.}\ \bibnamefont {Sivarajan}},\ and\ \bibinfo {author} {\bibfnamefont {N.}~\bibnamefont {Weiner}},\ }\href {https://doi.org/10.1103/PhysRevD.108.023520} {\bibfield  {journal} {\bibinfo  {journal} {Phys. Rev. D}\ }\textbf {\bibinfo {volume} {108}},\ \bibinfo {pages} {023520} (\bibinfo {year} {2023})},\ \Eprint {https://arxiv.org/abs/2207.03500} {arXiv:2207.03500 [astro-ph.CO]} \BibitemShut {NoStop}%
\bibitem [{\citenamefont {G\'omez-Valent}\ \emph {et~al.}(2022)\citenamefont {G\'omez-Valent}, \citenamefont {Zheng}, \citenamefont {Amendola}, \citenamefont {Wetterich},\ and\ \citenamefont {Pettorino}}]{Gomez-Valent:2022bku}%
  \BibitemOpen
  \bibfield  {author} {\bibinfo {author} {\bibfnamefont {A.}~\bibnamefont {G\'omez-Valent}}, \bibinfo {author} {\bibfnamefont {Z.}~\bibnamefont {Zheng}}, \bibinfo {author} {\bibfnamefont {L.}~\bibnamefont {Amendola}}, \bibinfo {author} {\bibfnamefont {C.}~\bibnamefont {Wetterich}},\ and\ \bibinfo {author} {\bibfnamefont {V.}~\bibnamefont {Pettorino}},\ }\href {https://doi.org/10.1103/PhysRevD.106.103522} {\bibfield  {journal} {\bibinfo  {journal} {Phys. Rev. D}\ }\textbf {\bibinfo {volume} {106}},\ \bibinfo {pages} {103522} (\bibinfo {year} {2022})},\ \Eprint {https://arxiv.org/abs/2207.14487} {arXiv:2207.14487 [astro-ph.CO]} \BibitemShut {NoStop}%
\bibitem [{\citenamefont {Odintsov}\ and\ \citenamefont {Oikonomou}(2022{\natexlab{b}})}]{Odintsov:2022umu}%
  \BibitemOpen
  \bibfield  {author} {\bibinfo {author} {\bibfnamefont {S.~D.}\ \bibnamefont {Odintsov}}\ and\ \bibinfo {author} {\bibfnamefont {V.~K.}\ \bibnamefont {Oikonomou}},\ }\href {https://doi.org/10.1209/0295-5075/ac8a13} {\bibfield  {journal} {\bibinfo  {journal} {EPL}\ }\textbf {\bibinfo {volume} {139}},\ \bibinfo {pages} {59003} (\bibinfo {year} {2022}{\natexlab{b}})},\ \Eprint {https://arxiv.org/abs/2208.07972} {arXiv:2208.07972 [gr-qc]} \BibitemShut {NoStop}%
\bibitem [{\citenamefont {Ge}\ \emph {et~al.}(2023)\citenamefont {Ge}, \citenamefont {Cyr-Racine},\ and\ \citenamefont {Knox}}]{Ge:2022qws}%
  \BibitemOpen
  \bibfield  {author} {\bibinfo {author} {\bibfnamefont {F.}~\bibnamefont {Ge}}, \bibinfo {author} {\bibfnamefont {F.-Y.}\ \bibnamefont {Cyr-Racine}},\ and\ \bibinfo {author} {\bibfnamefont {L.}~\bibnamefont {Knox}},\ }\href {https://doi.org/10.1103/PhysRevD.107.023517} {\bibfield  {journal} {\bibinfo  {journal} {Phys. Rev. D}\ }\textbf {\bibinfo {volume} {107}},\ \bibinfo {pages} {023517} (\bibinfo {year} {2023})},\ \Eprint {https://arxiv.org/abs/2210.16335} {arXiv:2210.16335 [astro-ph.CO]} \BibitemShut {NoStop}%
\bibitem [{\citenamefont {Schiavone}\ \emph {et~al.}(2023)\citenamefont {Schiavone}, \citenamefont {Montani},\ and\ \citenamefont {Bombacigno}}]{Schiavone:2022wvq}%
  \BibitemOpen
  \bibfield  {author} {\bibinfo {author} {\bibfnamefont {T.}~\bibnamefont {Schiavone}}, \bibinfo {author} {\bibfnamefont {G.}~\bibnamefont {Montani}},\ and\ \bibinfo {author} {\bibfnamefont {F.}~\bibnamefont {Bombacigno}},\ }\href {https://doi.org/10.1093/mnrasl/slad041} {\bibfield  {journal} {\bibinfo  {journal} {Mon. Not. Roy. Astron. Soc.}\ }\textbf {\bibinfo {volume} {522}},\ \bibinfo {pages} {L72} (\bibinfo {year} {2023})},\ \Eprint {https://arxiv.org/abs/2211.16737} {arXiv:2211.16737 [gr-qc]} \BibitemShut {NoStop}%
\bibitem [{\citenamefont {Brinckmann}\ \emph {et~al.}(2023)\citenamefont {Brinckmann}, \citenamefont {Chang}, \citenamefont {Du},\ and\ \citenamefont {LoVerde}}]{Brinckmann:2022ajr}%
  \BibitemOpen
  \bibfield  {author} {\bibinfo {author} {\bibfnamefont {T.}~\bibnamefont {Brinckmann}}, \bibinfo {author} {\bibfnamefont {J.~H.}\ \bibnamefont {Chang}}, \bibinfo {author} {\bibfnamefont {P.}~\bibnamefont {Du}},\ and\ \bibinfo {author} {\bibfnamefont {M.}~\bibnamefont {LoVerde}},\ }\href {https://doi.org/10.1103/PhysRevD.107.123517} {\bibfield  {journal} {\bibinfo  {journal} {Phys. Rev. D}\ }\textbf {\bibinfo {volume} {107}},\ \bibinfo {pages} {123517} (\bibinfo {year} {2023})},\ \Eprint {https://arxiv.org/abs/2212.13264} {arXiv:2212.13264 [astro-ph.CO]} \BibitemShut {NoStop}%
\bibitem [{\citenamefont {Khodadi}\ and\ \citenamefont {Schreck}(2023)}]{Khodadi:2023ezj}%
  \BibitemOpen
  \bibfield  {author} {\bibinfo {author} {\bibfnamefont {M.}~\bibnamefont {Khodadi}}\ and\ \bibinfo {author} {\bibfnamefont {M.}~\bibnamefont {Schreck}},\ }\href {https://doi.org/10.1016/j.dark.2023.101170} {\bibfield  {journal} {\bibinfo  {journal} {Phys. Dark Univ.}\ }\textbf {\bibinfo {volume} {39}},\ \bibinfo {pages} {101170} (\bibinfo {year} {2023})},\ \Eprint {https://arxiv.org/abs/2301.03883} {arXiv:2301.03883 [gr-qc]} \BibitemShut {NoStop}%
\bibitem [{\citenamefont {Kumar}\ \emph {et~al.}(2023)\citenamefont {Kumar}, \citenamefont {Nunes}, \citenamefont {Pan},\ and\ \citenamefont {Yadav}}]{Kumar:2023bqj}%
  \BibitemOpen
  \bibfield  {author} {\bibinfo {author} {\bibfnamefont {S.}~\bibnamefont {Kumar}}, \bibinfo {author} {\bibfnamefont {R.~C.}\ \bibnamefont {Nunes}}, \bibinfo {author} {\bibfnamefont {S.}~\bibnamefont {Pan}},\ and\ \bibinfo {author} {\bibfnamefont {P.}~\bibnamefont {Yadav}},\ }\href {https://doi.org/10.1016/j.dark.2023.101281} {\bibfield  {journal} {\bibinfo  {journal} {Phys. Dark Univ.}\ }\textbf {\bibinfo {volume} {42}},\ \bibinfo {pages} {101281} (\bibinfo {year} {2023})},\ \Eprint {https://arxiv.org/abs/2301.07897} {arXiv:2301.07897 [astro-ph.CO]} \BibitemShut {NoStop}%
\bibitem [{\citenamefont {Ben-Dayan}\ and\ \citenamefont {Kumar}(2023)}]{Ben-Dayan:2023rgt}%
  \BibitemOpen
  \bibfield  {author} {\bibinfo {author} {\bibfnamefont {I.}~\bibnamefont {Ben-Dayan}}\ and\ \bibinfo {author} {\bibfnamefont {U.}~\bibnamefont {Kumar}},\ }\href {https://doi.org/10.1088/1475-7516/2023/12/047} {\bibfield  {journal} {\bibinfo  {journal} {JCAP}\ }\textbf {\bibinfo {volume} {12}},\ \bibinfo {pages} {047}},\ \Eprint {https://arxiv.org/abs/2302.00067} {arXiv:2302.00067 [astro-ph.CO]} \BibitemShut {NoStop}%
\bibitem [{\citenamefont {Yadav}(2023)}]{Yadav:2023yyb}%
  \BibitemOpen
  \bibfield  {author} {\bibinfo {author} {\bibfnamefont {V.}~\bibnamefont {Yadav}},\ }\href {https://doi.org/10.1016/j.dark.2023.101365} {\bibfield  {journal} {\bibinfo  {journal} {Phys. Dark Univ.}\ }\textbf {\bibinfo {volume} {42}},\ \bibinfo {pages} {101365} (\bibinfo {year} {2023})},\ \Eprint {https://arxiv.org/abs/2306.16135} {arXiv:2306.16135 [astro-ph.CO]} \BibitemShut {NoStop}%
\bibitem [{\citenamefont {Sharma}\ \emph {et~al.}(2024)\citenamefont {Sharma}, \citenamefont {Das},\ and\ \citenamefont {Poulin}}]{Sharma:2023kzr}%
  \BibitemOpen
  \bibfield  {author} {\bibinfo {author} {\bibfnamefont {R.~K.}\ \bibnamefont {Sharma}}, \bibinfo {author} {\bibfnamefont {S.}~\bibnamefont {Das}},\ and\ \bibinfo {author} {\bibfnamefont {V.}~\bibnamefont {Poulin}},\ }\href {https://doi.org/10.1103/PhysRevD.109.043530} {\bibfield  {journal} {\bibinfo  {journal} {Phys. Rev. D}\ }\textbf {\bibinfo {volume} {109}},\ \bibinfo {pages} {043530} (\bibinfo {year} {2024})},\ \Eprint {https://arxiv.org/abs/2309.00401} {arXiv:2309.00401 [astro-ph.CO]} \BibitemShut {NoStop}%
\bibitem [{\citenamefont {Ramadan}\ \emph {et~al.}(2024)\citenamefont {Ramadan}, \citenamefont {Karwal},\ and\ \citenamefont {Sakstein}}]{Ramadan:2023ivw}%
  \BibitemOpen
  \bibfield  {author} {\bibinfo {author} {\bibfnamefont {O.~F.}\ \bibnamefont {Ramadan}}, \bibinfo {author} {\bibfnamefont {T.}~\bibnamefont {Karwal}},\ and\ \bibinfo {author} {\bibfnamefont {J.}~\bibnamefont {Sakstein}},\ }\href {https://doi.org/10.1103/PhysRevD.109.063525} {\bibfield  {journal} {\bibinfo  {journal} {Phys. Rev. D}\ }\textbf {\bibinfo {volume} {109}},\ \bibinfo {pages} {063525} (\bibinfo {year} {2024})},\ \Eprint {https://arxiv.org/abs/2309.08082} {arXiv:2309.08082 [astro-ph.CO]} \BibitemShut {NoStop}%
\bibitem [{\citenamefont {Fu}\ and\ \citenamefont {Wang}(2024)}]{Fu:2023tfo}%
  \BibitemOpen
  \bibfield  {author} {\bibinfo {author} {\bibfnamefont {C.}~\bibnamefont {Fu}}\ and\ \bibinfo {author} {\bibfnamefont {S.-J.}\ \bibnamefont {Wang}},\ }\href {https://doi.org/10.1103/PhysRevD.109.L041304} {\bibfield  {journal} {\bibinfo  {journal} {Phys. Rev. D}\ }\textbf {\bibinfo {volume} {109}},\ \bibinfo {pages} {L041304} (\bibinfo {year} {2024})},\ \Eprint {https://arxiv.org/abs/2310.12932} {arXiv:2310.12932 [astro-ph.CO]} \BibitemShut {NoStop}%
\bibitem [{\citenamefont {Efstathiou}\ \emph {et~al.}(2024)\citenamefont {Efstathiou}, \citenamefont {Rosenberg},\ and\ \citenamefont {Poulin}}]{Efstathiou:2023fbn}%
  \BibitemOpen
  \bibfield  {author} {\bibinfo {author} {\bibfnamefont {G.}~\bibnamefont {Efstathiou}}, \bibinfo {author} {\bibfnamefont {E.}~\bibnamefont {Rosenberg}},\ and\ \bibinfo {author} {\bibfnamefont {V.}~\bibnamefont {Poulin}},\ }\href {https://doi.org/10.1103/PhysRevLett.132.221002} {\bibfield  {journal} {\bibinfo  {journal} {Phys. Rev. Lett.}\ }\textbf {\bibinfo {volume} {132}},\ \bibinfo {pages} {221002} (\bibinfo {year} {2024})},\ \Eprint {https://arxiv.org/abs/2311.00524} {arXiv:2311.00524 [astro-ph.CO]} \BibitemShut {NoStop}%
\bibitem [{\citenamefont {Montani}\ \emph {et~al.}(2024)\citenamefont {Montani}, \citenamefont {Carlevaro},\ and\ \citenamefont {Dainotti}}]{Montani:2023ywn}%
  \BibitemOpen
  \bibfield  {author} {\bibinfo {author} {\bibfnamefont {G.}~\bibnamefont {Montani}}, \bibinfo {author} {\bibfnamefont {N.}~\bibnamefont {Carlevaro}},\ and\ \bibinfo {author} {\bibfnamefont {M.~G.}\ \bibnamefont {Dainotti}},\ }\href {https://doi.org/10.1016/j.dark.2024.101486} {\bibfield  {journal} {\bibinfo  {journal} {Phys. Dark Univ.}\ }\textbf {\bibinfo {volume} {44}},\ \bibinfo {pages} {101486} (\bibinfo {year} {2024})},\ \Eprint {https://arxiv.org/abs/2311.04822} {arXiv:2311.04822 [gr-qc]} \BibitemShut {NoStop}%
\bibitem [{\citenamefont {Stahl}\ \emph {et~al.}(2024)\citenamefont {Stahl}, \citenamefont {Famaey}, \citenamefont {Ibata}, \citenamefont {Hahn}, \citenamefont {Martinet},\ and\ \citenamefont {Montandon}}]{Stahl:2024stz}%
  \BibitemOpen
  \bibfield  {author} {\bibinfo {author} {\bibfnamefont {C.}~\bibnamefont {Stahl}}, \bibinfo {author} {\bibfnamefont {B.}~\bibnamefont {Famaey}}, \bibinfo {author} {\bibfnamefont {R.}~\bibnamefont {Ibata}}, \bibinfo {author} {\bibfnamefont {O.}~\bibnamefont {Hahn}}, \bibinfo {author} {\bibfnamefont {N.}~\bibnamefont {Martinet}},\ and\ \bibinfo {author} {\bibfnamefont {T.}~\bibnamefont {Montandon}},\ }\href {https://doi.org/10.1103/PhysRevD.110.063501} {\bibfield  {journal} {\bibinfo  {journal} {Phys. Rev. D}\ }\textbf {\bibinfo {volume} {110}},\ \bibinfo {pages} {063501} (\bibinfo {year} {2024})},\ \Eprint {https://arxiv.org/abs/2404.03244} {arXiv:2404.03244 [astro-ph.CO]} \BibitemShut {NoStop}%
\bibitem [{\citenamefont {Zhai}\ \emph {et~al.}(2023)\citenamefont {Zhai}, \citenamefont {Giar\`e}, \citenamefont {van~de Bruck}, \citenamefont {Di~Valentino}, \citenamefont {Mena},\ and\ \citenamefont {Nunes}}]{Zhai:2023yny}%
  \BibitemOpen
  \bibfield  {author} {\bibinfo {author} {\bibfnamefont {Y.}~\bibnamefont {Zhai}}, \bibinfo {author} {\bibfnamefont {W.}~\bibnamefont {Giar\`e}}, \bibinfo {author} {\bibfnamefont {C.}~\bibnamefont {van~de Bruck}}, \bibinfo {author} {\bibfnamefont {E.}~\bibnamefont {Di~Valentino}}, \bibinfo {author} {\bibfnamefont {O.}~\bibnamefont {Mena}},\ and\ \bibinfo {author} {\bibfnamefont {R.~C.}\ \bibnamefont {Nunes}},\ }\href {https://doi.org/10.1088/1475-7516/2023/07/032} {\bibfield  {journal} {\bibinfo  {journal} {JCAP}\ }\textbf {\bibinfo {volume} {07}},\ \bibinfo {pages} {032}},\ \Eprint {https://arxiv.org/abs/2303.08201} {arXiv:2303.08201 [astro-ph.CO]} \BibitemShut {NoStop}%
\bibitem [{\citenamefont {Garny}\ \emph {et~al.}(2024)\citenamefont {Garny}, \citenamefont {Niedermann}, \citenamefont {Rubira},\ and\ \citenamefont {Sloth}}]{Garny:2024ums}%
  \BibitemOpen
  \bibfield  {author} {\bibinfo {author} {\bibfnamefont {M.}~\bibnamefont {Garny}}, \bibinfo {author} {\bibfnamefont {F.}~\bibnamefont {Niedermann}}, \bibinfo {author} {\bibfnamefont {H.}~\bibnamefont {Rubira}},\ and\ \bibinfo {author} {\bibfnamefont {M.~S.}\ \bibnamefont {Sloth}},\ }\href {https://doi.org/10.1103/PhysRevD.110.023531} {\bibfield  {journal} {\bibinfo  {journal} {Phys. Rev. D}\ }\textbf {\bibinfo {volume} {110}},\ \bibinfo {pages} {023531} (\bibinfo {year} {2024})},\ \Eprint {https://arxiv.org/abs/2404.07256} {arXiv:2404.07256 [astro-ph.CO]} \BibitemShut {NoStop}%
\bibitem [{\citenamefont {Co}\ \emph {et~al.}(2024)\citenamefont {Co}, \citenamefont {Fernandez}, \citenamefont {Ghalsasi}, \citenamefont {Harigaya},\ and\ \citenamefont {Shelton}}]{Co:2024oek}%
  \BibitemOpen
  \bibfield  {author} {\bibinfo {author} {\bibfnamefont {R.~T.}\ \bibnamefont {Co}}, \bibinfo {author} {\bibfnamefont {N.}~\bibnamefont {Fernandez}}, \bibinfo {author} {\bibfnamefont {A.}~\bibnamefont {Ghalsasi}}, \bibinfo {author} {\bibfnamefont {K.}~\bibnamefont {Harigaya}},\ and\ \bibinfo {author} {\bibfnamefont {J.}~\bibnamefont {Shelton}},\ }\href {https://doi.org/10.1103/PhysRevD.110.083534} {\bibfield  {journal} {\bibinfo  {journal} {Phys. Rev. D}\ }\textbf {\bibinfo {volume} {110}},\ \bibinfo {pages} {083534} (\bibinfo {year} {2024})},\ \Eprint {https://arxiv.org/abs/2405.12268} {arXiv:2405.12268 [hep-ph]} \BibitemShut {NoStop}%
\bibitem [{\citenamefont {Toda}\ \emph {et~al.}(2024)\citenamefont {Toda}, \citenamefont {Giar\`e}, \citenamefont {\"Oz\"ulker}, \citenamefont {Di~Valentino},\ and\ \citenamefont {Vagnozzi}}]{Toda:2024ncp}%
  \BibitemOpen
  \bibfield  {author} {\bibinfo {author} {\bibfnamefont {Y.}~\bibnamefont {Toda}}, \bibinfo {author} {\bibfnamefont {W.}~\bibnamefont {Giar\`e}}, \bibinfo {author} {\bibfnamefont {E.}~\bibnamefont {\"Oz\"ulker}}, \bibinfo {author} {\bibfnamefont {E.}~\bibnamefont {Di~Valentino}},\ and\ \bibinfo {author} {\bibfnamefont {S.}~\bibnamefont {Vagnozzi}},\ }\href {https://doi.org/10.1016/j.dark.2024.101676} {\bibfield  {journal} {\bibinfo  {journal} {Phys. Dark Univ.}\ }\textbf {\bibinfo {volume} {46}},\ \bibinfo {pages} {101676} (\bibinfo {year} {2024})},\ \Eprint {https://arxiv.org/abs/2407.01173} {arXiv:2407.01173 [astro-ph.CO]} \BibitemShut {NoStop}%
\bibitem [{\citenamefont {Giar\`e}\ \emph {et~al.}(2024{\natexlab{a}})\citenamefont {Giar\`e}, \citenamefont {Zhai}, \citenamefont {Pan}, \citenamefont {Di~Valentino}, \citenamefont {Nunes},\ and\ \citenamefont {van~de Bruck}}]{Giare:2024ytc}%
  \BibitemOpen
  \bibfield  {author} {\bibinfo {author} {\bibfnamefont {W.}~\bibnamefont {Giar\`e}}, \bibinfo {author} {\bibfnamefont {Y.}~\bibnamefont {Zhai}}, \bibinfo {author} {\bibfnamefont {S.}~\bibnamefont {Pan}}, \bibinfo {author} {\bibfnamefont {E.}~\bibnamefont {Di~Valentino}}, \bibinfo {author} {\bibfnamefont {R.~C.}\ \bibnamefont {Nunes}},\ and\ \bibinfo {author} {\bibfnamefont {C.}~\bibnamefont {van~de Bruck}},\ }\href {https://doi.org/10.1103/PhysRevD.110.063527} {\bibfield  {journal} {\bibinfo  {journal} {Phys. Rev. D}\ }\textbf {\bibinfo {volume} {110}},\ \bibinfo {pages} {063527} (\bibinfo {year} {2024}{\natexlab{a}})},\ \Eprint {https://arxiv.org/abs/2404.02110} {arXiv:2404.02110 [astro-ph.CO]} \BibitemShut {NoStop}%
\bibitem [{\citenamefont {Percival}\ \emph {et~al.}(2007{\natexlab{a}})\citenamefont {Percival}, \citenamefont {Cole}, \citenamefont {Eisenstein}, \citenamefont {Nichol}, \citenamefont {Peacock}, \citenamefont {Pope},\ and\ \citenamefont {Szalay}}]{Percival:2007yw}%
  \BibitemOpen
  \bibfield  {author} {\bibinfo {author} {\bibfnamefont {W.~J.}\ \bibnamefont {Percival}}, \bibinfo {author} {\bibfnamefont {S.}~\bibnamefont {Cole}}, \bibinfo {author} {\bibfnamefont {D.~J.}\ \bibnamefont {Eisenstein}}, \bibinfo {author} {\bibfnamefont {R.~C.}\ \bibnamefont {Nichol}}, \bibinfo {author} {\bibfnamefont {J.~A.}\ \bibnamefont {Peacock}}, \bibinfo {author} {\bibfnamefont {A.~C.}\ \bibnamefont {Pope}},\ and\ \bibinfo {author} {\bibfnamefont {A.~S.}\ \bibnamefont {Szalay}},\ }\href {https://doi.org/10.1111/j.1365-2966.2007.12268.x} {\bibfield  {journal} {\bibinfo  {journal} {Mon. Not. Roy. Astron. Soc.}\ }\textbf {\bibinfo {volume} {381}},\ \bibinfo {pages} {1053} (\bibinfo {year} {2007}{\natexlab{a}})},\ \Eprint {https://arxiv.org/abs/0705.3323} {arXiv:0705.3323 [astro-ph]} \BibitemShut {NoStop}%
\bibitem [{\citenamefont {Giar\`e}(2024)}]{Giare:2024akf}%
  \BibitemOpen
  \bibfield  {author} {\bibinfo {author} {\bibfnamefont {W.}~\bibnamefont {Giar\`e}},\ }\href {https://doi.org/10.1103/PhysRevD.109.123545} {\bibfield  {journal} {\bibinfo  {journal} {Phys. Rev. D}\ }\textbf {\bibinfo {volume} {109}},\ \bibinfo {pages} {123545} (\bibinfo {year} {2024})},\ \Eprint {https://arxiv.org/abs/2404.12779} {arXiv:2404.12779 [astro-ph.CO]} \BibitemShut {NoStop}%
\bibitem [{\citenamefont {Akarsu}\ \emph {et~al.}(2024)\citenamefont {Akarsu}, \citenamefont {De~Felice}, \citenamefont {Di~Valentino}, \citenamefont {Kumar}, \citenamefont {Nunes}, \citenamefont {\"Oz\"ulker}, \citenamefont {Vazquez},\ and\ \citenamefont {Yadav}}]{Akarsu:2024eoo}%
  \BibitemOpen
  \bibfield  {author} {\bibinfo {author} {\bibfnamefont {O.}~\bibnamefont {Akarsu}}, \bibinfo {author} {\bibfnamefont {A.}~\bibnamefont {De~Felice}}, \bibinfo {author} {\bibfnamefont {E.}~\bibnamefont {Di~Valentino}}, \bibinfo {author} {\bibfnamefont {S.}~\bibnamefont {Kumar}}, \bibinfo {author} {\bibfnamefont {R.~C.}\ \bibnamefont {Nunes}}, \bibinfo {author} {\bibfnamefont {E.}~\bibnamefont {\"Oz\"ulker}}, \bibinfo {author} {\bibfnamefont {J.~A.}\ \bibnamefont {Vazquez}},\ and\ \bibinfo {author} {\bibfnamefont {A.}~\bibnamefont {Yadav}},\ }\href {https://doi.org/10.1103/PhysRevD.110.103527} {\bibfield  {journal} {\bibinfo  {journal} {Phys. Rev. D}\ }\textbf {\bibinfo {volume} {110}},\ \bibinfo {pages} {103527} (\bibinfo {year} {2024})},\ \Eprint {https://arxiv.org/abs/2406.07526} {arXiv:2406.07526 [astro-ph.CO]} \BibitemShut {NoStop}%
\bibitem [{\citenamefont {Giar\`e}\ \emph {et~al.}(2024{\natexlab{b}})\citenamefont {Giar\`e}, \citenamefont {Sabogal}, \citenamefont {Nunes},\ and\ \citenamefont {Di~Valentino}}]{Giare:2024smz}%
  \BibitemOpen
  \bibfield  {author} {\bibinfo {author} {\bibfnamefont {W.}~\bibnamefont {Giar\`e}}, \bibinfo {author} {\bibfnamefont {M.~A.}\ \bibnamefont {Sabogal}}, \bibinfo {author} {\bibfnamefont {R.~C.}\ \bibnamefont {Nunes}},\ and\ \bibinfo {author} {\bibfnamefont {E.}~\bibnamefont {Di~Valentino}},\ }\href {https://doi.org/10.1103/PhysRevLett.133.251003} {\bibfield  {journal} {\bibinfo  {journal} {Phys. Rev. Lett.}\ }\textbf {\bibinfo {volume} {133}},\ \bibinfo {pages} {251003} (\bibinfo {year} {2024}{\natexlab{b}})},\ \Eprint {https://arxiv.org/abs/2404.15232} {arXiv:2404.15232 [astro-ph.CO]} \BibitemShut {NoStop}%
\bibitem [{\citenamefont {Pogosian}\ \emph {et~al.}(2020)\citenamefont {Pogosian}, \citenamefont {Zhao},\ and\ \citenamefont {Jedamzik}}]{Pogosian:2020ded}%
  \BibitemOpen
  \bibfield  {author} {\bibinfo {author} {\bibfnamefont {L.}~\bibnamefont {Pogosian}}, \bibinfo {author} {\bibfnamefont {G.-B.}\ \bibnamefont {Zhao}},\ and\ \bibinfo {author} {\bibfnamefont {K.}~\bibnamefont {Jedamzik}},\ }\href {https://doi.org/10.3847/2041-8213/abc6a8} {\bibfield  {journal} {\bibinfo  {journal} {Astrophys. J. Lett.}\ }\textbf {\bibinfo {volume} {904}},\ \bibinfo {pages} {L17} (\bibinfo {year} {2020})},\ \Eprint {https://arxiv.org/abs/2009.08455} {arXiv:2009.08455 [astro-ph.CO]} \BibitemShut {NoStop}%
\bibitem [{\citenamefont {Staicova}(2023)}]{Staicova:2023jic}%
  \BibitemOpen
  \bibfield  {author} {\bibinfo {author} {\bibfnamefont {D.}~\bibnamefont {Staicova}},\ }\href {https://doi.org/10.22323/1.436.0188} {\bibfield  {journal} {\bibinfo  {journal} {PoS}\ }\textbf {\bibinfo {volume} {CORFU2022}},\ \bibinfo {pages} {188} (\bibinfo {year} {2023})},\ \Eprint {https://arxiv.org/abs/2303.11271} {arXiv:2303.11271 [astro-ph.CO]} \BibitemShut {NoStop}%
\bibitem [{\citenamefont {Specogna}\ \emph {et~al.}(2025)\citenamefont {Specogna}, \citenamefont {Adil}, \citenamefont {Ozulker}, \citenamefont {Di~Valentino}, \citenamefont {Nunes}, \citenamefont {Akarsu},\ and\ \citenamefont {Sen}}]{Specogna:2025guo}%
  \BibitemOpen
  \bibfield  {author} {\bibinfo {author} {\bibfnamefont {E.}~\bibnamefont {Specogna}}, \bibinfo {author} {\bibfnamefont {S.~A.}\ \bibnamefont {Adil}}, \bibinfo {author} {\bibfnamefont {E.}~\bibnamefont {Ozulker}}, \bibinfo {author} {\bibfnamefont {E.}~\bibnamefont {Di~Valentino}}, \bibinfo {author} {\bibfnamefont {R.~C.}\ \bibnamefont {Nunes}}, \bibinfo {author} {\bibfnamefont {O.}~\bibnamefont {Akarsu}},\ and\ \bibinfo {author} {\bibfnamefont {A.~A.}\ \bibnamefont {Sen}},\ }\href@noop {} {\bibfield  {journal} {\bibinfo  {journal} {to be published}\ } (\bibinfo {year} {2025})},\ \Eprint {https://arxiv.org/abs/2504.17859} {arXiv:2504.17859 [gr-qc]} \BibitemShut {NoStop}%
\bibitem [{\citenamefont {Menci}\ \emph {et~al.}(2024)\citenamefont {Menci}, \citenamefont {Adil}, \citenamefont {Mukhopadhyay}, \citenamefont {Sen},\ and\ \citenamefont {Vagnozzi}}]{Menci:2024rbq}%
  \BibitemOpen
  \bibfield  {author} {\bibinfo {author} {\bibfnamefont {N.}~\bibnamefont {Menci}}, \bibinfo {author} {\bibfnamefont {S.~A.}\ \bibnamefont {Adil}}, \bibinfo {author} {\bibfnamefont {U.}~\bibnamefont {Mukhopadhyay}}, \bibinfo {author} {\bibfnamefont {A.~A.}\ \bibnamefont {Sen}},\ and\ \bibinfo {author} {\bibfnamefont {S.}~\bibnamefont {Vagnozzi}},\ }\href {https://doi.org/10.1088/1475-7516/2024/07/072} {\bibfield  {journal} {\bibinfo  {journal} {JCAP}\ }\textbf {\bibinfo {volume} {07}},\ \bibinfo {pages} {072}},\ \Eprint {https://arxiv.org/abs/2401.12659} {arXiv:2401.12659 [astro-ph.CO]} \BibitemShut {NoStop}%
\bibitem [{\citenamefont {Adil}\ \emph {et~al.}(2024)\citenamefont {Adil}, \citenamefont {Akarsu}, \citenamefont {Di~Valentino}, \citenamefont {Nunes}, \citenamefont {\"Oz\"ulker}, \citenamefont {Sen},\ and\ \citenamefont {Specogna}}]{Adil:2023exv}%
  \BibitemOpen
  \bibfield  {author} {\bibinfo {author} {\bibfnamefont {S.~A.}\ \bibnamefont {Adil}}, \bibinfo {author} {\bibfnamefont {O.}~\bibnamefont {Akarsu}}, \bibinfo {author} {\bibfnamefont {E.}~\bibnamefont {Di~Valentino}}, \bibinfo {author} {\bibfnamefont {R.~C.}\ \bibnamefont {Nunes}}, \bibinfo {author} {\bibfnamefont {E.}~\bibnamefont {\"Oz\"ulker}}, \bibinfo {author} {\bibfnamefont {A.~A.}\ \bibnamefont {Sen}},\ and\ \bibinfo {author} {\bibfnamefont {E.}~\bibnamefont {Specogna}},\ }\href {https://doi.org/10.1103/PhysRevD.109.023527} {\bibfield  {journal} {\bibinfo  {journal} {Phys. Rev. D}\ }\textbf {\bibinfo {volume} {109}},\ \bibinfo {pages} {023527} (\bibinfo {year} {2024})},\ \Eprint {https://arxiv.org/abs/2306.08046} {arXiv:2306.08046 [astro-ph.CO]} \BibitemShut {NoStop}%
\bibitem [{\citenamefont {Abdalla}\ \emph {et~al.}(2022)\citenamefont {Abdalla} \emph {et~al.}}]{Abdalla:2022yfr}%
  \BibitemOpen
  \bibfield  {author} {\bibinfo {author} {\bibfnamefont {E.}~\bibnamefont {Abdalla}} \emph {et~al.},\ }\href {https://doi.org/10.1016/j.jheap.2022.04.002} {\bibfield  {journal} {\bibinfo  {journal} {JHEAp}\ }\textbf {\bibinfo {volume} {34}},\ \bibinfo {pages} {49} (\bibinfo {year} {2022})},\ \Eprint {https://arxiv.org/abs/2203.06142} {arXiv:2203.06142 [astro-ph.CO]} \BibitemShut {NoStop}%
\bibitem [{\citenamefont {Di~Valentino}\ \emph {et~al.}(2025{\natexlab{a}})\citenamefont {Di~Valentino} \emph {et~al.}}]{DiValentino:2025sru}%
  \BibitemOpen
  \bibfield  {author} {\bibinfo {author} {\bibfnamefont {E.}~\bibnamefont {Di~Valentino}} \emph {et~al.},\ }\href@noop {} {\bibfield  {journal} {\bibinfo  {journal} {Physics of the Dark Universe}\ } (\bibinfo {year} {2025}{\natexlab{a}})},\ \Eprint {https://arxiv.org/abs/2504.01669} {arXiv:2504.01669 [astro-ph.CO]} \BibitemShut {NoStop}%
\bibitem [{\citenamefont {Alam}\ \emph {et~al.}(2017)\citenamefont {Alam} \emph {et~al.}}]{alam}%
  \BibitemOpen
  \bibfield  {author} {\bibinfo {author} {\bibfnamefont {S.}~\bibnamefont {Alam}} \emph {et~al.} (\bibinfo {collaboration} {BOSS}),\ }\href {https://doi.org/10.1093/mnras/stx721} {\bibfield  {journal} {\bibinfo  {journal} {Mon. Not. Roy. Astron. Soc.}\ }\textbf {\bibinfo {volume} {470}},\ \bibinfo {pages} {2617} (\bibinfo {year} {2017})},\ \Eprint {https://arxiv.org/abs/1607.03155} {arXiv:1607.03155 [astro-ph.CO]} \BibitemShut {NoStop}%
\bibitem [{\citenamefont {Ata}\ \emph {et~al.}(2018)\citenamefont {Ata} \emph {et~al.}}]{Ata:2017dya}%
  \BibitemOpen
  \bibfield  {author} {\bibinfo {author} {\bibfnamefont {M.}~\bibnamefont {Ata}} \emph {et~al.} (\bibinfo {collaboration} {eBOSS}),\ }\href {https://doi.org/10.1093/mnras/stx2630} {\bibfield  {journal} {\bibinfo  {journal} {Mon. Not. Roy. Astron. Soc.}\ }\textbf {\bibinfo {volume} {473}},\ \bibinfo {pages} {4773} (\bibinfo {year} {2018})},\ \Eprint {https://arxiv.org/abs/1705.06373} {arXiv:1705.06373 [astro-ph.CO]} \BibitemShut {NoStop}%
\bibitem [{\citenamefont {Adame}\ \emph {et~al.}(2025)\citenamefont {Adame} \emph {et~al.}}]{desicollab}%
  \BibitemOpen
  \bibfield  {author} {\bibinfo {author} {\bibfnamefont {A.~G.}\ \bibnamefont {Adame}} \emph {et~al.} (\bibinfo {collaboration} {DESI}),\ }\href {https://doi.org/10.1088/1475-7516/2025/02/021} {\bibfield  {journal} {\bibinfo  {journal} {JCAP}\ }\textbf {\bibinfo {volume} {02}},\ \bibinfo {pages} {021}},\ \Eprint {https://arxiv.org/abs/2404.03002} {arXiv:2404.03002 [astro-ph.CO]} \BibitemShut {NoStop}%
\bibitem [{\citenamefont {Anselmi}\ \emph {et~al.}(2023)\citenamefont {Anselmi}, \citenamefont {Starkman},\ and\ \citenamefont {Renzi}}]{Anselmi2}%
  \BibitemOpen
  \bibfield  {author} {\bibinfo {author} {\bibfnamefont {S.}~\bibnamefont {Anselmi}}, \bibinfo {author} {\bibfnamefont {G.~D.}\ \bibnamefont {Starkman}},\ and\ \bibinfo {author} {\bibfnamefont {A.}~\bibnamefont {Renzi}},\ }\href {https://doi.org/10.1103/PhysRevD.107.123506} {\bibfield  {journal} {\bibinfo  {journal} {Phys. Rev. D}\ }\textbf {\bibinfo {volume} {107}},\ \bibinfo {pages} {123506} (\bibinfo {year} {2023})},\ \Eprint {https://arxiv.org/abs/2205.09098} {arXiv:2205.09098 [astro-ph.CO]} \BibitemShut {NoStop}%
\bibitem [{\citenamefont {Anselmi}\ \emph {et~al.}(2019)\citenamefont {Anselmi}, \citenamefont {Corasaniti}, \citenamefont {Sanchez}, \citenamefont {Starkman}, \citenamefont {Sheth},\ and\ \citenamefont {Zehavi}}]{Anselmi1}%
  \BibitemOpen
  \bibfield  {author} {\bibinfo {author} {\bibfnamefont {S.}~\bibnamefont {Anselmi}}, \bibinfo {author} {\bibfnamefont {P.-S.}\ \bibnamefont {Corasaniti}}, \bibinfo {author} {\bibfnamefont {A.~G.}\ \bibnamefont {Sanchez}}, \bibinfo {author} {\bibfnamefont {G.~D.}\ \bibnamefont {Starkman}}, \bibinfo {author} {\bibfnamefont {R.~K.}\ \bibnamefont {Sheth}},\ and\ \bibinfo {author} {\bibfnamefont {I.}~\bibnamefont {Zehavi}},\ }\href {https://doi.org/10.1103/PhysRevD.99.123515} {\bibfield  {journal} {\bibinfo  {journal} {Phys. Rev. D}\ }\textbf {\bibinfo {volume} {99}},\ \bibinfo {pages} {123515} (\bibinfo {year} {2019})},\ \Eprint {https://arxiv.org/abs/1811.12312} {arXiv:1811.12312 [astro-ph.CO]} \BibitemShut {NoStop}%
\bibitem [{\citenamefont {O'Dwyer}\ \emph {et~al.}(2020)\citenamefont {O'Dwyer}, \citenamefont {Anselmi}, \citenamefont {Starkman}, \citenamefont {Corasaniti}, \citenamefont {Sheth},\ and\ \citenamefont {Zehavi}}]{Anselmi3}%
  \BibitemOpen
  \bibfield  {author} {\bibinfo {author} {\bibfnamefont {M.}~\bibnamefont {O'Dwyer}}, \bibinfo {author} {\bibfnamefont {S.}~\bibnamefont {Anselmi}}, \bibinfo {author} {\bibfnamefont {G.~D.}\ \bibnamefont {Starkman}}, \bibinfo {author} {\bibfnamefont {P.-S.}\ \bibnamefont {Corasaniti}}, \bibinfo {author} {\bibfnamefont {R.~K.}\ \bibnamefont {Sheth}},\ and\ \bibinfo {author} {\bibfnamefont {I.}~\bibnamefont {Zehavi}},\ }\href {https://doi.org/10.1103/PhysRevD.101.083517} {\bibfield  {journal} {\bibinfo  {journal} {Phys. Rev. D}\ }\textbf {\bibinfo {volume} {101}},\ \bibinfo {pages} {083517} (\bibinfo {year} {2020})},\ \Eprint {https://arxiv.org/abs/1910.10698} {arXiv:1910.10698 [astro-ph.CO]} \BibitemShut {NoStop}%
\bibitem [{\citenamefont {Carvalho}\ \emph {et~al.}(2016)\citenamefont {Carvalho}, \citenamefont {Bernui}, \citenamefont {Benetti}, \citenamefont {Carvalho},\ and\ \citenamefont {Alcaniz}}]{carvalho}%
  \BibitemOpen
  \bibfield  {author} {\bibinfo {author} {\bibfnamefont {G.~C.}\ \bibnamefont {Carvalho}}, \bibinfo {author} {\bibfnamefont {A.}~\bibnamefont {Bernui}}, \bibinfo {author} {\bibfnamefont {M.}~\bibnamefont {Benetti}}, \bibinfo {author} {\bibfnamefont {J.~C.}\ \bibnamefont {Carvalho}},\ and\ \bibinfo {author} {\bibfnamefont {J.~S.}\ \bibnamefont {Alcaniz}},\ }\href {https://doi.org/10.1103/PhysRevD.93.023530} {\bibfield  {journal} {\bibinfo  {journal} {Phys. Rev. D}\ }\textbf {\bibinfo {volume} {93}},\ \bibinfo {pages} {023530} (\bibinfo {year} {2016})},\ \Eprint {https://arxiv.org/abs/1507.08972} {arXiv:1507.08972 [astro-ph.CO]} \BibitemShut {NoStop}%
\bibitem [{\citenamefont {Carvalho}\ \emph {et~al.}(2020)\citenamefont {Carvalho}, \citenamefont {Bernui}, \citenamefont {Benetti}, \citenamefont {Carvalho}, \citenamefont {de~Carvalho},\ and\ \citenamefont {Alcaniz}}]{Carvalho2017}%
  \BibitemOpen
  \bibfield  {author} {\bibinfo {author} {\bibfnamefont {G.~C.}\ \bibnamefont {Carvalho}}, \bibinfo {author} {\bibfnamefont {A.}~\bibnamefont {Bernui}}, \bibinfo {author} {\bibfnamefont {M.}~\bibnamefont {Benetti}}, \bibinfo {author} {\bibfnamefont {J.~C.}\ \bibnamefont {Carvalho}}, \bibinfo {author} {\bibfnamefont {E.}~\bibnamefont {de~Carvalho}},\ and\ \bibinfo {author} {\bibfnamefont {J.~S.}\ \bibnamefont {Alcaniz}},\ }\href {https://doi.org/10.1016/j.astropartphys.2020.102432} {\bibfield  {journal} {\bibinfo  {journal} {Astropart. Phys.}\ }\textbf {\bibinfo {volume} {119}},\ \bibinfo {pages} {102432} (\bibinfo {year} {2020})},\ \Eprint {https://arxiv.org/abs/1709.00271} {arXiv:1709.00271 [astro-ph.CO]} \BibitemShut {NoStop}%
\bibitem [{\citenamefont {de~Carvalho}\ \emph {et~al.}(2018)\citenamefont {de~Carvalho}, \citenamefont {Bernui}, \citenamefont {Carvalho}, \citenamefont {Novaes},\ and\ \citenamefont {Xavier}}]{Carvalho2018}%
  \BibitemOpen
  \bibfield  {author} {\bibinfo {author} {\bibfnamefont {E.}~\bibnamefont {de~Carvalho}}, \bibinfo {author} {\bibfnamefont {A.}~\bibnamefont {Bernui}}, \bibinfo {author} {\bibfnamefont {G.~C.}\ \bibnamefont {Carvalho}}, \bibinfo {author} {\bibfnamefont {C.~P.}\ \bibnamefont {Novaes}},\ and\ \bibinfo {author} {\bibfnamefont {H.~S.}\ \bibnamefont {Xavier}},\ }\href {https://doi.org/10.1088/1475-7516/2018/04/064} {\bibfield  {journal} {\bibinfo  {journal} {JCAP}\ }\textbf {\bibinfo {volume} {1804}}\bibfield  {number} {\bibinfo  {number} { (04)},\ \bibinfo {pages} {064}},\ }\Eprint {https://arxiv.org/abs/1709.00113} {arXiv:1709.00113 [astro-ph.CO]} \BibitemShut {NoStop}%
\bibitem [{\citenamefont {Beutler}\ \emph {et~al.}(2011{\natexlab{b}})\citenamefont {Beutler}, \citenamefont {Blake}, \citenamefont {Colless}, \citenamefont {Jones}, \citenamefont {Staveley-Smith}, \citenamefont {Campbell}, \citenamefont {Parker}, \citenamefont {Saunders},\ and\ \citenamefont {Watson}}]{Beutler:2011hx}%
  \BibitemOpen
  \bibfield  {author} {\bibinfo {author} {\bibfnamefont {F.}~\bibnamefont {Beutler}}, \bibinfo {author} {\bibfnamefont {C.}~\bibnamefont {Blake}}, \bibinfo {author} {\bibfnamefont {M.}~\bibnamefont {Colless}}, \bibinfo {author} {\bibfnamefont {D.~H.}\ \bibnamefont {Jones}}, \bibinfo {author} {\bibfnamefont {L.}~\bibnamefont {Staveley-Smith}}, \bibinfo {author} {\bibfnamefont {L.}~\bibnamefont {Campbell}}, \bibinfo {author} {\bibfnamefont {Q.}~\bibnamefont {Parker}}, \bibinfo {author} {\bibfnamefont {W.}~\bibnamefont {Saunders}},\ and\ \bibinfo {author} {\bibfnamefont {F.}~\bibnamefont {Watson}},\ }\href {https://doi.org/10.1111/j.1365-2966.2011.19250.x} {\bibfield  {journal} {\bibinfo  {journal} {Mon. Not. Roy. Astron. Soc.}\ }\textbf {\bibinfo {volume} {416}},\ \bibinfo {pages} {3017} (\bibinfo {year} {2011}{\natexlab{b}})},\ \Eprint {https://arxiv.org/abs/1106.3366} {arXiv:1106.3366 [astro-ph.CO]} \BibitemShut {NoStop}%
\bibitem [{\citenamefont {Ross}\ \emph {et~al.}(2015)\citenamefont {Ross} \emph {et~al.}}]{Ross:2014qpa}%
  \BibitemOpen
  \bibfield  {author} {\bibinfo {author} {\bibfnamefont {A.~J.}\ \bibnamefont {Ross}} \emph {et~al.},\ }\href {https://doi.org/10.1093/mnras/stv154} {\bibfield  {journal} {\bibinfo  {journal} {Mon. Not. Roy. Astron. Soc.}\ }\textbf {\bibinfo {volume} {449}},\ \bibinfo {pages} {835} (\bibinfo {year} {2015})},\ \Eprint {https://arxiv.org/abs/1409.3242} {arXiv:1409.3242 [astro-ph.CO]} \BibitemShut {NoStop}%
\bibitem [{\citenamefont {Eisenstein}\ and\ \citenamefont {Hu}(1998)}]{Eisenstein1998}%
  \BibitemOpen
  \bibfield  {author} {\bibinfo {author} {\bibfnamefont {D.~J.}\ \bibnamefont {Eisenstein}}\ and\ \bibinfo {author} {\bibfnamefont {W.}~\bibnamefont {Hu}},\ }\href {https://doi.org/10.1086/305424} {\bibfield  {journal} {\bibinfo  {journal} {Astrophys. J.}\ }\textbf {\bibinfo {volume} {496}},\ \bibinfo {pages} {605} (\bibinfo {year} {1998})},\ \Eprint {https://arxiv.org/abs/astro-ph/9709112} {arXiv:astro-ph/9709112 [astro-ph]} \BibitemShut {NoStop}%
\bibitem [{\citenamefont {Eisenstein}\ \emph {et~al.}(2005)\citenamefont {Eisenstein} \emph {et~al.}}]{SDSS:2005xqv}%
  \BibitemOpen
  \bibfield  {author} {\bibinfo {author} {\bibfnamefont {D.~J.}\ \bibnamefont {Eisenstein}} \emph {et~al.} (\bibinfo {collaboration} {SDSS}),\ }\href {https://doi.org/10.1086/466512} {\bibfield  {journal} {\bibinfo  {journal} {Astrophys. J.}\ }\textbf {\bibinfo {volume} {633}},\ \bibinfo {pages} {560} (\bibinfo {year} {2005})},\ \Eprint {https://arxiv.org/abs/astro-ph/0501171} {arXiv:astro-ph/0501171} \BibitemShut {NoStop}%
\bibitem [{\citenamefont {Percival}\ \emph {et~al.}(2007{\natexlab{b}})\citenamefont {Percival} \emph {et~al.}}]{Percival:2006gt}%
  \BibitemOpen
  \bibfield  {author} {\bibinfo {author} {\bibfnamefont {W.~J.}\ \bibnamefont {Percival}} \emph {et~al.},\ }\href {https://doi.org/10.1086/510772} {\bibfield  {journal} {\bibinfo  {journal} {Astrophys. J.}\ }\textbf {\bibinfo {volume} {657}},\ \bibinfo {pages} {51} (\bibinfo {year} {2007}{\natexlab{b}})}\BibitemShut {NoStop}%
\bibitem [{\citenamefont {Ross}\ \emph {et~al.}(2013)\citenamefont {Ross} \emph {et~al.}}]{BOSS:2012vpn}%
  \BibitemOpen
  \bibfield  {author} {\bibinfo {author} {\bibfnamefont {A.~J.}\ \bibnamefont {Ross}} \emph {et~al.} (\bibinfo {collaboration} {BOSS}),\ }\href {https://doi.org/10.1093/mnras/sts094} {\bibfield  {journal} {\bibinfo  {journal} {Mon. Not. Roy. Astron. Soc.}\ }\textbf {\bibinfo {volume} {428}},\ \bibinfo {pages} {1116} (\bibinfo {year} {2013})},\ \Eprint {https://arxiv.org/abs/1208.1491} {arXiv:1208.1491 [astro-ph.CO]} \BibitemShut {NoStop}%
\bibitem [{\citenamefont {Raichoor}\ \emph {et~al.}(2024)\citenamefont {Raichoor} \emph {et~al.}}]{DESI:2023tpn}%
  \BibitemOpen
  \bibfield  {author} {\bibinfo {author} {\bibfnamefont {A.}~\bibnamefont {Raichoor}} \emph {et~al.} (\bibinfo {collaboration} {DESI}),\ }\href {https://doi.org/10.3847/1538-3881/ad19c8} {\bibfield  {journal} {\bibinfo  {journal} {Astron. J.}\ }\textbf {\bibinfo {volume} {167}},\ \bibinfo {pages} {62} (\bibinfo {year} {2024})}\BibitemShut {NoStop}%
\bibitem [{\citenamefont {Cole}\ \emph {et~al.}(2005)\citenamefont {Cole} \emph {et~al.}}]{Cole:2005sx}%
  \BibitemOpen
  \bibfield  {author} {\bibinfo {author} {\bibfnamefont {S.}~\bibnamefont {Cole}} \emph {et~al.} (\bibinfo {collaboration} {2dFGRS}),\ }\href {https://doi.org/10.1111/j.1365-2966.2005.09318.x} {\bibfield  {journal} {\bibinfo  {journal} {Mon. Not. Roy. Astron. Soc.}\ }\textbf {\bibinfo {volume} {362}},\ \bibinfo {pages} {505} (\bibinfo {year} {2005})}\BibitemShut {NoStop}%
\bibitem [{\citenamefont {Landy}\ and\ \citenamefont {Szalay}(1993)}]{landy}%
  \BibitemOpen
  \bibfield  {author} {\bibinfo {author} {\bibfnamefont {S.~D.}\ \bibnamefont {Landy}}\ and\ \bibinfo {author} {\bibfnamefont {A.~S.}\ \bibnamefont {Szalay}},\ }\href {https://doi.org/10.1086/172900} {\bibfield  {journal} {\bibinfo  {journal} {Astrophys. J.}\ }\textbf {\bibinfo {volume} {412}},\ \bibinfo {pages} {64} (\bibinfo {year} {1993})}\BibitemShut {NoStop}%
\bibitem [{\citenamefont {Kazin}\ \emph {et~al.}(2011)\citenamefont {Kazin}, \citenamefont {Sánchez},\ and\ \citenamefont {Blanton}}]{Kazin_2011}%
  \BibitemOpen
  \bibfield  {author} {\bibinfo {author} {\bibfnamefont {E.~A.}\ \bibnamefont {Kazin}}, \bibinfo {author} {\bibfnamefont {A.~G.}\ \bibnamefont {Sánchez}},\ and\ \bibinfo {author} {\bibfnamefont {M.~R.}\ \bibnamefont {Blanton}},\ }\href {https://doi.org/10.1111/j.1365-2966.2011.19962.x} {\bibfield  {journal} {\bibinfo  {journal} {Monthly Notices of the Royal Astronomical Society}\ }\textbf {\bibinfo {volume} {419}},\ \bibinfo {pages} {3223–3243} (\bibinfo {year} {2011})}\BibitemShut {NoStop}%
\bibitem [{\citenamefont {Kazin}\ \emph {et~al.}(2013)\citenamefont {Kazin} \emph {et~al.}}]{Kazin:2013rxa}%
  \BibitemOpen
  \bibfield  {author} {\bibinfo {author} {\bibfnamefont {E.~A.}\ \bibnamefont {Kazin}} \emph {et~al.},\ }\href {https://doi.org/10.1093/mnras/stt1261} {\bibfield  {journal} {\bibinfo  {journal} {Mon. Not. Roy. Astron. Soc.}\ }\textbf {\bibinfo {volume} {435}},\ \bibinfo {pages} {64} (\bibinfo {year} {2013})},\ \Eprint {https://arxiv.org/abs/1303.4391} {arXiv:1303.4391 [astro-ph.CO]} \BibitemShut {NoStop}%
\bibitem [{\citenamefont {Sánchez}\ \emph {et~al.}(2011)\citenamefont {Sánchez} \emph {et~al.}}]{sanchez}%
  \BibitemOpen
  \bibfield  {author} {\bibinfo {author} {\bibfnamefont {E.}~\bibnamefont {Sánchez}} \emph {et~al.},\ }\href {https://doi.org/10.1111/j.1365-2966.2010.17679.x} {\bibfield  {journal} {\bibinfo  {journal} {Mon. Not. Roy. Astron. Soc.}\ }\textbf {\bibinfo {volume} {411}},\ \bibinfo {pages} {277} (\bibinfo {year} {2011})},\ \Eprint {https://arxiv.org/abs/1006.3226} {arXiv:1006.3226 [astro-ph.CO]} \BibitemShut {NoStop}%
\bibitem [{\citenamefont {{du Mas des Bourboux}}\ \emph {et~al.}(2017)\citenamefont {{du Mas des Bourboux}} \emph {et~al.}}]{Bourboux:2017cbm}%
  \BibitemOpen
  \bibfield  {author} {\bibinfo {author} {\bibfnamefont {H.}~\bibnamefont {{du Mas des Bourboux}}} \emph {et~al.},\ }\href {https://doi.org/10.1051/0004-6361/201731731} {\bibfield  {journal} {\bibinfo  {journal} {Astron. Astrophys.}\ }\textbf {\bibinfo {volume} {608}},\ \bibinfo {pages} {A130} (\bibinfo {year} {2017})},\ \Eprint {https://arxiv.org/abs/1708.02225} {arXiv:1708.02225 [astro-ph.CO]} \BibitemShut {NoStop}%
\bibitem [{\citenamefont {Evslin}\ \emph {et~al.}(2018)\citenamefont {Evslin}, \citenamefont {Sen},\ and\ \citenamefont {Ruchika}}]{evslin}%
  \BibitemOpen
  \bibfield  {author} {\bibinfo {author} {\bibfnamefont {J.}~\bibnamefont {Evslin}}, \bibinfo {author} {\bibfnamefont {A.~A.}\ \bibnamefont {Sen}},\ and\ \bibinfo {author} {\bibnamefont {Ruchika}},\ }\href {https://doi.org/10.1103/PhysRevD.97.103511} {\bibfield  {journal} {\bibinfo  {journal} {Phys. Rev. D}\ }\textbf {\bibinfo {volume} {97}},\ \bibinfo {pages} {103511} (\bibinfo {year} {2018})},\ \Eprint {https://arxiv.org/abs/1711.01051} {arXiv:1711.01051 [astro-ph.CO]} \BibitemShut {NoStop}%
\bibitem [{\citenamefont {Scolnic}\ \emph {et~al.}(2018)\citenamefont {Scolnic} \emph {et~al.}}]{Scolnic}%
  \BibitemOpen
  \bibfield  {author} {\bibinfo {author} {\bibfnamefont {D.~M.}\ \bibnamefont {Scolnic}} \emph {et~al.},\ }\href {https://doi.org/10.3847/1538-4357/aab9bb} {\bibfield  {journal} {\bibinfo  {journal} {Astrophys. J.}\ }\textbf {\bibinfo {volume} {859}},\ \bibinfo {pages} {101} (\bibinfo {year} {2018})},\ \Eprint {https://arxiv.org/abs/1710.00845} {arXiv:1710.00845 [astro-ph.CO]} \BibitemShut {NoStop}%
\bibitem [{\citenamefont {Jones}\ and\ \citenamefont {Scolnic}(2018)}]{sntable}%
  \BibitemOpen
  \bibfield  {author} {\bibinfo {author} {\bibfnamefont {D.}~\bibnamefont {Jones}}\ and\ \bibinfo {author} {\bibfnamefont {D.}~\bibnamefont {Scolnic}},\ }\href {https://doi.org/10.17909/T95Q4X} {\bibinfo {title} {Catalogs of cosmologically useful type ia supernovae from pan-starrs ("ps1cosmo")}},\ \bibinfo {howpublished} {MAST Archive at Space Telescope Science Institute} (\bibinfo {year} {2018}),\ \bibinfo {note} {available at \url{https://archive.stsci.edu/doi/resolve/resolve.html?doi=10.17909/T95Q4X}}\BibitemShut {NoStop}%
\bibitem [{\citenamefont {Pietrzyński}\ \emph {et~al.}(2019)\citenamefont {Pietrzyński} \emph {et~al.}}]{piet}%
  \BibitemOpen
  \bibfield  {author} {\bibinfo {author} {\bibfnamefont {G.}~\bibnamefont {Pietrzyński}} \emph {et~al.},\ }\href {https://doi.org/10.1038/s41586-019-0999-4} {\bibfield  {journal} {\bibinfo  {journal} {Nature}\ }\textbf {\bibinfo {volume} {567}},\ \bibinfo {pages} {200} (\bibinfo {year} {2019})},\ \Eprint {https://arxiv.org/abs/1903.08096} {arXiv:1903.08096 [astro-ph.GA]} \BibitemShut {NoStop}%
\bibitem [{\citenamefont {Reid}\ \emph {et~al.}(2019)\citenamefont {Reid} \emph {et~al.}}]{reid}%
  \BibitemOpen
  \bibfield  {author} {\bibinfo {author} {\bibfnamefont {M.~J.}\ \bibnamefont {Reid}} \emph {et~al.},\ }\href {https://doi.org/10.3847/2041-8213/ab552d} {\bibfield  {journal} {\bibinfo  {journal} {Astrophys. J. Lett.}\ }\textbf {\bibinfo {volume} {886}},\ \bibinfo {pages} {L27} (\bibinfo {year} {2019})},\ \Eprint {https://arxiv.org/abs/1910.03357} {arXiv:1910.03357 [astro-ph.GA]} \BibitemShut {NoStop}%
\bibitem [{\citenamefont {Riess}\ \emph {et~al.}(2021{\natexlab{a}})\citenamefont {Riess} \emph {et~al.}}]{riess21}%
  \BibitemOpen
  \bibfield  {author} {\bibinfo {author} {\bibfnamefont {A.~G.}\ \bibnamefont {Riess}} \emph {et~al.},\ }\href {https://doi.org/10.3847/2041-8213/abdbaf} {\bibfield  {journal} {\bibinfo  {journal} {Astrophys. J.}\ }\textbf {\bibinfo {volume} {908}},\ \bibinfo {pages} {L6} (\bibinfo {year} {2021}{\natexlab{a}})},\ \Eprint {https://arxiv.org/abs/2012.08534} {arXiv:2012.08534 [astro-ph.CO]} \BibitemShut {NoStop}%
\bibitem [{\citenamefont {Lindegren}\ \emph {et~al.}(2021{\natexlab{a}})\citenamefont {Lindegren} \emph {et~al.}}]{lind20a}%
  \BibitemOpen
  \bibfield  {author} {\bibinfo {author} {\bibfnamefont {L.}~\bibnamefont {Lindegren}} \emph {et~al.},\ }\href {https://doi.org/10.1051/0004-6361/202039709} {\bibfield  {journal} {\bibinfo  {journal} {Astron. Astrophys.}\ }\textbf {\bibinfo {volume} {649}},\ \bibinfo {pages} {A2} (\bibinfo {year} {2021}{\natexlab{a}})},\ \Eprint {https://arxiv.org/abs/2012.01742} {arXiv:2012.01742 [astro-ph.IM]} \BibitemShut {NoStop}%
\bibitem [{\citenamefont {Lindegren}\ \emph {et~al.}(2021{\natexlab{b}})\citenamefont {Lindegren} \emph {et~al.}}]{lind20b}%
  \BibitemOpen
  \bibfield  {author} {\bibinfo {author} {\bibfnamefont {L.}~\bibnamefont {Lindegren}} \emph {et~al.},\ }\href {https://doi.org/10.1051/0004-6361/202039653} {\bibfield  {journal} {\bibinfo  {journal} {Astron. Astrophys.}\ }\textbf {\bibinfo {volume} {649}},\ \bibinfo {pages} {A4} (\bibinfo {year} {2021}{\natexlab{b}})},\ \Eprint {https://arxiv.org/abs/2012.03380} {arXiv:2012.03380 [astro-ph.IM]} \BibitemShut {NoStop}%
\bibitem [{\citenamefont {Lemos}\ \emph {et~al.}(2023)\citenamefont {Lemos}, \citenamefont {Ruchika}, \citenamefont {Carvalho},\ and\ \citenamefont {Alcaniz}}]{lemos}%
  \BibitemOpen
  \bibfield  {author} {\bibinfo {author} {\bibfnamefont {T.}~\bibnamefont {Lemos}}, \bibinfo {author} {\bibnamefont {Ruchika}}, \bibinfo {author} {\bibfnamefont {J.~C.}\ \bibnamefont {Carvalho}},\ and\ \bibinfo {author} {\bibfnamefont {J.}~\bibnamefont {Alcaniz}},\ }\href {https://doi.org/10.1140/epjc/s10052-023-11651-3} {\bibfield  {journal} {\bibinfo  {journal} {Eur. Phys. J. C}\ }\textbf {\bibinfo {volume} {83}},\ \bibinfo {pages} {495} (\bibinfo {year} {2023})},\ \Eprint {https://arxiv.org/abs/2303.15066} {arXiv:2303.15066 [astro-ph.CO]} \BibitemShut {NoStop}%
\bibitem [{\citenamefont {Aiola}\ \emph {et~al.}(2020)\citenamefont {Aiola} \emph {et~al.}}]{aiola}%
  \BibitemOpen
  \bibfield  {author} {\bibinfo {author} {\bibfnamefont {S.}~\bibnamefont {Aiola}} \emph {et~al.} (\bibinfo {collaboration} {ACT}),\ }\href {https://doi.org/10.1088/1475-7516/2020/12/047} {\bibfield  {journal} {\bibinfo  {journal} {JCAP}\ }\textbf {\bibinfo {volume} {12}},\ \bibinfo {pages} {047}},\ \Eprint {https://arxiv.org/abs/2007.07288} {arXiv:2007.07288 [astro-ph.CO]} \BibitemShut {NoStop}%
\bibitem [{\citenamefont {Riess}\ \emph {et~al.}(2019)\citenamefont {Riess}, \citenamefont {Casertano}, \citenamefont {Yuan}, \citenamefont {Macri},\ and\ \citenamefont {Scolnic}}]{Reiss2019}%
  \BibitemOpen
  \bibfield  {author} {\bibinfo {author} {\bibfnamefont {A.~G.}\ \bibnamefont {Riess}}, \bibinfo {author} {\bibfnamefont {S.}~\bibnamefont {Casertano}}, \bibinfo {author} {\bibfnamefont {W.}~\bibnamefont {Yuan}}, \bibinfo {author} {\bibfnamefont {L.~M.}\ \bibnamefont {Macri}},\ and\ \bibinfo {author} {\bibfnamefont {D.}~\bibnamefont {Scolnic}},\ }\href {https://doi.org/10.3847/1538-4357/ab1422} {\bibfield  {journal} {\bibinfo  {journal} {Astrophys. J.}\ }\textbf {\bibinfo {volume} {876}},\ \bibinfo {pages} {85} (\bibinfo {year} {2019})},\ \Eprint {https://arxiv.org/abs/1903.07603} {arXiv:1903.07603 [astro-ph.CO]} \BibitemShut {NoStop}%
\bibitem [{\citenamefont {Riess}\ \emph {et~al.}(2021{\natexlab{b}})\citenamefont {Riess}, \citenamefont {Casertano}, \citenamefont {Yuan}, \citenamefont {Bowers}, \citenamefont {Macri}, \citenamefont {Zinn},\ and\ \citenamefont {Scolnic}}]{Reiss2021}%
  \BibitemOpen
  \bibfield  {author} {\bibinfo {author} {\bibfnamefont {A.~G.}\ \bibnamefont {Riess}}, \bibinfo {author} {\bibfnamefont {S.}~\bibnamefont {Casertano}}, \bibinfo {author} {\bibfnamefont {W.}~\bibnamefont {Yuan}}, \bibinfo {author} {\bibfnamefont {J.~B.}\ \bibnamefont {Bowers}}, \bibinfo {author} {\bibfnamefont {L.}~\bibnamefont {Macri}}, \bibinfo {author} {\bibfnamefont {J.~C.}\ \bibnamefont {Zinn}},\ and\ \bibinfo {author} {\bibfnamefont {D.}~\bibnamefont {Scolnic}},\ }\href {https://doi.org/10.3847/2041-8213/abdbaf} {\bibfield  {journal} {\bibinfo  {journal} {Astrophys. J. Lett.}\ }\textbf {\bibinfo {volume} {908}},\ \bibinfo {pages} {L6} (\bibinfo {year} {2021}{\natexlab{b}})},\ \Eprint {https://arxiv.org/abs/2012.08534} {arXiv:2012.08534 [astro-ph.CO]} \BibitemShut {NoStop}%
\bibitem [{\citenamefont {Pesce}\ \emph {et~al.}(2020)\citenamefont {Pesce} \emph {et~al.}}]{Masers}%
  \BibitemOpen
  \bibfield  {author} {\bibinfo {author} {\bibfnamefont {D.~W.}\ \bibnamefont {Pesce}} \emph {et~al.},\ }\href {https://doi.org/10.3847/2041-8213/ab75f0} {\bibfield  {journal} {\bibinfo  {journal} {Astrophys. J. Lett.}\ }\textbf {\bibinfo {volume} {891}},\ \bibinfo {pages} {L1} (\bibinfo {year} {2020})},\ \Eprint {https://arxiv.org/abs/2001.09213} {arXiv:2001.09213 [astro-ph.CO]} \BibitemShut {NoStop}%
\bibitem [{\citenamefont {Kourkchi}\ \emph {et~al.}(2020)\citenamefont {Kourkchi}, \citenamefont {Tully}, \citenamefont {Anand}, \citenamefont {Courtois}, \citenamefont {Dupuy}, \citenamefont {Neill}, \citenamefont {Rizzi},\ and\ \citenamefont {Seibert}}]{TFR}%
  \BibitemOpen
  \bibfield  {author} {\bibinfo {author} {\bibfnamefont {E.}~\bibnamefont {Kourkchi}}, \bibinfo {author} {\bibfnamefont {R.~B.}\ \bibnamefont {Tully}}, \bibinfo {author} {\bibfnamefont {G.~S.}\ \bibnamefont {Anand}}, \bibinfo {author} {\bibfnamefont {H.~M.}\ \bibnamefont {Courtois}}, \bibinfo {author} {\bibfnamefont {A.}~\bibnamefont {Dupuy}}, \bibinfo {author} {\bibfnamefont {J.~D.}\ \bibnamefont {Neill}}, \bibinfo {author} {\bibfnamefont {L.}~\bibnamefont {Rizzi}},\ and\ \bibinfo {author} {\bibfnamefont {M.}~\bibnamefont {Seibert}},\ }\href {https://doi.org/10.3847/1538-4357/ab901c} {\bibfield  {journal} {\bibinfo  {journal} {Astrophys. J.}\ }\textbf {\bibinfo {volume} {896}},\ \bibinfo {pages} {3} (\bibinfo {year} {2020})},\ \Eprint {https://arxiv.org/abs/2004.14499} {arXiv:2004.14499 [astro-ph.GA]} \BibitemShut {NoStop}%
\bibitem [{\citenamefont {D'Amico}\ \emph {et~al.}(2020)\citenamefont {D'Amico}, \citenamefont {Gleyzes}, \citenamefont {Kokron}, \citenamefont {Markovic}, \citenamefont {Senatore}, \citenamefont {Zhang}, \citenamefont {Beutler},\ and\ \citenamefont {Gil-Marín}}]{BOSS}%
  \BibitemOpen
  \bibfield  {author} {\bibinfo {author} {\bibfnamefont {G.}~\bibnamefont {D'Amico}}, \bibinfo {author} {\bibfnamefont {J.}~\bibnamefont {Gleyzes}}, \bibinfo {author} {\bibfnamefont {N.}~\bibnamefont {Kokron}}, \bibinfo {author} {\bibfnamefont {K.}~\bibnamefont {Markovic}}, \bibinfo {author} {\bibfnamefont {L.}~\bibnamefont {Senatore}}, \bibinfo {author} {\bibfnamefont {P.}~\bibnamefont {Zhang}}, \bibinfo {author} {\bibfnamefont {F.}~\bibnamefont {Beutler}},\ and\ \bibinfo {author} {\bibfnamefont {H.}~\bibnamefont {Gil-Marín}},\ }\href {https://doi.org/10.1088/1475-7516/2020/05/005} {\bibfield  {journal} {\bibinfo  {journal} {JCAP}\ }\textbf {\bibinfo {volume} {05}},\ \bibinfo {pages} {005}},\ \Eprint {https://arxiv.org/abs/1909.05271} {arXiv:1909.05271 [astro-ph.CO]} \BibitemShut {NoStop}%
\bibitem [{\citenamefont {Efstathiou}(2021)}]{Efstathiou}%
  \BibitemOpen
  \bibfield  {author} {\bibinfo {author} {\bibfnamefont {G.}~\bibnamefont {Efstathiou}},\ }\href {https://doi.org/10.1093/mnras/stab1588} {\bibfield  {journal} {\bibinfo  {journal} {Mon. Not. Roy. Astron. Soc.}\ }\textbf {\bibinfo {volume} {505}},\ \bibinfo {pages} {3866} (\bibinfo {year} {2021})},\ \Eprint {https://arxiv.org/abs/2103.08723} {arXiv:2103.08723 [astro-ph.CO]} \BibitemShut {NoStop}%
\bibitem [{\citenamefont {Foreman-Mackey}\ \emph {et~al.}(2013)\citenamefont {Foreman-Mackey}, \citenamefont {Hogg}, \citenamefont {Lang},\ and\ \citenamefont {Goodman}}]{ForemanMackey:2012ig}%
  \BibitemOpen
  \bibfield  {author} {\bibinfo {author} {\bibfnamefont {D.}~\bibnamefont {Foreman-Mackey}}, \bibinfo {author} {\bibfnamefont {D.~W.}\ \bibnamefont {Hogg}}, \bibinfo {author} {\bibfnamefont {D.}~\bibnamefont {Lang}},\ and\ \bibinfo {author} {\bibfnamefont {J.}~\bibnamefont {Goodman}},\ }\href {https://doi.org/10.1086/670067} {\bibfield  {journal} {\bibinfo  {journal} {Publ. Astron. Soc. Pac.}\ }\textbf {\bibinfo {volume} {125}},\ \bibinfo {pages} {306} (\bibinfo {year} {2013})},\ \Eprint {https://arxiv.org/abs/1202.3665} {arXiv:1202.3665 [astro-ph.IM]} \BibitemShut {NoStop}%
\bibitem [{\citenamefont {Bernal}\ \emph {et~al.}(2016)\citenamefont {Bernal}, \citenamefont {Verde},\ and\ \citenamefont {Riess}}]{Bernal:2016gxb}%
  \BibitemOpen
  \bibfield  {author} {\bibinfo {author} {\bibfnamefont {J.~L.}\ \bibnamefont {Bernal}}, \bibinfo {author} {\bibfnamefont {L.}~\bibnamefont {Verde}},\ and\ \bibinfo {author} {\bibfnamefont {A.~G.}\ \bibnamefont {Riess}},\ }\href {https://doi.org/10.1088/1475-7516/2016/10/019} {\bibfield  {journal} {\bibinfo  {journal} {JCAP}\ }\textbf {\bibinfo {volume} {10}},\ \bibinfo {pages} {019}},\ \Eprint {https://arxiv.org/abs/1607.05617} {arXiv:1607.05617 [astro-ph.CO]} \BibitemShut {NoStop}%
\bibitem [{\citenamefont {Di~Valentino}\ \emph {et~al.}(2025{\natexlab{b}})\citenamefont {Di~Valentino} \emph {et~al.}}]{CosmoVerse:2025txj}%
  \BibitemOpen
  \bibfield  {author} {\bibinfo {author} {\bibfnamefont {E.}~\bibnamefont {Di~Valentino}} \emph {et~al.} (\bibinfo {collaboration} {CosmoVerse}),\ }\href@noop {} {\bibfield  {journal} {\bibinfo  {journal} {accepted for publication in PDU}\ } (\bibinfo {year} {2025}{\natexlab{b}})},\ \Eprint {https://arxiv.org/abs/2504.01669} {arXiv:2504.01669 [astro-ph.CO]} \BibitemShut {NoStop}%
\bibitem [{\citenamefont {Ruchika}\ \emph {et~al.}(2025)\citenamefont {Ruchika}, \citenamefont {Musso}, \citenamefont {Alam},\ and\ \citenamefont {Seth}}]{Ruchika_bao}%
  \BibitemOpen
  \bibfield  {author} {\bibinfo {author} {\bibnamefont {Ruchika}}, \bibinfo {author} {\bibfnamefont {M.}~\bibnamefont {Musso}}, \bibinfo {author} {\bibfnamefont {S.}~\bibnamefont {Alam}},\ and\ \bibinfo {author} {\bibfnamefont {R.}~\bibnamefont {Seth}}} (\bibinfo {year} {2025}),\ \bibinfo {note} {unpublished manuscript}\BibitemShut {NoStop}%
\bibitem [{\citenamefont {Chevallier}\ and\ \citenamefont {Polarski}(2001)}]{Polarski}%
  \BibitemOpen
  \bibfield  {author} {\bibinfo {author} {\bibfnamefont {M.}~\bibnamefont {Chevallier}}\ and\ \bibinfo {author} {\bibfnamefont {D.}~\bibnamefont {Polarski}},\ }\href {https://doi.org/10.1142/S0218271801000822} {\bibfield  {journal} {\bibinfo  {journal} {Int. J. Mod. Phys. D}\ }\textbf {\bibinfo {volume} {10}},\ \bibinfo {pages} {213} (\bibinfo {year} {2001})},\ \Eprint {https://arxiv.org/abs/gr-qc/0009008} {arXiv:gr-qc/0009008 [gr-qc]} \BibitemShut {NoStop}%
\bibitem [{\citenamefont {Nunes}\ \emph {et~al.}(2020)\citenamefont {Nunes}, \citenamefont {Bonilla}, \citenamefont {Pan},\ and\ \citenamefont {Saridakis}}]{Nunes1}%
  \BibitemOpen
  \bibfield  {author} {\bibinfo {author} {\bibfnamefont {R.~C.}\ \bibnamefont {Nunes}}, \bibinfo {author} {\bibfnamefont {A.}~\bibnamefont {Bonilla}}, \bibinfo {author} {\bibfnamefont {S.}~\bibnamefont {Pan}},\ and\ \bibinfo {author} {\bibfnamefont {E.~N.}\ \bibnamefont {Saridakis}},\ }\href {https://doi.org/10.1093/mnras/staa2036} {\bibfield  {journal} {\bibinfo  {journal} {Mon. Not. Roy. Astron. Soc.}\ }\textbf {\bibinfo {volume} {497}},\ \bibinfo {pages} {2133} (\bibinfo {year} {2020})}\BibitemShut {NoStop}%
\bibitem [{\citenamefont {Nunes}\ and\ \citenamefont {Bernui}(2020)}]{Nunes2}%
  \BibitemOpen
  \bibfield  {author} {\bibinfo {author} {\bibfnamefont {R.~C.}\ \bibnamefont {Nunes}}\ and\ \bibinfo {author} {\bibfnamefont {A.}~\bibnamefont {Bernui}},\ }\href {https://doi.org/10.1140/epjc/s10052-020-08601-8} {\bibfield  {journal} {\bibinfo  {journal} {Eur. Phys. J. C}\ }\textbf {\bibinfo {volume} {80}},\ \bibinfo {pages} {1025} (\bibinfo {year} {2020})},\ \Eprint {https://arxiv.org/abs/2008.03259} {arXiv:2008.03259 [astro-ph.CO]} \BibitemShut {NoStop}%
\bibitem [{\citenamefont {G\'omez-Valent}\ \emph {et~al.}(2024)\citenamefont {G\'omez-Valent}, \citenamefont {Favale}, \citenamefont {Migliaccio},\ and\ \citenamefont {Sen}}]{Gomez-Valent:2023uof}%
  \BibitemOpen
  \bibfield  {author} {\bibinfo {author} {\bibfnamefont {A.}~\bibnamefont {G\'omez-Valent}}, \bibinfo {author} {\bibfnamefont {A.}~\bibnamefont {Favale}}, \bibinfo {author} {\bibfnamefont {M.}~\bibnamefont {Migliaccio}},\ and\ \bibinfo {author} {\bibfnamefont {A.~A.}\ \bibnamefont {Sen}},\ }\href {https://doi.org/10.1103/PhysRevD.109.023525} {\bibfield  {journal} {\bibinfo  {journal} {Phys. Rev. D}\ }\textbf {\bibinfo {volume} {109}},\ \bibinfo {pages} {023525} (\bibinfo {year} {2024})},\ \Eprint {https://arxiv.org/abs/2309.07795} {arXiv:2309.07795 [astro-ph.CO]} \BibitemShut {NoStop}%
\bibitem [{\citenamefont {Favale}\ \emph {et~al.}(2024)\citenamefont {Favale} \emph {et~al.}}]{favale}%
  \BibitemOpen
  \bibfield  {author} {\bibinfo {author} {\bibfnamefont {A.}~\bibnamefont {Favale}} \emph {et~al.},\ }\href@noop {} {\bibfield  {journal} {\bibinfo  {journal} {arXiv e-prints}\ } (\bibinfo {year} {2024})},\ \Eprint {https://arxiv.org/abs/2405.12142} {arXiv:2405.12142 [astro-ph.CO]} \BibitemShut {NoStop}%
\bibitem [{\citenamefont {Chen}\ \emph {et~al.}(2019)\citenamefont {Chen}, \citenamefont {Huang},\ and\ \citenamefont {Wang}}]{Chen:2018dbv}%
  \BibitemOpen
  \bibfield  {author} {\bibinfo {author} {\bibfnamefont {L.}~\bibnamefont {Chen}}, \bibinfo {author} {\bibfnamefont {Q.-G.}\ \bibnamefont {Huang}},\ and\ \bibinfo {author} {\bibfnamefont {K.}~\bibnamefont {Wang}},\ }\href {https://doi.org/10.1088/1475-7516/2019/02/028} {\bibfield  {journal} {\bibinfo  {journal} {JCAP}\ }\textbf {\bibinfo {volume} {02}},\ \bibinfo {pages} {028}},\ \Eprint {https://arxiv.org/abs/1808.05724} {arXiv:1808.05724 [astro-ph.CO]} \BibitemShut {NoStop}%
\end{thebibliography}%

\end{document}